\documentclass[prd, twocolumn, superscriptaddress, nofootinbib, notitlepage, floatfix, preprintnumbers]{revtex4-1}
\usepackage{xcolor}
\usepackage{graphicx}
\usepackage{wrapfig}
\usepackage{amsmath}
\usepackage{amsfonts}
\usepackage{amssymb}
\usepackage{multirow}
\usepackage{slashed}
\usepackage{physics}
\usepackage[colorlinks=true, linkcolor=blue, citecolor=blue, urlcolor=blue]{hyperref}
\newcommand\befs{\begin{figure*}}
\newcommand\eefs[1]{\label{fig:#1}\end{figure*}}
\newcommand\bef{\begin{figure}}
\newcommand\eef[1]{\label{fig:#1}\end{figure}}
\newcommand\beq{\begin{equation}}
\newcommand\eeq[1]{\label{#1}\end{equation}}
\newcommand\beqa{\begin{eqnarray}}
\newcommand\eeqa[1]{\label{#1}\end{eqnarray}}
\newcommand\bet{\begin{table}}
\newcommand\eet[1]{\label{tb:#1}\end{table}}
\newcommand\bets{\begin{table*}}
\newcommand\eets[1]{\label{tb:#1}\end{table*}}
\newcommand\fgn[1]{Fig.\ \ref{fig:#1}}
\newcommand\eqn[1]{Eq.\ (\ref{#1})}
\newcommand\scn[1]{Section \ref{sec:#1}}
\newcommand\apx[1]{Appendix \ref{sec:#1}}

\begin{document}
\widetext

\title{
Quark distribution inside a pion in many-flavor (2 + 1)-dimensional QCD
using lattice computations: UV listens to IR
}
\author{Nikhil Karthik}
\email{nkarthik.work@gmail.com}
\affiliation{Department of Physics, College of William \& Mary, 300 Ukrop Way, Williamsburg, VA 23185, USA}
\affiliation{Thomas Jefferson National Accelerator Facility, Newport News, VA 23606, USA}
\begin{abstract}
    We study the changes in the short-distance quark structure of
    the Nambu-Goldstone boson when the long-distance symmetry-breaking
    scales are depleted controllably.  We achieve this by studying
    the valence Parton Distribution Function (PDF) of pion in 2+1 dimensional two-color QCD, with
    the number $N$ of massless quarks as the tunable parameter that
    slides the theory from being strongly broken for $N=0$ to the
    conformal window for $N>4$, where the theory is gapped by the
    fixed finite volume.  We perform our study non-perturbatively
    using lattice simulations with $N=0,2,4,8$ flavors of nearly
    massless two-component Wilson-Dirac sea quarks and employ the
    leading-twist formalism (LaMET/SDF) to compute the 
    PDF of pion at a fixed valence mass.
    We find that the relative variations in the first few PDF moments
    are only mild compared to the changes in decay constant, but the
    shape of the reconstructed $x$-dependent PDF itself dramatically
    changes from being broad in the scale-broken sector to being
    sharply peaked in the near-conformal region, best reflected in
    PDF shape observables such as the cumulants.
\end{abstract}

\preprint{JLAB-THY-21-3258}
\date{\today} 
\maketitle

\section{Introduction}\label{sec:intro}

QCD offers a unified theoretical description of the mass-gapped
hadron spectrum in its infrared limit, and of the asymptotically
free quarks and gluons in the short-distance limit. While the
perturbative facet of QCD has been subject to stringent tests in
collider experiments~\cite{dissertori2003quantum}, the only
first-principle field theoretic description of the low-energy
hadronic physics with a controlled continuum limit comes from the
numerical lattice QCD computations.  Even though lattice QCD
reproduces the low-energy behavior of QCD precisely (c.f.,
~\cite{Liu:2016kbb,Padmanath:2019wid,Briceno:2017max} and references
therein), a good theoretical understanding of QCD by abstracting
and characterizing the cause of the complex nonperturbative low-energy
features to few relevant aspects of the theory is sought after.
A fascinating aspect of QCD is the very fact that there
is a non-zero mass-gap in this classically conformal field theory.
Understanding how length-scale emerges in QCD in terms of the
short-distance dynamics of the partons inside the proton and other
hadrons is one way to approach this problem, and will be a key
question that will be studied in the Electron-Ion
Collider~\cite{Accardi:2012qut}.  A starting point in this approach
is to break-down the energy-momentum tensor~\cite{Ji:1994av,Hatta:2020iin}
into the quark and gluon momentum fractions
(c.f.,~\cite{Alexandrou:2017oeh,Alexandrou:2020sml} for recent
results for such a break-down of proton momentum fraction), and the
trace-anomaly parts. This approach thus closely ties the understanding
of emergence of scale in the infrared to the parton distribution
functions (PDF), $f(x)$ of hadrons and their moments, with $x$ being 
the momentum fraction of hadrons carried by a parton.

One of the low-energy scale generating mechanism is the spontaneous
chiral symmetry breaking (SSB) that leads to a dimensionful quark
condensate, with the pion being its Nambu-Goldstone (NG) boson. The
natural question then is whether we can learn about the SSB physics
by studying the quark and gluon structure of pion, which has to be
constrained in such a way as to make it exactly massless in the
chiral limit.  Therefore, not surprisingly, the PDF of pion has been
determined from multiple analyses of experimental data with increasing
levels of sophisticated analysis techniques, processes being included
and at different perturbative
orders~\cite{Badier:1983mj,Betev:1985pf,Conway:1989fs,Owens:1984zj,Sutton:1991ay,Gluck:1991ey,
Gluck:1999xe, Wijesooriya:2005ir,Aicher:2010cb}.  Of special
theoretical interest has been the valence quark distribution and
its puzzling $(1-x)^\beta$ large-$x$ behavior --- whether the value
of $\beta\approx 1$ or $\ge 2$.  These aspects have been extensively
studied through many model
calculations~\cite{Brodsky:1973kr,Aicher:2010cb,Nguyen:2011jy,Chen:2016sno,Cui:2020tdf,Roberts:2020udq,deTeramond:2018ecg,RuizArriola:2002wr,Broniowski:2017wbr,Lan:2019rba,Barry:2018ort,Novikov:2020snp,
Bednar:2018mtf}.  With the rapid progress in the leading-twist
perturbative matching formalisms (LaMET~\cite{Ji:2013dva,Ji:2014gla},
SDF~\cite{Radyushkin:2017cyf,Orginos:2017kos,Braun:2007wv}, good
lattice cross-section using current-current
correlators~\cite{Ma:2014jla,Ma:2017pxb}, and see
reviews~\cite{Constantinou:2020pek,Zhao:2018fyu,Cichy:2018mum,Monahan:2018euv,Ji:2020ect}),
the lattice QCD computations of the pion PDF have been able to weigh
in on the large-$x$
behavior~\cite{Gao:2020ito,Chen:2018fwa,Izubuchi:2019lyk,Joo:2019bzr,Lin:2020ssv,Sufian:2019bol,Sufian:2020vzb}.
While the lattice findings seem to lean closer to $\beta\approx 1$,
the studies~\cite{Gao:2020ito,Sufian:2020vzb} found that variations
in alternate analysis methods could make the results consistent
with 2 as well.  Thus, our understanding of the pion PDF is still
evolving, and will be guided further by some of the upcoming
experiments~\cite{Aguilar:2019teb,Denisov:2018unj} as well as the
future lattice computations.

The aim of this paper is to make use of the leading-twist
formalism and extend it to lattice computations of PDF in a family of
QCD-like theories, with the degree of infrared scale-breaking varying
within the family. By studying how and which aspects of the quark
structure inside the Nambu-Goldstone boson (which we simply call
as the pion) change because of variations in
the infrared scale, induced by the choice of members
in the family of theories, we aim to understand the correlations
between the quark structure of pion and the long-distance
vacuum structure. By such observations on how the PDFs evolve to their
functional forms in the strongly-broken theories as one slowly
turns-on the infrared scales from near-zero values, it also gives us a 
new viewpoint of the nonperturbative origin of the parton distribution.

The model system we choose to work with is the 2+1 dimensional
two-color ($N_c=2$) QCD coupled to even number $N$ of two-component
massless Dirac fermions in a parity-invariant manner.  In a previous
study~\cite{Karthik:2018nzf}, we found the global flavor symmetry
in this system to be spontaneously broken for $N \lesssim 4$, and
conformal in the infrared for $N \ge 8$ with nontrivial infrared
mass anomalous dimension. In order for us to meaningfully talk about
the ground-state with pion quantum number in the infrared across
both the scale-broken and conformal regimes, we perform this study
in a fixed box size and at finite valence quark mass; as an upshot,
an artificially produced mass-gapped system from an underlying CFT
serves as a scientific control to compare a naturally mass-gapped
scale-broken QCD-like system with.  

While reducing the computational
cost of exploratory studies as the present one, the 2+1 dimensional
gauge theories coupled to massless fermions by themselves are being
studied for their unexpected dual
relationships~\cite{Seiberg:2016gmd,Karch:2016sxi,Choi:2018tuh,Sharon:2018apk},
as well as for their condensed-matter physics
applications~\cite{Gazit:2018vsa,Song:2018ial,Ma:2020pjs}, especially
the SU(2) theory being relevant to spin liquids.  As an alternative
proposal to understanding the infrared mass gap to arise from
quark-gluon parton dynamics, the identification of few symmetry-breaking
operators, such as four-Fermi operators and monopole operators that
are naively irrelevant in the UV Gaussian fixed point, but become
relevant in an interacting infrared fixed-point as the cause of the
mass-gap below the conformal window is being pursued in 2+1
dimensions~\cite{Chester:2016ref,Chester:2016wrc,Pufu:2013vpa}.
Thus, the confluence of the recent developments in studying the
quark structure of hadrons using lattice computations, with the
fast pace of theoretical developments in 2+1 dimensional field
theories in the infrared, seems to be a promising avenue to understand
the confinement, symmetry breaking and the quark-gluon interactions
leading to them. We should also point to previous applications of
the LaMET/SDF methodology used in this paper to other QCD-like
systems in
Refs.~\cite{Ji:2018waw,Jia:2018qee,DelDebbio:2020cbz,Kock:2020frx}.

The plan of the paper is as follows. In \scn{3dtheory}, we describe
the aspects of parity-invariant 2+1 dimensional SU(2) theory in the
continuum that are relevant for this paper. In \scn{sop}, we describe
the set-up of the calculation and state the problem addressed in
this paper precisely.  In \scn{ope}, we explain the leading twist
methodology that is used in this paper to obtain PDFs on the lattice,
and also explain its differences from 3+1 dimensional version. In
\scn{latdetail}, we explain the lattice setup and computational
techniques, and in \scn{analysis}, we explain the extraction of
boosted pion bilocal matrix element. These two sections can be
skipped if one is not interested in the technical details. In
\scn{results}, we present the results.

\section{Three-dimensional parity-invariant SU(2) theory and its symmetries}\label{sec:3dtheory}

The zero temperature system is defined on three-dimensional Euclidean
torus of physical extents $\ell_1\times\ell_2\times \ell_3$ with
$\ell_3\gg \ell_1,\ell_2$.  Here, we are using the convention that
$\mu=1,2$ are the spatial $x,y$-directions, and $\mu=3$ is the
temporal $t$-direction.  We will refer to the the aspect ratio of
the spatial slice as $\zeta = \ell_2/\ell_1$.  The parity-invariant
three dimensional QCD consists of SU(2) valued gauge fields coupled
to $N$ flavors of two-component fermions.  Writing the action as
$S=S_g+S_f$, the gauge action is
\beq
S_g = \frac{1}{4g^2}\sum_{\mu,\nu=1}^3 \int d^3x {\rm Tr}F_{\mu\nu}^2,
\eeq{cgaction}
with $F$ being the SU(2) algebra valued field strength. The important
difference from the 3+1 dimensional QCD is that the gauge
coupling $g^2$ has mass dimension 1, making the theory
super-renormalizable. This makes the scale-setting simpler, as
one simply needs to measure all dimensionful quantities in the
fundamental units of $g^2$. Once UV regulated,
the continuum limit is simply obtained by removing the regulator
at the fixed values of quantities in units of $g^2$. Particularly
for this work, it will also greatly simply our computation of the
PDF compared to 3+1 dimensions. The dimensionful nature of
$g^2$ also makes the theory trivially asymptotically free.

The SU(2) gauge fields $a_\mu$ are coupled to an even number $N=2n$
flavors of Dirac fermions, which are two-component spinors, in a
parity-invariant manner; in order to make the action parity-invariant,
$n$ of the fermion flavors have mass $+m$ and the other $n$ have
mass $-m$. We will refer to the fermions with positive mass as $u$
and those with negative mass as $d$. This is a deliberate choice
to be analogous to the light flavors in 3+1 dimensional QCD. Throughout
this paper, we will simply refer to the two-component Dirac fermions
as {\sl quarks}, to make the connection to the real-world $3+1$
dimensional QCD easier. The $N=2n$ flavor parity-invariant continuum
action is
\beq
S_f = \sum_{i=1}^{n}\overline{u}_i(\slashed{D}+m) u_i + \sum_{i=1}^n \overline{d}_i(\slashed{D}-m) d_i;\quad \slashed{D}=\sum_{\mu=1}^3\sigma_\mu(\partial_\mu+i a_\mu)
\eeq{contferm}
with $\sigma_\mu$ being the three $2\times2$ Pauli matrices. In 2+1
dimensions, the continuum Dirac operator is anti-Hermitian, and
therefore, one can rewrite the Dirac operator that the $d$-quarks
couple to as $-\slashed{D}^\dagger$. This will be the form of the
lattice regulated fermion action. We will use the value of $n$ as
a tunable knob to control the infrared fate of the theory.

The explicitly massive theory has a global Sp$(n)\times$Sp$(n)$
symmetry~\cite{Magnea:1999iv} (with the symplectic group being
special for SU(2) gauge group due to it being pseudo-real. For other
generic color, it becomes U$(n)\times$U$(n)$ symmetry). At the
massless point, the theory has a larger Sp$(N)$ symmetry, which
gets spontaneously broken to Sp$(n)\times$Sp$(n)$ symmetry when $N$
is smaller than some critical flavor
$N<N^*$~\cite{Pisarski:1984dj,Vafa:1984xh}. The conformal window
extends for all $N$ above $N^*$.  Indications of non-zero value of
$N^*$ have been seen in previous large-$N$ Schwinger-Dyson
Equation study~\cite{Appelquist:1989tc} and in $\epsilon$-expansion
calculation~\cite{Goldman:2016wwk}.  In a previous lattice
study~\cite{Karthik:2018nzf} to determine $N^*$, we found that it
is likely for $N^*$ to lie somewhere between 4 and 6, with $N=8$
showing strong evidences in the finite-size scaling of low-lying
Dirac eigen values for being scale-invariant in the infrared, with
mass-anomalous dimension $\gamma_m\approx 0.4$.  In the scale-broken
side for $N< N^*$, the theory develops a scalar condensate, $\Sigma
\equiv \left\langle \overline{u}_i u_i - \overline{d}_i d_i\right\rangle$
that sets the infrared scale even after the box size taken to
infinity, and sets the scale for the mass-gaps in the theory; the
hadronic content in the SU(2) theory are mesons of the type
$\overline{q}q$ and diquarks (``baryons") of the type $q^T\tau_2
q$ with $\tau_2$ being the Pauli matrix in color space. In the
scale-broken sector, there will be $4n^2$ Nambu-Goldstone (NG)
modes. Of these, the $2n^2$ NG modes will be the mesons
\beq
\pi^+_{ij} = \overline{d}_j u_i; \quad \pi^-_{ij} = \overline{u}_j d_i
\eeq{pidef}
which we simply refer to as {\sl pions} in this theory, that couple
to the conserved flavor currents ${\cal A}_{\mu,ij}(x)= \overline{d}_j
\sigma_\mu u_i(x)$. Associated with the symmetry-breaking, there
is the pion decay constant~\footnote{The mass-dimension of $F_\pi$
in $d$ space-time dimensions is $(d-2)/2$; $1/2$ in $d=3$. We will
therefore consider $F_\pi^2$ to be the IR scale in the subsequent
sections},
\beq
\langle0| {\cal A}_{\mu,ij}^\mp(0)|\pi^\pm_{ij}; p_\mu\rangle \equiv -i p_\mu F_\pi,
\eeq{fpidef}
with an on-shell pion at momentum $p=(p_1,p_2,E)$.  The remaining
set of NG modes will be of the diquark type $u^T_i \sigma_2\tau_2
d_j$ and  $d^T_i \sigma_2\tau_2 u_j$ that couple to the corresponding
conserved currents. Since these conserved currents and extra NG
modes are very special to the SU(2) theory, we simply consider the
pions $\pi^+$ and $\pi^-$ that exists for any number of color, and
roughly belong to the U$(2n)\to$ U$(n)\times$U$(n)$ symmetry breaking
part of the enlarged Sp$(2n)\to$ Sp$(n)\times$Sp$(n)$ symmetry-breaking
pattern for SU(2) gauge theory~\footnote{At the level of correlators
after Wick contraction of fermions, the diquark correlators can be
seen to be degenerate with the that of mesons.}.

Before ending this section dealing with the system in the continuum,
we discuss the subtlety with parity symmetry in the theory. The spatial
parity ${\cal P}$ acts as
\beqa
x=(x_1,x_2,x_3) &\to& x' = (-x_1,x_2,x_3),\cr
[a_1(x),a_2(x),a_3(x)] &\to& [-a_1(x'),a_2(x'), a_3(x')],\cr
[u_i(x), d_i(x)] &\to& [\sigma_1 u(x'), \sigma_1 d(x') ],\cr
[\overline{u}_i(x), \overline{d}_i(x) ] &\to& [-\overline{u}_i(x')\sigma_1, -\overline{d}_i(x')\sigma_1].
\eeqa{sparity}
For a single two-component Dirac fermion, $N=1$, this symmetry is
broken by the mass term and also becomes anomalous in the massless
limit~\cite{Niemi:1983rq,Redlich:1983dv}. While it appears as though
this is not a symmetry of~\eqn{contferm} with even $N$, in fact,
it is a symmetry once the fermions are integrated out, or the
symmetry can be made more obvious by performing a parity ${\cal P}$
transformation along with a pairwise flavor permutation, ${\cal G}:
[u_i(x), d_i(x)]\to [d_i(x'),u_i(x')]$. This ${\cal G}{\cal P}$
operation is usually referred to as the parity in the literature on
parity-invariant theories in 2+1 dimensions.  Since we are interested
in hadron spectroscopy, it is easier to consider the usual notion
of spatial parity ${\cal P}$ above, and whether bilinears are odd
or even under it.  As in 3+1 dimensions, the pions $\pi$ and the
corresponding current ${\cal A}_\mu$ in 2+1 dimensions are
pseudo-scalars and axial-vectors under the spatial parity ${\cal
P}$ (However under ${\cal G}{\cal P}$, the bilinears can be linearly
combined to become even under it, but this does not play any role
in our further discussions.)

\section{Description of method and statement of the problem }\label{sec:sop}

We propose to study the internal quark structure of NG boson as a
function of the varying vacuum structure by varying the number of
massless fermion flavors, wherein the theory moves from being
scale-broken to being conformal in the infrared. In this section
we first discuss how to prepare a well-defined massive valence pion
on top of a vacuum containing massless sea quarks, and then
define its valence PDF which we will use to characterize the pion
quark structure.

\subsection{Setting-up the computation such as to ensure non-zero mass-gaps for all $N$}

In the scale-broken side of small $N$, the infrared content of the
theory are the mass-gapped hadrons, with the typical gaps, denoted
by $M_H$, set by the IR scales such as the condensate $\Sigma$, the
decay constant $F_\pi$, and in the case of pure-gauge theory, the
string tension $\sigma$. These non-zero gaps survive the thermodynamic
and massless quark limit.  On the other hand, in the conformal side
of the theory, the eigenstates of the Hamiltonian are gappless and
continuous in the thermodynamic limit.  We need to deform the theory
to introduce a mass-gapped spectrum in order for us to address the
quark structure of a distinct ground state for any number of flavors.
Such a deformation would only be a sub-leading correction in the
scale-broken side, but will be the leading contribution in the
conformal side.  One can introduce the non-zero mass-gap (1) by
studying the theory at finite spatial volume, where the box size
$\ell$ sets the infrared scale~\cite{Thomson:2016ttt}. That is, the
mass-gaps become $M_H(\ell,\zeta) = c(\zeta)/\ell$ where $\zeta$
is the aspect ratio of the two-dimensional spatial torus, and $c$
is some function of $\zeta$.  (2) by introducing finite quark mass
$m$ in the theory~\cite{DelDebbio:2010ze}, so that all the masses
receive finite mass corrections with $M_H(m)$.

In this work, we will take a hybrid approach, which is easier to
implement in a lattice calculation, than being theoretically pristine.
In order to capture the effect of $N$ flavors of fermions without
them getting decoupled in the infrared, we sample the gauge
configurations in the theory coupled to $N$ massless fermions in a
finite spatial volume, $\ell^2$ with $\zeta=1$. In the lattice
theory terminology, the sea quarks are massless. On the gauge fields
sampled this way, we construct pion states $\pi_{ij}$ built out of
$u_i$ and $d_j$ quarks which have a finite quark mass $m$, which
is tuned so that the mass of pion for any flavor $N$ in $\ell^2$
spatial volume is an arbitrary chosen value. In the lattice
terminology, the valence pion is made massive by tuning the mass
of the valence quark masses to non-zero values. This approach has
the advantages that (a) the depletion of the infrared scales is
preserved due to the presence of massless fermions, and (b) the
pion is massive even in the scale-broken side which makes computation
of matrix elements feasible without large periodicity
effects~\cite{Gao:2020ito}, which is a technical boon.

We chose the mass of the valence pion, $M^{\rm val}_\pi=0.53 g^2$,
and kept this fixed across all $N$.  The reason for this value being
that the pion is light enough to have the chiral properties in the
scale-broken side, whereas in the conformal side, it will ensure
that the mass  $M^{\rm val}_\pi(m,\ell,\zeta)$, which is now a
function of valence quark mass and spatial volume, is dominated by
the finite $m$.  This preference is because we will use hadrons
boosted in the $x$-direction in our computation of PDF, which will
cause a Lorentz {\sl expansion} of extent $\ell_1$ to $\gamma\ell_1$
in the rest frame of that state, which effectively will decrease
the aspect ratio $\zeta$ to $\zeta/\gamma$ in that state's
rest-frame~\cite{Rummukainen:1995vs}. Our choice of $M_\pi^{\rm
val}$ is to minimize the effect of this variation in $\zeta$ on the
energy-momentum dispersion for the ground state in the IR conformal
sector.  One could also use lattices with small $\zeta$ (i.e.,
$\ell_1\gg \ell_2$) to begin with, but we realized it post facto
and intend to improve the calculation with $\zeta<1$.

\subsection{Definition of $u-d$ and valence PDFs}
Having described the preparation of valence pion state of mass
$M_\pi^{\rm val}$ above, we now specify how to study its valence
PDF, $f_v(x)$, which we will use to characterize the UV quark
structure of the pion.  To define a valence PDF, we should first
consider the $u-d$ PDF of the $\pi^+_{ij}=u_i\overline{d}_j$ pion
which has a well defined operator definition as
\beqa
f_{u_i-d_j}(x) &\equiv& \int \frac{d\xi^-}{4\pi} e^{-i x \xi^- P^+} \langle \pi_{ij}; P| O_{\sigma^+}|  \pi_{ij}; P \rangle,\cr
O_{\sigma^+}(\xi) &=& \sum_{k=1}^n\bigg{(} \overline{u}_k(\xi^-)\sigma^+ W_+(\xi_-,0) u_k(0) 
\cr&&\quad\quad\quad- \overline{d}_k(\xi_-)\sigma^+ W_+(\xi^-,0) d_k(0)\bigg{)}.\cr &&\quad
\eeqa{pdfdef}
Here, the light-cone coordinates $\xi^\pm = (x_3\pm x_1)/\sqrt{2}$,
$\sigma^\pm=(\sigma_3\pm \sigma_1)/\sqrt{2}$ and $W_+(\xi_-,0)$ is
the straight Wilson-line along the light-cone connecting the quark
and anti-quark that are displaced by $\xi^-$.  Roughly speaking,
the bilocal operator $O_{\sigma^+}$ counts the number of massless
$u$ type quark minus the number of $d$ type quark moving long the
light-cone, each carrying a fraction $x$ of the momentum $P^+$.  We
have written the operator $O$ formally to be a singlet in the
unbroken Sp$(n)\times$Sp$(n)$ symmetry but non-singlet in the full
Sp$(2n)$ symmetry. For practical purposes, we can simply speak of
the operator of the type $\overline{u}_i u_i - \overline{d}_j d_j$
for a pion of type $\pi_{ij}$.  Since the magnitude of the quark
masses are all the same as $|m|$, we will drop the indices $i,j$
from $\pi_{ij}$ and $f_{u_i-d_j}(x)$.  The $u-d$ PDF has support
from $x\in [-1,1]$.  The charge conjugation symmetry and the ${\cal
G}{\cal P}$ symmetry ensures that $f_{u-d}(x)=f_{u-d}(-x)$. Thus,
we can write,
\beq
f_{u-d}(x) = \begin{cases}0.5 f_v(x)\quad{\rm for\ } x>0,\cr 0.5 f_v(-x) \quad{\rm for\ } x<0.\end{cases}
\eeq{valdef}
This defines for us the valence PDF $f_v(x)$ of pion defined in
$x\in[0,1]$~\footnote{By defining antiquark distribution
$f_{\overline{q}}(x)=-f_q(-x)$, and using the same symmetry argument, one can
see that $f_v(x)= f_u(x)-f_{\overline{u}}(x)$ for $x>0$.}. Their moments
are defined as 
\beq
\langle x^n \rangle_{u-d}=\int_{-1}^1 x^n f_{u-d}(x) dx;\quad \langle x^n \rangle_v=\int_0^1 x^n f_v(x) dx,
\eeq{momdefine}
respectively.  The even moments $\langle x^{2k}\rangle_{u-d}=\langle
x^{2k}\rangle_v$, but for the odd ones, $\langle x^{2k-1}\rangle_{u-d}=0$
whereas $\langle x^{2k-1}\rangle_v\ne 0$.  Since we can only determine
$f_{u-d}$ via the well-defined operator definition above, we will
be inferring properties of $f_v(x)$ indirectly from $f_{u-d}(x)$
in this paper.

With the set-up and key quantities defined, the precise questions
we want to address are the following. As we increase $N$, the IR
scales will vanish, and can be quantified by how $F_\pi$ decreases.
Is $f_v(x)$ of the pion sensitive to the changes in the symmetry-broken
vacuum  given its role as the NG boson? If so, to what degree the
PDF changes with $F_\pi$ and what aspects of the pion valence PDF
and its moments are sensitive to these changes?

\section{Leading-twist OPE in a planar world}\label{sec:ope}

The defining equation for the PDF involving the quark and antiquark
separated on the light-cone is given in \eqn{pdfdef}. Instead, one
can take the matrix element
\beq
2P^+ {\cal M}(\xi^-,P^+) \equiv  \langle \pi; P| O_{\sigma^+}|  \pi; P \rangle,
\eeq{itddef}
as the defining central object, which is also called as Ioffe-time
distribution~\cite{Braun:1994jq}, 
and one can define the moments $\langle x^k\rangle_{u-d}$
of $u-d$ PDF through its expansion as a function $\nu=P^+\xi^-$, referred to as 
the Ioffe time in the literature,
\beqa
&&{\cal M}(\xi^-,P^+)= {\cal M}(\nu)=\sum_{k=0}^\infty \frac{(-i\xi^- P^+)^k}{k!}  \langle x^k\rangle_{u-d}\qquad{\rm with},\cr
&&\langle \pi; P|\left[\overline{u}\sigma^+ (iD^+)^k u-(u\leftrightarrow d)\right]| \pi; P \rangle \equiv 2P^+(P^+)^k \langle x^k\rangle_{u-d}.\cr&&\quad
\eeqa{opeitd}
Only even $u-d$ moments are non-vanishing in the above equation for
the pion.  Since, the 2+1 dimensional QCD is super-renormalizable,
owing to the dimensionful coupling, there are no UV divergences
once the theory is regularized.  Therefore, there are no additional
scales $\mu$ entering the matrix elements defining $\langle x^k\rangle$
in the equation above, unlike in 3+1 dimensions.  Hence, one can
talk of the PDF without referencing an $\overline{\rm MS}$
renormalization scale of the PDF in 2+1 dimensions, as is the case
in the super-renormalizable 1+1 dimensional QCD as well.

A brute-force Monte Carlo computation of ${\cal M}(\xi^-P^+)$ via simulation in Euclidean space-time is
difficult due to the unequal time separation in the operator
$O_{\sigma^+}$ evaluated within a hadron state (however, there is no fundamental theoretical issue in 
performing the Wick rotation from the Minkowski to Euclidean space-time~\cite{Carlson:2017gpk,Briceno:2017cpo}).  Computing 
the matrix elements of the local operators~\cite{Martinelli:1987zd}
defining the PDF moments in \eqn{opeitd} is one possibility. Another recent
method~\cite{Ji:2013dva,Radyushkin:2017cyf}, which has been proven
to be very successful in 3+1d, is to compute the following equal
time bilocal matrix element of the pion boosted with a momentum
$P=(P_1,0,E(P_1))$,
\beqa
&& 2E {\cal M}^B(z_1,P_1)  \equiv   \langle \pi; P| O_{\sigma_3}(z_1)|  \pi; P \rangle;\qquad{\rm where},\cr
&& O_{\sigma_3}(z_1) = \overline{u}(0)\sigma_3 W_{\hat{1}}(0,z) u(z) - (u\leftrightarrow d),
\eeqa{pitddef}
containing a purely spatial displacement $z=(z_1,0,0)$ of the quark
and anti-quark.  The operator now has $\sigma_3$, along the
$t$-direction instead of the $\sigma^+$ present in \eqn{pdfdef}.
The straight Wilson line along the $x$-direction joining the quark
and antiquark is denoted as $W_{\hat{1}}$.  This equal time matrix
element in 3+1 dimensions has been called quasi-PDF matrix element,
pseudo-ITD matrix element or, simply as Ioffe-time Distribution in
the literature.  In this paper, we simply refer to the equal time
correlation above as the bilocal
quark bilinear matrix element (or simply as bilocal matrix elements)
due to its central role in both quasi- and pseudo- approaches to PDF
from lattice, and the present work can be viewed from any perspective
the reader wants to approach it with.  The OPE of the above equal
time bilocal matrix element~\cite{Izubuchi:2018srq}, arranged by
twist, gives
\beqa
{\cal M}^B(z_1,P_1)&=&\sum_{k=0}^\infty \frac{(-i z_1 P_1)^k}{k!} \langle x^k\rangle_{u-d}\cr&&+{\cal O}\bigg{(}(g^2 z_1)^2, (F^2_\pi z_1)^2, (M_\pi^{\rm val}z_1)^2\bigg{)},
\eeqa{biope}
with the first term being at leading twist, and the rest are higher
twist corrections due to the non-vanishing $P^2$ and $z^2$ present
off the light-cone, unlike in the similar expression \eqn{opeitd}
on the light-cone.  A similar OPE expansion formalism to relate the equal-time and light-cone matrix elements 
was considered before in the context of current-current correlators~\cite{Braun:2007wv}.
The leading twist term is exactly the same as
the one in \eqn{opeitd}. In 3+1 dimensions, the similar
expression~\cite{Izubuchi:2018srq,Orginos:2017kos} will involve a
matching Wilson-coefficient $c_n(z^2\mu^2)$ which is 1 at tree-level
and the terms higher order in coupling capture $\log(z^2\mu^2)$
divergence in the limit of $z^2\to 0$.  In the above expression for
2+1 dimensions, the Wilson coefficients take their tree level value
$c_n=1$, and there are no higher order perturbative corrections to
this tree-level value at leading twist, making it exact at leading
twist.  This peculiarity in 2+1 dimension arises because the coupling
$g^2$ has mass dimension 1, which means that a perturbative correction
to $c_n$ increases the twist of the term occurring in the OPE by
1. Therefore, we have discarded such higher-order terms as $(g^2
z_1)^2$ higher-twist corrections.  Along with such corrections,
there could be other genuine higher twist corrections coming from
higher-dimensional operators that occur in the OPE, which we have
denoted by a $(F^2_\pi z)^2$ corrections.  Even at leading twist,
there will be target mass
corrections~\cite{Chen:2016utp,Radyushkin:2017ffo} coming from the
trace terms which bring factors of $P^2=(M_\pi^{\rm val})^2$.  We
have denoted these as the $(M_\pi^{\rm val}z_1)^2$ corrections
above.

In the above discussion, we have been a little cavalier about the
Wilson-line.  In 3+1 dimensions, the self energy divergence of the
Wilson-loop causes a non-perturbative $\exp(-c z_1/a)$ suppression
of Wilson-line as a function of $z$~\cite{Dotsenko:1979wb,Ishikawa:2017faj,Ji:2017oey}. 
The non-perturbative renormalization
of the bilocal operator removes this nonperturbative $z_1$ dependence.
In 2+1 dimensions, there will instead be $\exp(-c' g^2 z_1)$ behavior
as $g^2$ is dimensionful; one way to justify this is to see that
the set of 1-loop diagram in real-space that contributes to the
$\alpha_s (z_1 a)/a^2$ behavior of the bare Wilson-line in
3+1 dimensions, now contributes $g^2 (z_1 a)/a$; where, the
factor of $(z_1 a)$ in both the dimensions comes from the integral measure
when the end-points of the gluon loop on the
Wilson-line become nearly coincident, whereas, the other $1/a^2$
factor in 3+1 dimensions comes from $|z|^{-2}$ gauge field
propagator, and similarly the $1/a$  factor in 2+1 dimensions comes
from the corresponding $|z|^{-1}$ gauge field propagator. The
residual $\exp(-c' g^2 z)$ effect of the Wilson-loop insertion,
which is hadron momentum independent, is then a non-perturbative
higher-twist effect, which we remove by forming ratio as done in
3+1 dimensions~\cite{Orginos:2017kos}, namely
\beq
\tilde{\cal M}(z_1,P_1)=\left(\frac{{\cal M}^B(z_1,P_1)}{{\cal M}^B(z_1,0)}\right)\left(\frac{{\cal M}^B(0,0)}{{\cal M}^B(0,P_1)}\right),
\eeq{ritd}
which we expect to converge to leading-twist expansion in \eqn{biope}
better in a moderate range of $z_1$ and $P_1$. The reason for the
second factor in the above equation to ensure $z_1=0$ matrix element
is 1 by definition, since it is the charge of the pion.  From the
OPE, one can obtain the light-front Ioffe-time Distribution ${\cal
M}$ from the Euclidean construction $\tilde{\cal M}$ in the limit,
\beq
{\cal M}(\nu) = \lim_{\substack{z_1\to 0, P_1\to \infty\cr P_1 z_1 = \nu}} \tilde{\cal M}(z_1,P_1).
\eeq{lim}
In practice however, we will simply be looking at a set of data
that spans a range of $z_1$ and $P_1$.  If the leading twist expansion
works, we expect a scaling $\tilde{\cal M}(z_1,P_1) = \tilde{\cal
M}(z_1 P_1)$ for all $z_1$ and the range of $P_1$ where the scaling
violations from $z_1^2$-type higher twist corrections in \eqn{biope}
are negligible. Based on fits of \eqn{biope} to a subset of data
where the leading twist OPE works the best, we will be able to infer
${\cal M}(\nu)$, and the PDF and its moments.

\section{Lattice methodology and technical details}\label{sec:latdetail}
In this section, we detail the lattice regularization of 2+1
dimensional QCD in a parity-invariant manner, the gauge field
statistics, and the construction of correlators required to build
the pion bilocal matrix element.

We regulate the system defined on $\ell_1\times \ell_2 \times \ell_3$
on a periodic $L_1\times L_2\times L_3$ lattice with isotropic
lattice spacing $a=\ell_\mu/L_\mu$. In this paper, we will be using
$28\times28\times48$ lattices.  The basic gluon object in the
computation are the SU(2) gauge-links, $U_{\mu,x}$ connecting the
lattice site $x$ to $x+\hat\mu$.  The gauge action is the lattice
regulated Wilson single plaquette action,
\beq
S_g = -\frac{\beta}{2}\sum_{\mu>\nu=1}^3\sum_x {\rm Re\ Tr} P_{\mu\nu}(x);\quad \beta=\frac{4}{g^2 a},
\eeq{gaction}
where $P_{\mu\nu}(x)$ is the SU(2) valued plaquette at lattice site
$x=(x_1,x_2,x_3)$. Periodic boundary condition is imposed on all
three directions (an explicit anti-periodic boundary condition in
the temporal direction is superfluous as $-1$ is part of the SU(2)
gauge group). We will use a single fixed lattice spacing $\beta=9.3333$
in this work.  Our choice is based on an observation in the study
of glueballs in 2+1 dimensional pure-gauge SU(2)
theory~\cite{Teper:1998te}, where $L_1=L_2=28$ at this lattice
spacing was found to be close to the thermodynamic limit.

The gauge fields are coupled to a system of $N=2n$ massless fermions,
which we regulate by using two-component Wilson-Dirac
fermions~\cite{Karthik:2018nzf,Karthik:2015sgq,Karthik:2015sza},
defined using the regulated Dirac operator,
\beq
\slashed{D}_w = \slashed{D}_n+ B + m_w,
\eeq{wdirac}
where $\slashed{D}_n$ is the naive Dirac operator,
\beq
\slashed{D}_n = \frac{1}{2} \sum_{\mu=1}^3\sigma_\mu\left(U^{(n)}_{\mu,x}\delta_{x+\hat\mu,y}-U^{(n)\dagger}_{\mu,x-\hat\mu}\delta_{x-\hat\mu,y}\right),
\eeq{naive}
$B$ is the Wilson term,
\beq
B = -3\delta_{x,y}+\frac{1}{2}\sum_{\mu=1}^3 \left(U^{(n)}_{\mu,x}\delta_{x+\hat\mu,y}+U^{(n)\dagger}_{\mu,x-\hat\mu}\delta_{x-\hat\mu,y}\right),
\eeq{wop}
and $m_w$ is the Wilson fermion mass in lattice units. The lattice
fermion is coupled to the gauge fields via $n$-step Stout
smeared~\cite{Morningstar:2003gk} gauge links, $U^{(n)}_{\mu,x}$,
in order to reduce the lattice artifacts coming from irrelevant UV
fluctuations~\cite{Capitani:2006ni,Gupta:2013vha}, with the
identification $U^{(0)}_{\mu,x}=U_{\mu,x}$. We used 1-step stout
smeared links in the Wilson-Dirac operator with optimal value
$\epsilon=0.65$ for the smearing parameter.  The lattice regulated
action that is exactly invariant under spatial parity is
\beq
S_f = \sum_{i=1}^{n}\overline{u}_i\slashed{D}_w u_i - \sum_{i=1}^n \overline{d}_i\slashed{D}^\dagger_w d_i,
\eeq{latfermact}
making the partition function,
\beq
Z = \int [dU] \det\left(\slashed{D}_w\slashed{D}_w^\dagger\right)^{n} e^{-S_g},
\eeq{partfunc}
with a positive definite measure that can be simulated with Monte
Carlo algorithms.  The theory only has an Sp$(n)\times$Sp$(n)$
symmetry even when $m_w$ is tuned to the massless point, and the
full Sp$(2n)$ symmetry will be recovered in the continuum limit.

\subsection{Gauge field generation}
We studied the theories with $N=0,2,4,8$ of approximately massless
Wilson-Dirac (sea) quarks at a fixed lattice spacing corresponding
to $\beta=9.3333$, and using fixed $28^2\times 48$ lattices.  We
generated gauge configurations using the standard Hybrid Monte Carlo
algorithm~\cite{Duane:1987de} using $n$ copies of Gaussian noise
vectors to sample the determinant
$\det\left(\slashed{D}_w\slashed{D}_w^\dagger\right)$.  We tuned
the value of the Wilson mass $m_w$ to the approximate massless point
such that the smallest Dirac eigenvalue $\Lambda_1(m_w)$ has a
minimum at the tuned $m_w$ in the finite fixed volume. Since the
Dirac eigenvalues are gapped in finite volume, the eigenvalues
occurring are not zero at the massless point, and hence makes the
HMC tractable.  For $N=2,4,8$, the values of sea quark mass
$m_w=-0.06836, -0.06513, -0.06060$ respectively.  The details of
the tuning are given in~\cite{Karthik:2018nzf}.  We used Omelyan
integrator~\cite{PhysRevE.65.056706} for the molecular dynamics
(MD) evolution.  The analytical results on the fermion force
calculation for the MD evolution are given, for example,
in~\cite{Morningstar:2003gk,Karthik:2014nta}.  We dynamically tuned
the step size of the integrator such that the acceptance rate was
at least 85\%; in practice the average acceptance was typically
90\% at the different $N$. For thermalization, we discarded the
first 400 trajectories in each stream that were started from random
configurations. After that, gauge configurations every 5 trajectories
were stored and used for correlator measurements.  This way, we
generated 24.5k, 25.2k, 27.3k and 30.2k configurations for $N=0,2,4,8$
flavor respectively. The autocorrelation time is less than 5
trajectories, and to be safe, we used jack-knife blocks of bin size
larger than 20 configurations ($\sim 100$ trajectories).

\subsection{Choice of valence quark masses to create a massive valence pion}

Using the configuration generated with near massless sea quarks,
we tuned the valence Wilson-Dirac quark mass to produce a valence
pion ground state of mass $M_\pi^{\rm val}=0.53 g^2$ in units of
$g^2$, or $M_\pi^{\rm val}a=0.2265$ in lattice units. We performed
this tuning for all $N=0,2,4,8$ flavors, so that the valence pion
mass is held fixed. We performed this analysis by scanning a set
of $m_w$ and interpolating the $M_\pi^{\rm val}$ dependence on
$m_w$, and zero-in on the exact tuned mass value. This way, we found
the tuned valence quark mass corresponding to $0.53 g^2$ pion mass
to be $m^{\rm val}_w=-0.02, -0.01875, -0.0235$ and $-0.051$ for
$N=0,2,4$ and 8 respectively. In the subsequent computations to be
described below for the structure calculation, we used the above
mass in the Wilson fermion propagators $\slashed{D}^{-1}_w$. For
this choice of valence pion mass, the values of $e^{-M_\pi^{\rm
val} (aL_3-2t_s)}=0.0043$ for $t_s=12a$, implying only a small
periodicity effect of 0.4\% when operators are temporally separated
by $12$ lattice units.

\subsection{Two point function computations}
The first step is to find the ground and the excited state contributions
to the pion two-point function.  Since the leading twist formalism
demands boosted pion states, we construct pion sources that project
to definite momentum states. Namely, we construct the two point
functions,
\beqa
C_{\rm 2pt}(t_s;P_1)&=&\left\langle \pi_S(\mathbf{x_0},t_s) \pi^\dagger_S(\mathbf{P},0) \right\rangle,\quad{\rm with},\cr
\pi_S(\mathbf{P},t_s) &=& \sum_{\mathbf{x}} \overline{d}(\mathbf{x},t_s)u(\mathbf{x},t_s) e^{-i\mathbf{P}\cdot\mathbf{x}},
\eeqa{c2pt}
using smeared source $\pi_S(\mathbf{P},0)$ and smeared sink
$\pi_S(\mathbf{P},t_s)$ that project to momentum $\mathbf{P}=(P_1,0)$.
We chose a single source point $\mathbf{x}_0$ per configuration.
We use the momenta,
\beq
aP_1 = \frac{2\pi}{L_1} n_1,
\eeq{momlist}
for $n_1=0,1,2,3,4$. At the given fixed $\beta$, these momenta
correspond to $P_1/g^2 = 0.52, 1.05, 1.57. 2.09$ respectively in
units of coupling $g^2$.  In order to suppress the tower of excited
states, we use Wuppertal smeared quark sources~\cite{Gusken:1989ad}
to construct the pion source and sink. For this, we used 10 steps
of two-dimensional stout smeared links to construct the smearing
kernel with smearing parameter $\epsilon_{\rm 2d}=0.3$ in order to
smoothen the spatial links further. Through a set of tuning runs
at $P_1=0$, we found the optimal number of steps $n_{\rm wup}$ of
Wuppertal smearing to be 80 with Wuppertal smearing parameter
$\delta=0.6$ for $N=0,2$ flavors, for $N=4$ we found $(n_{\rm
wup}=120,\delta=0.6)$, and for $N=8$ we found  $(n_{\rm
wup}=160,\delta=0.6)$, reflecting an increasing effective radius
of pion with increasing number of flavors.  In order to increase
the overlap with the ground state at non-zero momenta, we used
boosted Wuppertal smearing~\cite{Bali:2016lva} built out of quark
sources that are boosted with a quark momentum $k_1=\zeta' P_1$
which then are used to construct the Wuppertal sources (with the
same smearing radius as at $P_1=0$).  We found the optimal boost
parameter $\zeta'$ for $n_1=1,2,3,4$ to be 0.8, 0.8, 0.7, 0.6
respectively.

The fermion contractions to evaluate \eqn{c2pt} are similar to the
3+1 dimensional case, with the thing to remember is $\langle d^a_x
\overline{d}^b_y\rangle =
\left([-\slashed{D}_w^\dagger]^{-1}\right)^{ab}_{xy}$ and $\langle
u^a_x \overline{u}^b_y\rangle =
\left([\slashed{D}_w]^{-1}\right)^{ab}_{xy}$. One can then simply
use $\left([-\slashed{D}_w^\dagger]^{-1}\right)_{xy} =
\left([-\slashed{D}_w]^{-1}\right)_{yx}^{\dagger_{cs}}$ with
$A^{\dagger_{cs}}$ meaning a conjugate transpose of $A$ only over
color-spin space, thereby halving the number of inversions, just
like in 3+1 dimensions. We used Conjugate Gradient algorithm for
inversion here (and also in the HMC) with a stopping criterion of
$10^{-10}$.

\subsection{Three point function computations}

The next important ingredient in the PDF computation is the three-point
function between the pion source, pion sink and the bilocal operator
${\cal O}_{\sigma_3}$,
\beq
C_{\rm 3pt}(t_s,\tau;P_1,z_1)\equiv\left\langle \pi_S(\mathbf{x_0},t_s) O_{\rm \sigma_3}(z_1;\tau) \pi^\dagger_S(\mathbf{P},0)\right\rangle,
\eeq{c3pt}
with the zero momentum projected bilocal operator that is inserted
at a time-slice $\tau$ between the pion source and sink at time-slice
$t_s$,
\beqa
&& O_{\sigma_3}(z_1;\tau) =\sum_{\mathbf{x}}\bigg{(} \overline{u}(x)\sigma_3 W_{\hat{1}}(x,x+{\cal L}) u(x+{\cal L})- \cr && \qquad\overline{d}(x)\sigma_3 W_{\hat{1}}(x,x+{\cal L}) d(x+{\cal L})\bigg{)};\quad x=(\mathbf{x},\tau).
\eeqa{opedef}
Here, the quark and antiquark are separated along the $x$-direction
by ${\cal L}=(z_1,0,0)$, and the operator is made gauge invariant
with the smeared Wilson line, $W_{\hat{1}} = \prod_{x'\in[x,x+{\cal L}]}
U_{1,x'}^{(n)}$.  We used only 2-level stout smeared links $
U_{1,x'}^{(2)}$ for this construction, so as to not risk the smearing
to spoil the UV physics. The pion source and sink are smeared using
the same set of parameters used for the corresponding two-point
function. It should be noted that the $u$ and $d$ quark operators
appearing in $O_{\sigma_3}$  are simple point operators.
\befs
\centering
\includegraphics[scale=0.55]{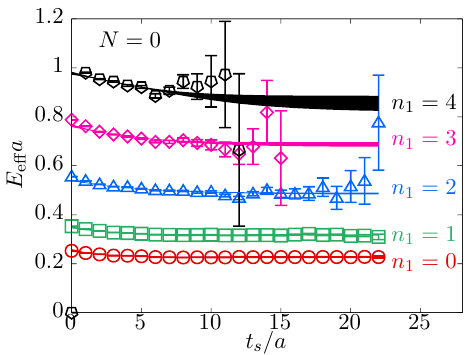}
\includegraphics[scale=0.55]{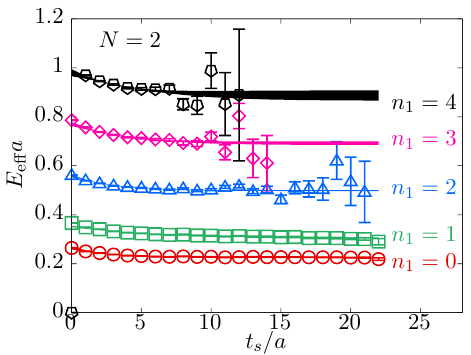}
\includegraphics[scale=0.55]{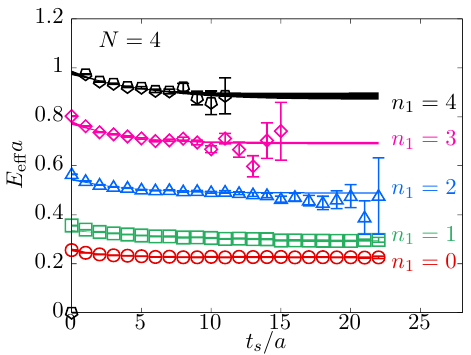}
\includegraphics[scale=0.55]{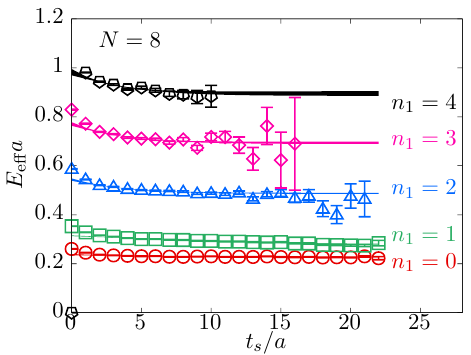}

\includegraphics[scale=0.55]{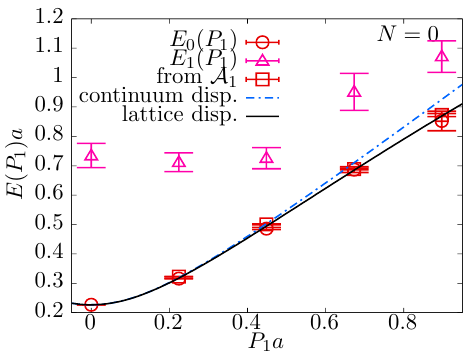}
\includegraphics[scale=0.55]{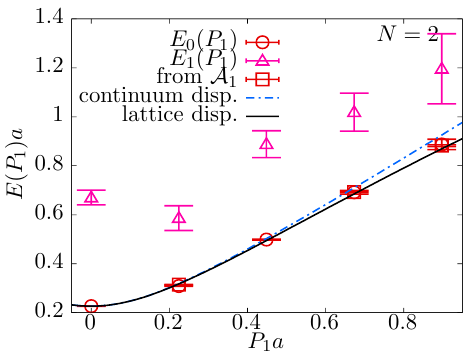}
\includegraphics[scale=0.55]{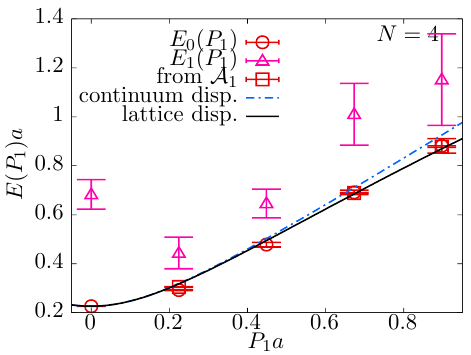}
\includegraphics[scale=0.55]{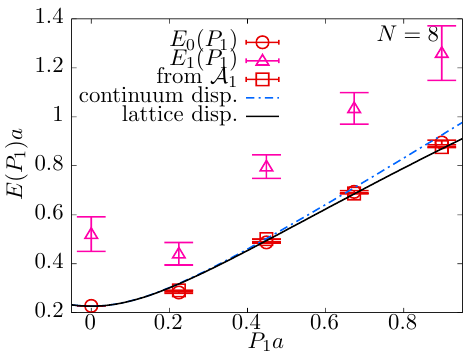}

\caption{
    The spectral content of pion two-point functions in $N=0,2,4,8$
    quark flavors from left to right.  (Top) The effective mass
    from pion two-point function at different momenta (different
    colored symbols) shown as a function of source-sink separations
    $t_s/a$ in lattice units. The bands are the expected effective
    mass from two-state fits to the two-point function.  (Bottom)
    The dispersion relation between ground state (red circles)  and
    the first excited state (purple triangle) energy on boosted
    momentum is shown. For comparison, the expected single particle
    dispersion in the continuum (blue dashed curve)  and on the
    lattice (black curve) are shown.  The ground state pion energy
    extracted from axial-vector ${\cal A}_1$ correlator are also shown at
    non-zero momenta.
}
\eefs{twoptana}

The contractions for the above three-point function were performed
using the sequential-source trick (c.f., appendix
of~\cite{Izubuchi:2019lyk} for details relevant to the bilocal
operator) to take care of the necessary Fourier summation over
two-dimensional time-slice at the sink. It should be noted that,
similar to the $u-d$ three-point function of the pion in 3+1
dimensions, the three-point function is purely real at all $P_1$.
Also, there are no fermion-line disconnected pieces; this comes
non-trivially at finite lattice spacing, by the parity invariance
of the action which guarantees that $\langle {\rm Tr}(\slashed{D}_w^{-1})
\rangle = -\langle {\rm Tr}({\slashed{D}^\dagger_w}^{-1}) \rangle$.
If one used 2+1 dimensional overlap fermions~\cite{Karthik:2016ppr},
the cancellation of disconnected piece would have been on each
configuration.

\subsection{Pion decay constant computations}

We will quantify the presence of infrared scale in the system using
the pion decay constant, $F_\pi$, defined in \eqn{fpidef}. We
extracted this matrix element using the axialvector-pion two-point
function (c.f.,~\cite{Mastropas:2014fsa}),
\beq
C_{\pi-{\cal A}}(t_s)\equiv \left\langle {\cal A}_3(\mathbf{x}_0,t_s) \pi^\dagger_S(\mathbf{P}=0,0)\right\rangle;\ {\cal A}_3(x)=\overline{d}(x)\sigma_3 u(x).
\eeq{fpi2pt}
The pion source was optimally Wuppertal smeared, whereas the current
is constructed out of point quark operators.  We will describe the
analysis of the two-point function leading to $F_\pi$ in a subsequent
section.

\section{Analysis of correlator data to obtain the pion bilocal matrix element \& $F_\pi$}\label{sec:analysis}

\subsection{The spectral content of pion correlator}

The two point function in~\eqn{c2pt} gives information on the
spectrum of states contributing to the pion quantum number, that
we will use to extract the ground state boosted bilocal matrix
element. In the top panels of \fgn{twoptana}, we have shown the
effective masses for the pion at all flavors as a function of
source-sink separation $t_s$, both in lattice units.  For each
flavor, the effective masses at the five momenta are shown by the
different symbols. First, one can see that the ground state displays
a well defined plateau for all $N$, even for $N=8$, thereby
demonstrating the effectiveness of gapping the spectrum by finite
valence quark mass and volume even in the otherwise conformal
infrared theories. We can see that the value of the ground state
mass has been tuned well to be $\approx 0.227a$ in all the theories,
which corresponds to $0.53g^2$ physical mass. The smallest $t_s$
from where one can see a well-defined plateau, at least for the
smallest three momenta, increases with momenta due to the decreasing
gap with the excited state with the boost.  However, we have tuned
the Wuppertal smearing and quark boost parameters precisely to
reduce the amplitudes of the excited state as best we could, and
the any observed deviation from the plateau at smaller $t_s$ was
the best we could reduce it to, without compromising on the noise
at larger $t_s$. Since the range of $t_s$ that we will use to analyze
the three-point function falls in the small $t_s$ region without
the plateau, we need to understand the spectral content of $C_{\rm
2pt}$ better.

\befs
\centering
\includegraphics[scale=0.42]{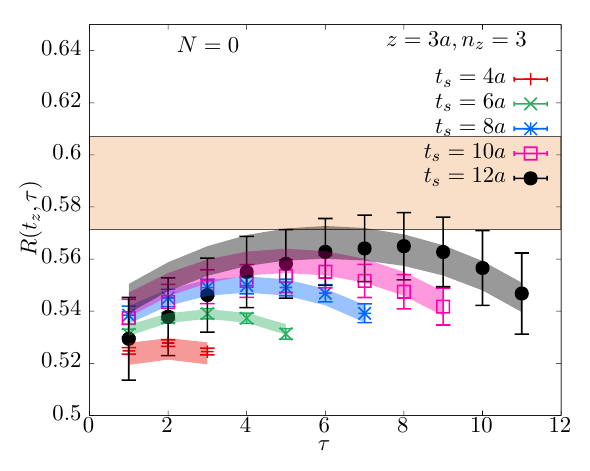}
\includegraphics[scale=0.42]{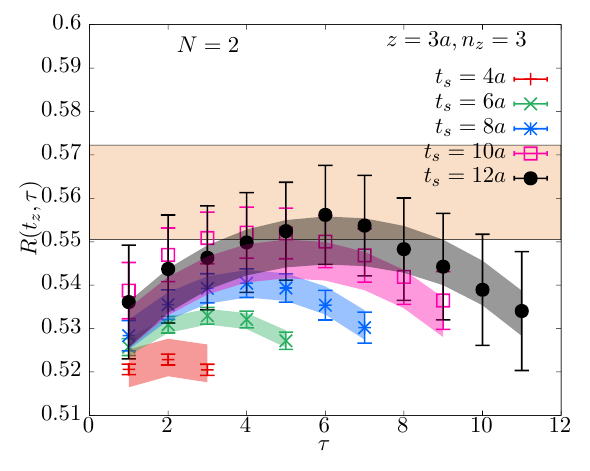}
\includegraphics[scale=0.42]{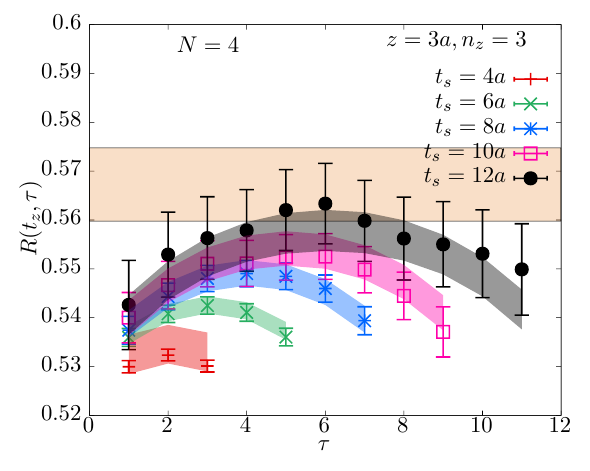}
\includegraphics[scale=0.42]{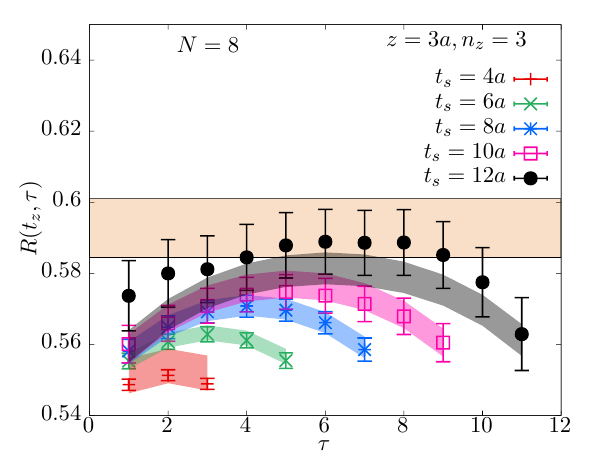}

\includegraphics[scale=0.42]{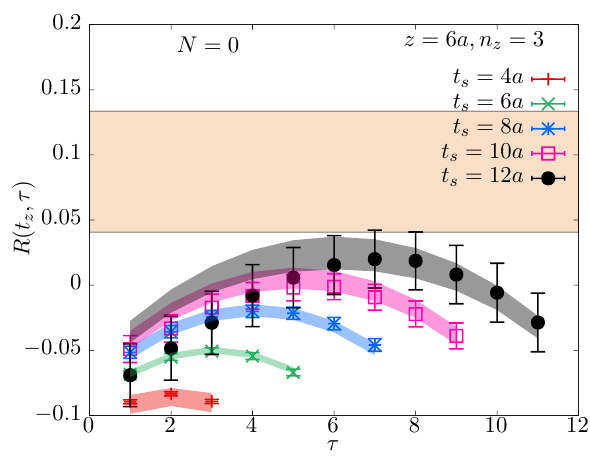}
\includegraphics[scale=0.42]{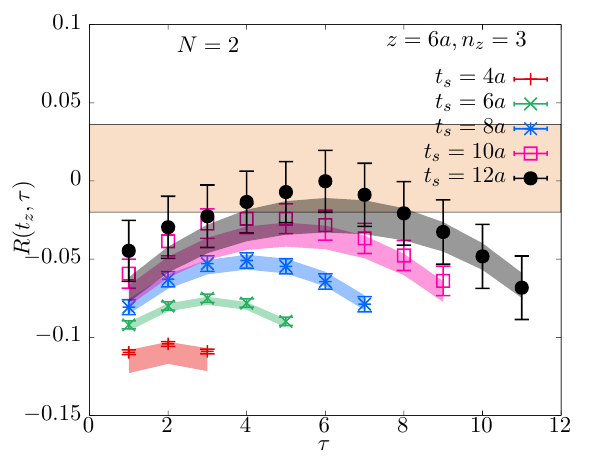}
\includegraphics[scale=0.42]{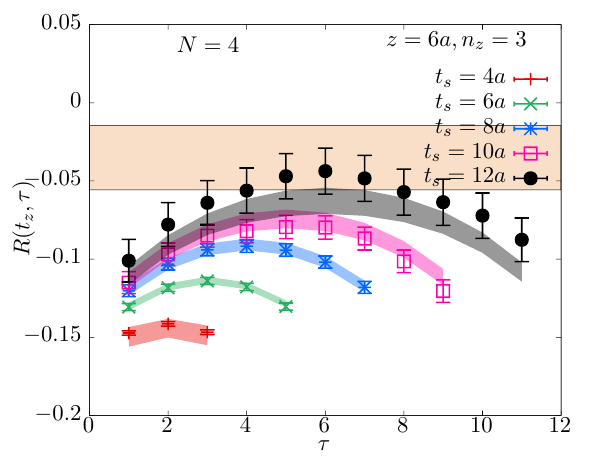}
\includegraphics[scale=0.42]{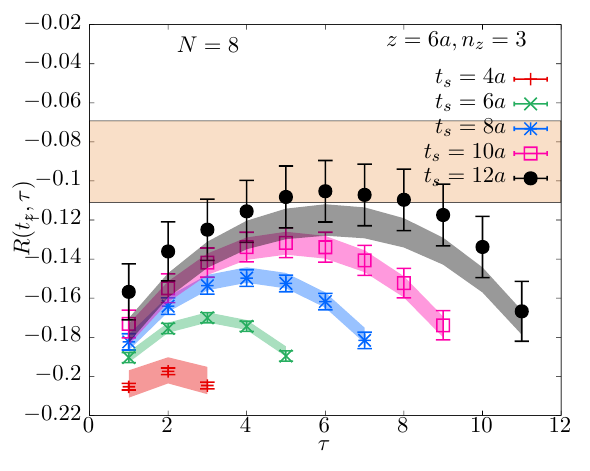}

\caption{
    The estimation of matrix element $h_{00}(z_1,P_1)$ by the two-state
    fits to the 
    ratio, $R(t_s,\tau)$. Some sample fits to the operator insertion
    point, $\tau$, and the source-sink separation, $t_s$, dependence
    of $R$ are shown in the different panels; the top panels are
    at $z_1=3a$ and the bottom ones are $z_1=6a$, at a momentum of
    $n_1=3$. For each $z_1$, the panels from left to right are from
    different number of flavors.
}
\eefs{extpol}

We take the spectral decomposition of $C_{\rm 2pt}$,
\beq
C_{\rm 2pt}(t_s;P_1) =\sum_{i=0}^{N_{\rm state}-1} |A_i|^2 \left( e^{-E_i(P_1) t_s} + e^{-E_i(P_1)\left(aL_3-t_s\right)}\right), 
\eeq{spectrum}
and truncate it at $N_{\rm state}=2$, which is referred to as
the two-state ansatz. We found that this is enough to describe the behavior
of $C_{\rm 2pt}(t_s)$ for all the $P_1$ used, even starting from
$t_s=2a$ and be able to reproduce the value of ground state $E_0$,
as obtained from one-state fit with the minimum $t_s>10a$.  The
uncertainly bands for the effective mass curves for the different
$P_1$ based on the two-state fits over the range
$t_s\in[3a, 24a]$ are also shown in \fgn{twoptana} along with the
data. The quality of the fits are seen to be good, which is also
reflected in $\chi^2/{\rm dof}\approx 1$ for the fits.  We repeated
the two-point function computation using $\langle {\cal A}_1(0) {\cal
A}_1(t_s)\rangle$ correlators also; at $P_1=0$, this gives the mass
of the axial-vector meson, but at non-zero momentum the lowest mass comes
from the pion due to the non-zero overlap $\sim P_1 F_\pi$ with the
lighter pion state. Thus, at non-zero momenta this gave a cross-check
on the determined ground-state values of the pion.

\befs
\centering
\includegraphics[scale=0.48]{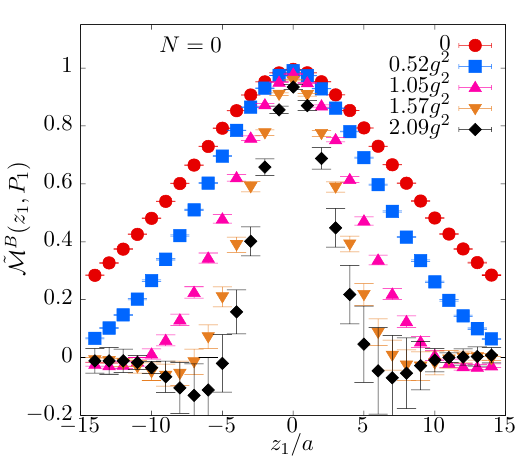}
\includegraphics[scale=0.48]{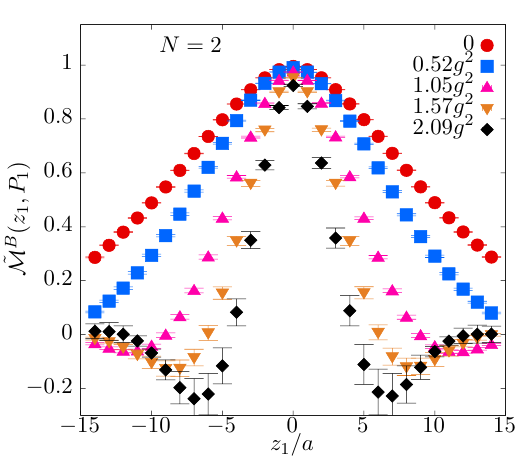}
\includegraphics[scale=0.48]{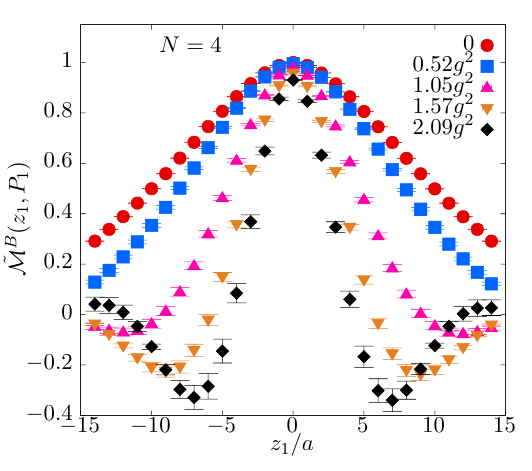}
\includegraphics[scale=0.48]{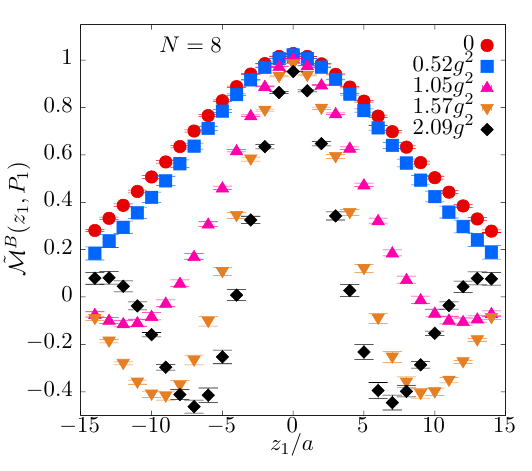}
\caption{
    The matrix elements at $N=0,2,4,8$ (left to right)  obtained
    by extrapolation are shown as a function of quark-antiquark
    separation $z_1/a$. The different colored symbols correspond to
    different fixed momenta $P_1$, which are shown in the legend in units
    of gauge coupling $g^2$.
}
\eefs{hb}

In the bottom panels of \fgn{twoptana}, we have shown the best fit
values of the ground state energy $E_0$ and the first excited state
$E_1$ as a function of boosted momentum $P_1$. The different panels
are again for the five different $N$.  We have compared the data
for $E_0(P_1)$ with the curves for the single particle dispersion
in the continuum, $E_0(P_1)=\sqrt{P_1^2+{M_\pi^{\rm val}}^2}$, shown
as the blue dashed curves. There is a slight discrepancy which
increases with $P_1$, as $P_1a\approx 1$ at the largest momentum
used. We can understand this by, instead comparing the data with
the lattice dispersion, $\cosh(E_0(P_1)a) = \cosh(M_\pi^{\rm
val}a)+1-\cos(P_1a)$. This lattice single particle dispersion curve
is shown as black continuous curve in the figures. The nice agreement
tells us that there are possible lattice corrections at the level
of $3-4\%$ at the highest momenta, which can be reduced in the
future by going to much finer lattices. But this effect will persist
for all the $N$, and therefore, we do not expect this to affect
variations as a function of $N$ that we are interested in.  As a
cross-check, we have also shown the values of ground-state masses
of pion as extracted from the axial-vector ${\cal A}_1$  correlator
at non-zero momenta, which can be seen to agree with the values
from the simple pion correlator. While the slight disagreement with the dispersion curve 
at higher momenta are understood as lattice spacing effect, a slight
disagreement at the level of 4\% is also seen at the smallest
non-zero momentum corresponding to $n_1=1$ for $N=4$ and 8. This
tells us that the valence pion mass for the near-conformal and
conformal theories mildly originate from aspect-ratio($\zeta$)-dependent
$1/\ell$ effect that we described in \scn{sop}, in spite of our
effort to use somewhat larger value of valence quark mass. As the
pion is boosted, the aspect ratio in the boosted frame decreases,
and causes the observed small discrepancy at the smallest non-zero
momentum. At the larger momentum, these aspect-ratio variations are
not important as the leading $E\propto P_1$ relativistic dependence
takes over. Therefore, as a precaution, we will avoid using $n_1=1$
momentum in our analysis of three-point function to avoid systematic
effects.  In a future computation, we aim to rectify this
by using lattices with $\zeta<1$.

In the next section, we will use the extracted energies and amplitudes
in the two-point function to determine the ground state bilocal
matrix element from the three-point functions.

\subsection{The extraction of the pion bilocal matrix element from
three-point function}\label{sec:pi3ptext}

The required ground state matrix element of the bilocal operator
can be obtained from the spectral decomposition of the three-point
function,
\beq
C_{\rm 3pt}(t_s,\tau;P_1,z_1) = \sum_{i,j=0}^{N_{\rm state}-1} A_i^* A_j h_{ij}(z_1,P_1) e^{-E_i(t_s-\tau)-E_j\tau},
\eeq{3ptspec}
with the amplitudes $A_i$ and energies $E_i$, being the same as
obtained from the two-point function analysis. The matrix
elements are the terms $h_{ij}=\langle E_i,P_1|{\cal
O}_{\sigma_3}(z_1)|E_j,P_1\rangle$. Therefore, the leading term
$h_{0,0}$ is the required ground state matrix matrix element
$\langle\pi;P_1|{\cal O}_{\sigma_3}(z_1)|\pi;P_1\rangle$.

We extracted this leading term by fitting the $t_s,\tau$ dependencies
of the actual $C_{\rm 3pt}$ data at various fixed $z_1$ and $P_1$
using the above spectral decomposition truncated to $N_{\rm state}=2$
(since we found that $N_{\rm state}=2$ was able to describe the
corresponding two-point function well even from small $t_s\approx
2-3a$). In practice, we constructed the ratio,
\beq
R(t_s,\tau)\equiv \frac{C_{\rm 3pt}(t_s,\tau)}{C_{\rm 2pt}(t_s)},
\eeq{c3c2ratio}
with the $P_1$ and $z_1$ arguments being the same for both numerator
and denominator, and hence notationally suppressed above. We then
fitted $R(t_s,\tau)$ using the ratio of expressions in \eqn{3ptspec}
and \eqn{spectrum}, with $h_{ij}$ as the fit parameters. We took
the values of the amplitudes $|A_i|$ and energies $E_i$ from the
two-state fit analysis of $C_{\rm 2pt}$ with the fit range over
$t_s\in[3a,24a]$. We used these $(A_i,E_i)$ from the same Jack-knife
blocks as used for the three-point function analysis. We performed
these fits over $\tau\in[2a,t_s-2a]$ to reduce larger excited state
effects for insertion closer to source and sink. Further, we used
all the data for $t_s/a \in[6,8], [6,10], [6,12], [6,12], [6,10]$
for momenta $n_1=0,1,2,3,4$ respectively; we skipped $t_s=10a,12a$
for $n_1=0$ in order to reduce the 0.4\% lattice periodicity effect
due to the smaller $E_0$, and similarly we skipped only $t_s=12a$
for the larger $n_1=1$ momentum.  We did not used $t_s=12a$ for
$n_1=4$ as the two-point function for $t_s>10a$ was very noisy.
While the extrapolated values were insensitive to changes in fitting
windows, we kept the fit systematic fixed for all $N$ so that even
if there is any unnoticed systematic error, it is unlikely to affect the
overall variations in the data as a function of $N$, which is our
interest in this paper.  Such a two-state fit to the ratio $R$
resulted in good fits for all $z_1$ and $P_1$. 

Some sample data for
$R$ along with the results from the fits are shown in \fgn{extpol}
for the case of momentum $n_1=3$. The top and the bottom panels are for
fixed $z_1=3a$ and $6a$ respectively, with the different $N$ shown
in the different columns. The fits, shown as bands, agree with the
data well for all $N$, and the extrapolated value is shown as the
horizontal band.  The ground state matrix element ${\cal
M}^B(z_1,P_1)=h_{00}(z_1,P_1)$ so extracted, are shown as a function
of $z_1/a$ in \fgn{hb}; with each panel for different $N$, and in
each panel, the data for different momenta differentiated by color
and symbols. The data has not been symmetrized by hand with respect
to $z_1$ and $-z_1$, so as to show that the symmetry emerges
automatically, which is a simple cross-check on the computation.
The local matrix element corresponding to 
$z_1=0$ should be precisely be 1 if we had used an exact
conserved current on the lattice, which we have not. So, one sees
the matrix element at $z=0$ to be slightly below 1 at $z=0$ and
this difference with 1 increases with larger momentum; we found
this lattice spacing effect to be of the kind $(a P_1)^2$, which
was also seen in 3+1 dimensional computation~\cite{Gao:2020ito}. We will see
that such effects are nicely canceled by an overall normalization
such that $z=0$ matrix elements are exactly 1; this is justified
since the information on the PDF comes from the variations in $z_1$
and $P_1$, and not by a fixed overall normalization.

\subsection{Determination of pion decay constant}
\bef
\centering
\includegraphics[scale=0.8]{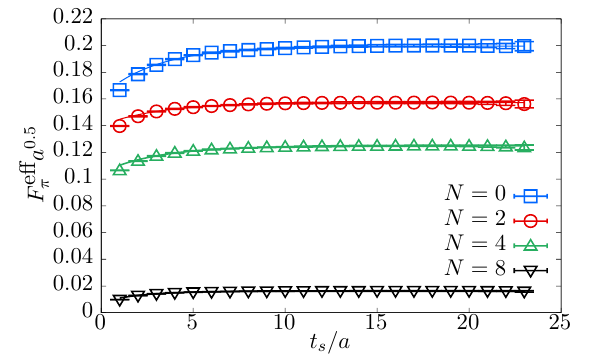}
\caption{
The determination of pion
decay constant $F_\pi$. The data points are the effective $F^{\rm
eff}_\pi(t_s)$ as determined from the $\langle {\cal A}_3 \pi\rangle(t_s)$
correlator, and the curves are fits to $F_\pi+A e^{-\delta m t_s}$.
}
\eef{fpiana}

We determined the pion decay constant through the spectral decomposition
of the correlator $C_{{\cal A}-\pi}$ in \eqn{fpi2pt} as,
\beqa
C_{{\cal A}-\pi}(t_s) &=&\frac{-F_\pi}{\sqrt{2 M_\pi^{\rm val}}} M_\pi^{\rm val} A_0 \left(e^{-M_\pi^{\rm val} t_s} - e^{-M_\pi^{\rm val} (a L_3-t_s)}\right)\cr&&\qquad +A'e^{-E't_s}+\ldots,
\eeqa{fpispec}
which follows from \eqn{fpidef} with $\mu=3$.  The factor of
$\sqrt{2M_\pi^{\rm val}}$ is to convert the lattice normalization
of state vectors to the relativistic one used in defining $F_\pi$.
The factor $A_0$ is the amplitude $\langle 0|\pi^S|\pi\rangle$,
which we take from the smeared-smeared pion two-point function at
zero momentum; we only determine the magnitude of $A_0$, and therefore
we are assuming there is no phase in $A_0$.  The correlator is
anti-periodic in the $t$-direction, which can be seen by a rotation
in the $xt$-plane; taking $(x_1,x_2,x_3)\to (-x_1,x_2,-x_3)$ along
with $u,d\to \sigma_2u,\sigma_2 d$.  The excited state contributions
captured by $e^{-E't_s}$. We fit the above functional form along
with a subleading excited state contribution to the $C_{{\cal A}-\pi}(t_s)$
correlator to determine $F_\pi$. Such fits worked well even starting
from $t_s=2a$ with the fitted value of $F_\pi$ independent of the
fit range. In \fgn{fpiana}, we show an effective $F_\pi^{\rm
eff}(t_s)$ obtained by inverting right-hand side of \eqn{fpispec}
without excited state term to get a $t_s$ dependent value of $F_\pi$.
The curves in the plot are the expectations for $F_\pi^{\rm eff}(t_s)$
from the excited state fits, which can be seen to perform well.
From this analysis, we find $F_\pi a^{0.5} =0.200(1), 0.1574(7),
0.1249(4), 0.0164(1)$ for $N=0,2,4,8$ in lattice units.

\section{Results}\label{sec:results}
We will present the results in the following logical order; first,
we will explain how we measure the presence of infrared scale, by
which we establish that the infrared scales are indeed depleted as
a function of number of massless fermion flavors $N$.  Then, we
will infer the Mellin moments of the PDF and reconstruct the PDF
based on a two-parameter model, and see how the PDF related quantities
change as a function of $N$.  This induces a correlation
between the infrared scale and the PDF parameters, which we look
for.

\subsection{The depletion of infrared scales}
\bef
\centering
\includegraphics[scale=0.8]{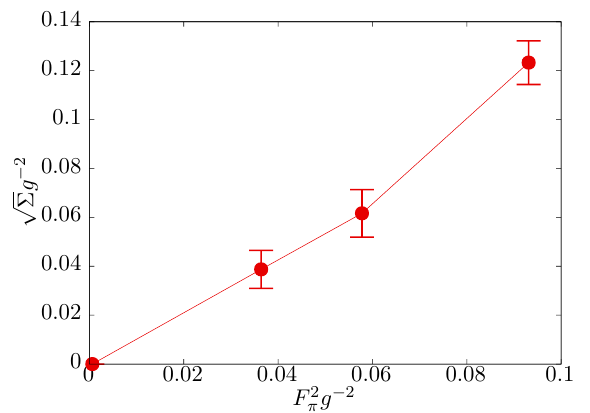}
\includegraphics[scale=0.8]{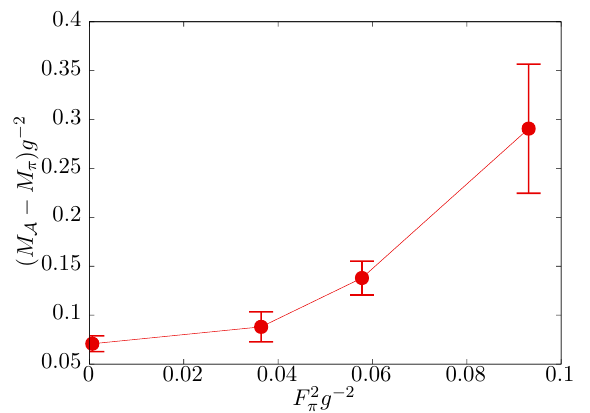}
\caption{
The changes to the infrared scales when the number of massless quark
flavor is changed. (Top panel) the quark condensate as a
function of decay constant, both implicitly depending on the number
of quark flavors.  (Bottom panel) a similar plot showing how
the mass gap between the pion and the axial vector changes as a
function of decay constant.
}
\eef{scales}

For the $N=0$ pure-gauge theory, the confining infrared can simply
be characterized by the string-tension, which takes a value
$\sqrt{\sigma}=0.335 g^2$ for the SU(2) theory~\cite{Teper:1998te}.
For theories with non-zero $N$, string-tension is not a good parameter
to use; instead we use the condensate $\Sigma$ and the decay constant
$F_\pi$.  In a previous study with R. Narayanan~\cite{Karthik:2018nzf},
we measured the scalar condensate $\Sigma$ that breaks the Sp$(N)$
global flavor symmetry, as a function of $N$ in the massless limit
of the theory. For this, we compared the finite-size scaling (FSS)
of the low-lying eigenvalues, $\lambda_i\propto z_i\Sigma^{-1}
\ell^{-3}$ behavior of the eigenvalues in an $\ell^3$ box, where
the proportionality constant $z_i$ are the eigenvalues of the random
matrix model from the non-chiral Gaussian Orthogonal
Ensemble~\cite{Verbaarschot:1994ip}, which shares the same symmetries
as the Dirac operator coupled to SU(2) gauge field in 2+1 dimension.
The coefficient $\Sigma$ is the condensate in the massless limit.
We found non-zero $\Sigma/g^4=0.0152(22),0.0038(12), 0.0025(7)$ and
$0.0(6)10^{-6}$ for $N=0,2,4,8$ flavor respectively. The $N=8$ and
12 theories were instead likely to be infrared conformal with
non-trivial mass anomalous dimensions $\gamma_m=0.38(8)$ and
$\gamma_m=0.48(6)$ respectively; that is, the Dirac eigenvalues
displayed its FSS as $\lambda_i\propto\ell^{-\gamma_m-1}$ rather
than an $\ell^{-3}$ FSS expected from SSB. Here, we should remark
that we found that it was also possible to describe the eigenvalue
FSS in the $N=4$ theory assuming a rather large $\gamma_m=0.6$ along
with additional subleading $1/\ell$ FSS corrections; however, in
light of the results on $F_\pi$ in this work, it appears that the
$N=4$ theory indeed is more likely to be scale-broken in the IR.
The $F_\pi$ we determined here are at finite volume and finite
valence pion mass, but it cannot change the non-zero $F_\pi$ result
for $N=4$ because of the following. One should notice that the very
small $F_\pi^2=6\times10^{-4} g^2$ for $N=8$ is most likely to arise
due to finite volume and valence quark mass, and hence it gives the
typical correction to $F_\pi$ due to these effects; the value of
$F_\pi$ for $N=4$ theory is $F_\pi^2= 0.036 g^2$, which is much
larger than those typical corrections, and hence justifying our
inference about the IR fate of $N=4$. Thus, our current understanding
about the IR fate of many-flavor SU(2) gauge theory is that $N=0,2,4$
are likely to be scale-broken, whereas the $N\ge 8$ are likely to
be conformal in the IR.

The depletion of all the infrared scales due to the monotonic
reduction in condensate is quite apparent. To see this, we plot
different mass-scales, all appropriately casted to have mass-dimension
1, one versus another.  In the top panel of \fgn{scales}, we plot
the mass-scale from condensate, $\sqrt{\Sigma}/g^2$, as a function
of another scale, $F_\pi^2/g^2$.  The two seem to be almost directly
proportional. Another infrared scale one could use is the mass-gap,
$M_{\cal A}-M_\pi^{\rm val}$ between the pion and the axial-vector
meson. In the bottom panel of \fgn{scales}, we correlate this
mass-gap with $F_\pi^2$. Again, it is clear than the mass-splitting
also shrinks with the other diminishing, perhaps a more fundamental
scale, $F_\pi$.  The mass-splitting does not go to zero even for
$N=8$ most likely because of the finite fixed volume and the quark
mass. The one-to-one positive correlation between the infrared
scales also suggests that we can now make the number of flavors
implicit, and simply ask for the effect of reducing one infrared
scale on another, as done in \fgn{scales}. As one would have expected,
a factor reduction in $F_\pi$ induces a reduction in other scales
by a similar factor. We now apply this perspective to quark structure
of pion, where the effect is not obvious.

\subsection{Response of the pion PDF to changes in infrared}
\bef
\centering
\includegraphics[scale=0.8]{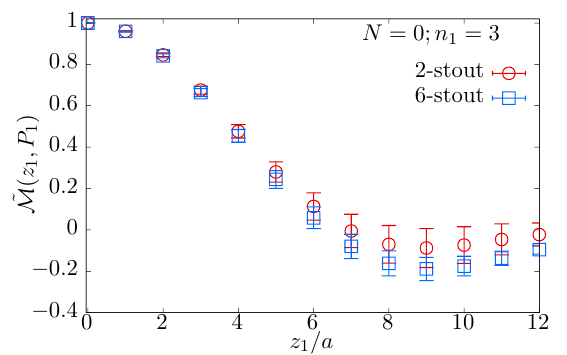}
\caption{A cross-check that at least an application of small number
of Stout smearing to the Wilson-line connecting quark-antiquark is
harmless. The pion bilocal matrix element at $n_1=3$ momentum in $N=0$ theory
with 2-stout and 6-stout smeared Wilson line insertions are compared and
shown to be consistent once the ratio is taken.}
\eef{stout}

The central object in our analysis of PDF is the equal time bilocal
matrix element $\tilde{\cal M}(z_1,P_1)$ in \eqn{ritd}, formed by
taking ratios of the matrix elements ${\cal M}^B(z_1,P_1)$ that we
obtained directly from the three-point function analysis.  We formed
these ratios to remove the presence of $\exp(-c' g^2 z_1)$ behavior
due to the Wilson-line which is present in the definition of the bilocal operator,
and hence ensure a better description by the OPE. In \apx{px0apx},
we describe features of ${\cal M}^B$ itself, and here we proceed
with using the improved $\tilde{\cal M}$.  Through its OPE
in~\eqn{biope}, $\tilde{\cal M}$ contains the leading twist terms
that relate to the pion PDF as well as contribution from operators
with higher-twist, which could have interesting physics in their
own right, but for our purposes here are contaminations.  We can
distill the leading-twist PDF terms in a practical lattice computation,
when  $z_1$ is small and $P_1$ is large, so that one has a finite
range of $z_1P_1$ which can simply be described the lead-twist part
of the OPE in the analysis.  Before going further, we need to first
make sure that the ratio in \eqn{ritd} indeed cancels any remnant
non-perturbative $z_1$-dependent factor from the usage of Wilson line
in the operator.  For this, we performed the computations of
$\tilde{\cal M}(z_1,P_1)$ with 2-stout and 6-stout smeared Wilson lines for
a sample case with $n_1=3$ momentum in the $N=0$ theory. The results
from the two are compared in \fgn{stout}, where one can see a good
agreement between the two, giving confidence that the results are
not stout smearing dependent, at least for few steps of it. The
results get less noisier when number of stout smearing steps
increases, but we use 2-stout in order to be conservative.

\befs
\centering
\includegraphics[scale=0.85]{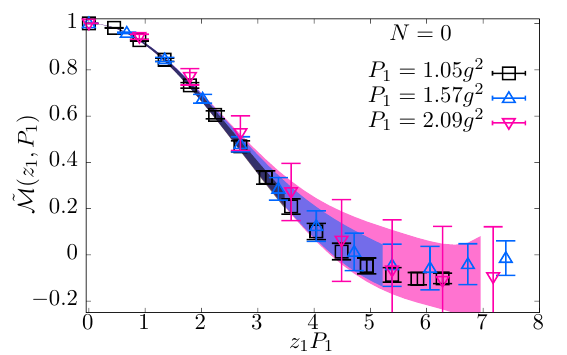}
\includegraphics[scale=0.85]{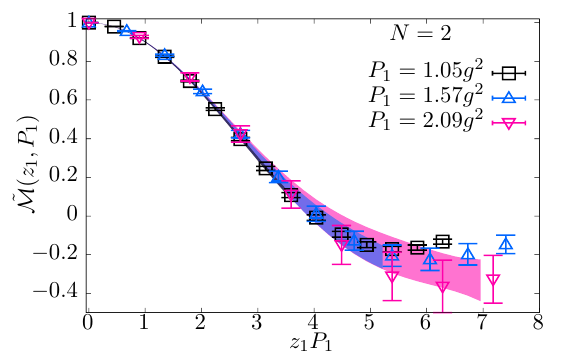}

\includegraphics[scale=0.85]{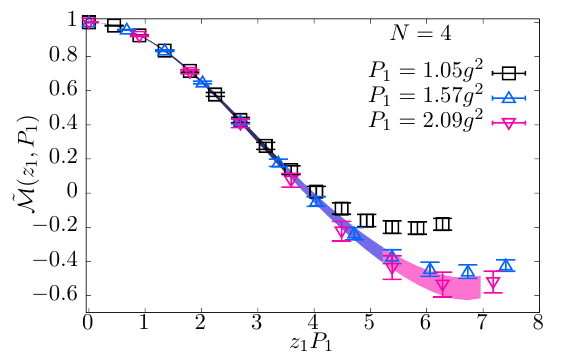}
\includegraphics[scale=0.85]{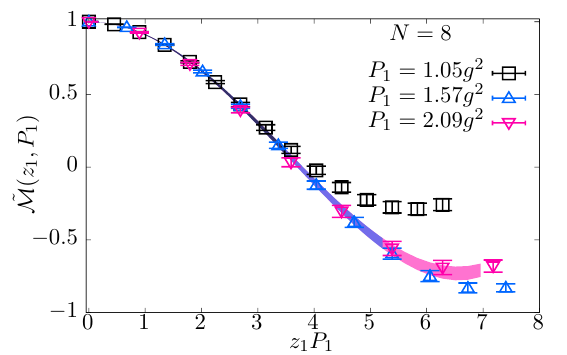}
\caption{
    The pion bilocal quark bilinear matrix elements, $\tilde{{\cal M}}(z_1,P_1)$
    from $N=0,2,4$ and 8 flavor theories are shown in the four panels. In each panel,
    the matrix elements from $P_1=1.05,1.57,2.09 g^2$ are put together and shown 
    as a function of $z_1 P_1$. The color and symbols differentiate the data at fixed 
    $P_1$. The bands are the expectations for $\tilde{{\cal M}}(z_1,P_1)$ based on 
    the fits to the leading twist expression in \eqn{opework}. The bands extend over 
    points included in the fit.
}
\eefs{itdfit}

We perform two types of analysis on $\tilde{\cal M}$, namely, a
model-independent determination of the even moments of the valence
quark PDF, and secondly, by model-dependent reconstruction of
$x$-dependent PDF (even the {\sl model-independent} determination can have additional systematic 
dependences on the modelling of higher-twist effects, lattice corrections, fit ranges, etc).
For both the ways, the starting point is the working
version of the leading twist OPE in \eqn{biope} with some unknown
higher-twist $z_1^2$ corrections, namely,
\beq
\tilde{\cal M}(z_1,P_1)=\frac{1+\left[\sum_{k=1}^{N_{\rm max}}(-1)^k \frac{(z_1 P_1)^{2k}}{(2k)!} \langle x^{2k}\rangle_v\right] + b z_1^2}{1+b z_1^2}.
\eeq{opework}
We have rewritten leading twist part of \eqn{biope} in a different
form above so that is clear that $\langle x^0\rangle=1$, $\tilde{\cal
M}$ is purely real and that only even valence PDF moments $\langle
x^{2k}\rangle_v$ appear. 
These are very specific properties of
$\tilde{\cal M}$ for a pion in 2+1 as well as 3+1 dimensions.  
We have modelled the higher-twist effect in such a way that the 
individual matrix elements ${\cal M}(z_1,P_1)$ and ${\cal M}(z_1,0)$, that 
enter the ratio for $\tilde{\cal M}(z_1,P_1)$, suffer from a momentum independent leading $b z_1^2$ higher-twist 
correction, which leads to the presence of similar term in the numerator and demoninator of \eqn{opework}. Effectively, this causes
a leading $(z_1 P_1)^2 z_1^2$ correction term in the ratio. 
The upper-cutoff of the sum $N_{\rm max}$ is infinity but for practical
implementation, we need to work with smaller $N_{\rm max}$ since
the data is only sensitive to some smaller powers $k$.  Here, the
moments of the PDF are the unknowns we are interested it, but we
will also fit the parameter $b$ to effectively take care any higher
twist $g^2z, F_\pi^2z$ corrections, and also any target mass
corrections.  We also tried adding lattice corrections of the form
$(aP_1)^2$ to the above functional form of OPE~\cite{Gao:2020ito} for $z_1>0$,
but such terms were found to be unnecessary and consistent with
zero well within errors. Therefore, we do not present such an
analysis here. Since this work is at a fixed lattice spacing, any overall $O(a)$
correction that is independent of $z_1$ and $P_1$ cannot quantified, but a correction such as $(a z_1)^2$ can be
absorbed with the $z_1^2$ correction that we have already added.

In any method of analysis, we need to choose the range of $z_1$ and
$P_1$ carefully, since we will not be taking either $z_1\to 0$ or
$P_1\to\infty$ limits, and instead we will simply be fitting the
data which spans a finite range of $z_1$ and $P_1$. First, we will
work with momenta $P_1/g^2\ge 1$ to make sure that for a given
separation $z_1$, a term like $(P_1 z_1)^k$ is larger than a similar
order term $(g^2 z_1)^k$. This leaves the momenta corresponding to
$n_1=2,3,4$.  Through this choice, we are also guaranteed that
$M_\pi^{\rm val}/P_1$ and $F_\pi^2/P_1$ corrections would also be
controlled. For the range of quark-antiquark separation $z_1$, we
have two choices; we might want $z_1 g^2 < 1$ or $z_1 F_\pi^2 < 1$,
where the first factor is simply due to the superficial dimensional
scale in the system and the second is the {\sl natural} infrared
scale.  We assume that the superficial scale will arise simply due
the $\exp(-c g^2 z)$-type Wilson-line term which we find to be
nicely canceled in the ratio $\tilde{\cal M}$. Due to the natural
infrared scales being at least a factor $10$ smaller than $g^2$ for
$N=0$, and even smaller for larger $N$, even a usage of $z=10a$
will only lead to $F_\pi^2 z=0.4$ in this system. Thus, we restricted
ourselves to $z\in[a,8a]$ and change the maximum $z_1$ to $6a$ and
$10a$ to check for the robustness of results. The justifications
for the used ranges of $z_1,P_1$, will also bear out in the data.

\subsubsection{Model-independent inferences}

In the model independent analysis~\cite{Karpie:2018zaz,Gao:2020ito},
we first fit \eqn{opework} to $\tilde{\cal M}(z_1,P_1)$ data over
the specified range of $z_1$ and $P_1$ with the even moments $\langle
x^{2k}\rangle_v$ being the fit parameters.  In addition, we also
fitted the high-twist parameter $b$ to take care leading higher
twist effects; but their values were consistent with zero, and when
we performed the fits without the higher-twist corrections, the
results were consistent with the one including it. Here, we keep
this correction nonetheless. Since the valence quark PDF is positive
(since the anti-$u$ quark and $d$ quark arises only radiatively in
$u\overline{d}$ pion , whereas the $u$ quark is present at tree-level
itself), it imposes a set of inequalities to be satisfied by the
moments as discussed in~\cite{Gao:2020ito}; with the important one
being $\langle x^{k}\rangle_v < \langle x^m\rangle_v$ for $k>m$.
We imposed these constraints in the fit using the methods discussed
in~\cite{Gao:2020ito}.  With such constraints, one can add as many
moments, $N_{\rm max}$, in the analysis without over-fitting the
data, except that it will result in many of the higher-moments,
which the data is not sensitive, to converge to zero. We found that
$N_{\rm max}=5$ was sufficient to describe the data in the range
we fitted, as we describe below.

\bef
\centering
\includegraphics[scale=0.95]{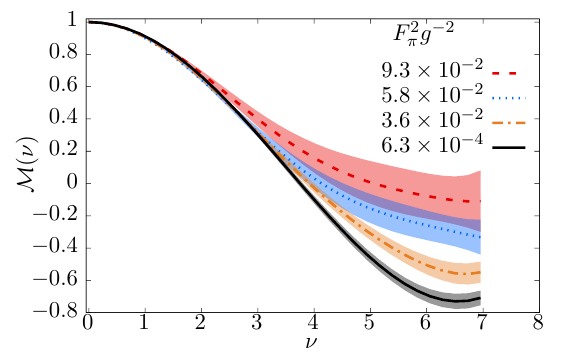}
\caption{The effect of reduction in $F^2_\pi$ on the 
bilocal matrix element (Ioffe-time distribution) ${\cal M}(\nu)$ is shown.
The bands are inferred from the fits to lead-twist expression
in \eqn{opework} and taking its $z_1\to 0, P_1\to\infty$ limit
at fixed $P_1 z_1=\nu$.
The matrix element is shown as a function
of $\nu=P^+\xi^-$ (Ioffe time). 
The different colored bands are
 at different $F_\pi^2 g^{-2}$.
}
\eef{allitd}

In \fgn{itdfit}, we show the data for $\tilde{\cal M}(z_1,P_1)$ as a
function of $z_1 P_1$; the data from different fixed pion momenta
$P_1$ are differentiated by the colored symbols. The four different
panels show the data from $N=0,2,4,8 $ flavors. The data can be
seen to fall on almost universal curves as a function of $z_1P_1$,
which demonstrates the dominance of the lead-twist part, as is 
essential for this work.
As seen by the early peeling-off of $P_1=1.05g^2$ data from the
higher momenta data beyond $P_1 z_1 > 4$ for $N=4,8$ suggests that
the leading twist dominance works better for $N=0$ than for $N=8$.
This could be because the natural higher-twist scale in the broken
phase is $F_\pi^2$ which is smaller than the natural scale corrections
$g^2 z$, and the finite-box scale, $z/\ell$, which could be important
in the conformal phase.  However, for the range of $z_1=6a,8a,10a$,
this ensuing higher twist effect is less important even for $P_1=1.05
g^2$, and definitely not important for higher momenta. This justifies
our choices of fit ranges and the reasoning we presented before.
The data gets increasingly precise with increasing $N$ because the
fluctuations in the gauge field decreases roughly as $1/\sqrt{N}$
for larger $N$.  The bands of various colors in \fgn{itdfit} are the
expectations for ${\cal M}(z_1,P_1)$ from the best fits from the
analysis; the colors match the corresponding color for the momenta
for the data. The bands cover the range of $P_1 z_1$ for each 
$P_1$ for a fixed range of $z_1$ up to $8a$, and for 
$P_1=1.05g^2$ this range is within
the point where the higher twist effects start becoming visible at this
lower momentum. The quality of fits are very good with
resulting $\chi^2/{\rm dof}\approx 0.6$ to 0.8.

\bef
\centering
\includegraphics[scale=0.8]{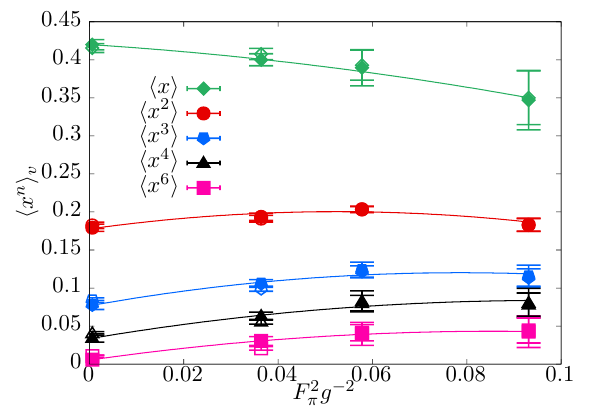}

\includegraphics[scale=0.8]{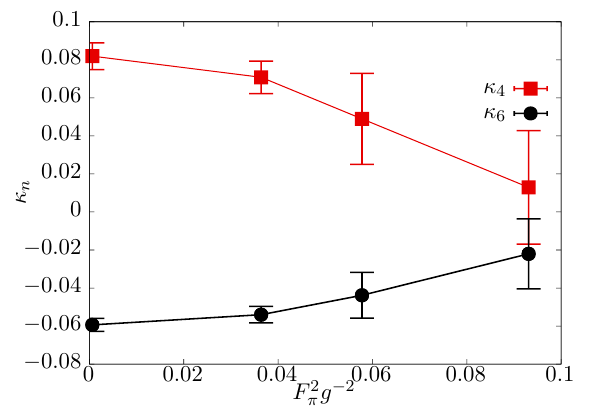}

\caption{(Top) The correlation between decay constant and the valence PDF moments.
The filled symbols are obtained from model-independent fits and the 
open ones from model-dependent PDF ansatz fits. For model-independent
fits only the even moments are directly obtained, whereas the odd moments
$\langle x\rangle_v$ and $\langle x^3\rangle_v$ were obtained indirectly
by definition in \eqn{solve}. The curves are quadratic fits in order to
interpolate the data. (Bottom)  
The correlation between decay constant and the cumulants $\kappa_4$ and 
$\kappa_6$ of $u-d$ PDF.
}
\eef{fpicorrel}

In \fgn{allitd}, we have taken the results of the above model-independent
fits to $\tilde{\cal M}(z_1,P_1)$ and extracted the light-front
bilocal matrix element, ${\cal M}(\nu)$, which is the Ioffe-time
Distribution (ITD). As we discussed, the variation of infrared
scales with $N$, induces a direct infrared scale dependence of
various quantities.  Therefore, in \fgn{allitd}, we have shown the
ITD dependence on the decay constant. This is the main result in
this paper, which we will process further and look at from various
angles.  As the infrared scale-breaking is made stronger, as reflected
in $F_\pi$, the corresponding valence quark ITD starts peeling off
from the large-$N$ conformal curve at shorter and shorter $\nu$.
In this process, however, the ITD remains almost universal up until
$\nu\approx 3$. This tells us that the lowest non-zero $u-d$ moment,
$\langle x^2\rangle_{u-d}= \langle x^2\rangle_v$, must remain quite
insensitive to the scale changes.  Thus, the effect of scale-breaking
seems to be encoded in the fall-off rate of the ITD for $\nu>3$
with an almost fixed lowest even moment.  We can infer simply that
this will reflect in the low-$x$ behavior, which is typically modeled
as a Regge-type $x^\alpha$ behavior, and also in the large-$x$,
$(1-x)^\beta$ behavior of the underlying valence PDF. This is
because, the tail of the ITD typically carries information on the
small-$x$ asymptotic of the PDF, whereas given the inference that
the lowest moment will be almost fixed, will induce a variation in
$\beta$ as well via the implicit relation $\langle
x^2\rangle_v(\alpha,\beta,\ldots)$, with $(\alpha,\beta,\ldots)$
being the parametrization of the shape of the PDF.

In the top panel of \fgn{fpicorrel}, we have plotted the $F_\pi^2$ dependence of the
first three even moments $\langle x^2\rangle_v, \langle x^4\rangle_v,
\langle x^6\rangle_v$, as obtained directly from the model-independent
analysis discussed above, using the closed symbols. Since we can
directly get only the even valence moments, we infer the odd moments
from  $\langle x^2\rangle_v$ and $\langle x^4\rangle_v$ by assuming
a two-parameter PDF ansatz,
\beq
f_v(x) = {\cal N} x^\alpha (1-x)^\beta,
\eeq{pdfmodel}
with normalization ${\cal N}$ to ensure $\langle x^0\rangle_v=1$, and simply
solve for $\alpha$ and $\beta$ through the two equations,
\beqa
\frac{\Gamma(3+\alpha)\Gamma(2+\alpha+\beta)}{\Gamma(1+\alpha)\Gamma(4+\alpha+\beta)}&=&\langle x^2\rangle_v,\cr
\frac{\Gamma(5+\alpha)\Gamma(2+\alpha+\beta)}{\Gamma(1+\alpha)\Gamma(6+\alpha+\beta)}&=&\langle x^4\rangle_v.
\eeqa{solve}
Through this we get $\langle x^{2k-1}\rangle_v(\alpha,\beta)$ by
this semi-model-dependent analysis. As a cross-check, this procedure
also predicts the even moment $\langle x^6\rangle(\alpha,\beta)$,
which we found to agree well with the actual value we obtained in
the model-independent analysis. These odd moments $\langle x\rangle_v$
and $ \langle x^3\rangle_v$ are also shown in \fgn{fpicorrel}. The
inferred value of $\langle x\rangle_v$, the fraction of pion mass
carried by a valence quark, seems to be $\approx 0.35$ in $N=0$
theory and increases to $\approx 0.45$ as $F_\pi$ decreases to zero.
Thus, even in the strongly confined regime of 2+1 dimensional SU(2)
QCD, about 30\% of pion mass is carried by gluons and sea quarks,
which one might want to contrast with the 3+1 dimensional QCD where
this fraction is about $\approx 55$\%, at a scale of $3$
GeV~\cite{Barry:2018ort}, and decreases further as the scale
approaches $\Lambda_{\rm QCD}$. Thus, it might be that the
scale-independent value of moments in 2+1 dimensions has to be
compared with PDFs in 3+1 dimensions determined at typical
non-perturbative hadronic scales to serve as good analogues. We
interpolated the data with a quadratic in $F^2_\pi g^{-2}$, which
are shown as the curves in \fgn{fpicorrel}. It is quite striking
how the individual moments themselves weakly depend on $F_\pi$.
One should contrast this behavior with the commensurate dependence
of other IR quantities to this decrease in $F_\pi$.  As we inferred
from the ITD itself, $\langle x^2\rangle_v$ seems to be the least
sensitive to changes in $F_\pi$.

\befs
\centering
\includegraphics[scale=0.6]{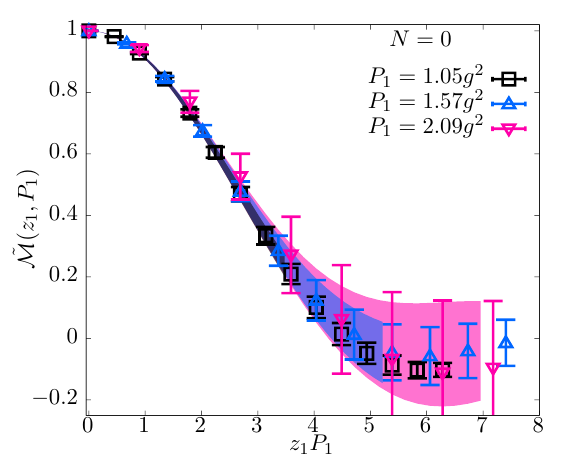}
\includegraphics[scale=0.6]{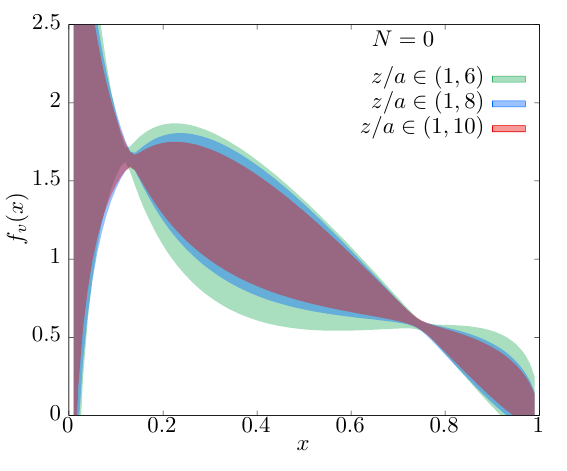}
\includegraphics[scale=0.6]{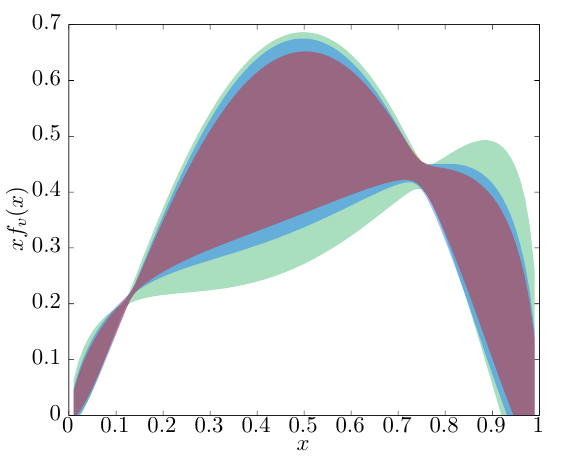}

\includegraphics[scale=0.6]{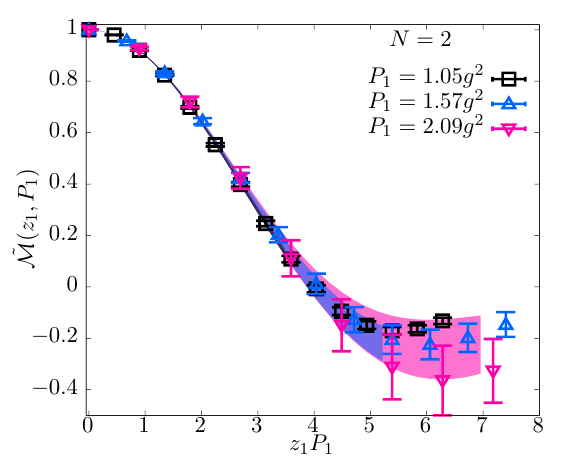}
\includegraphics[scale=0.6]{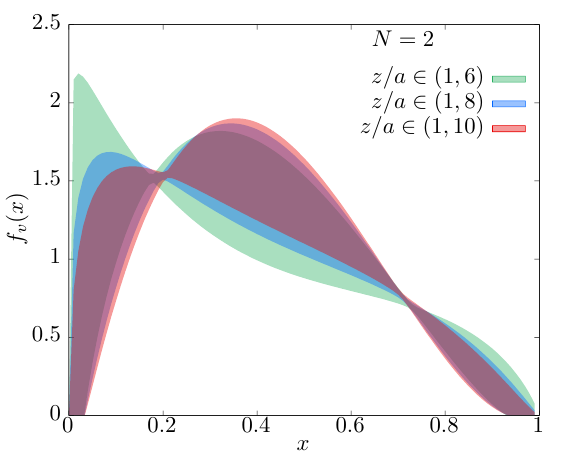}
\includegraphics[scale=0.6]{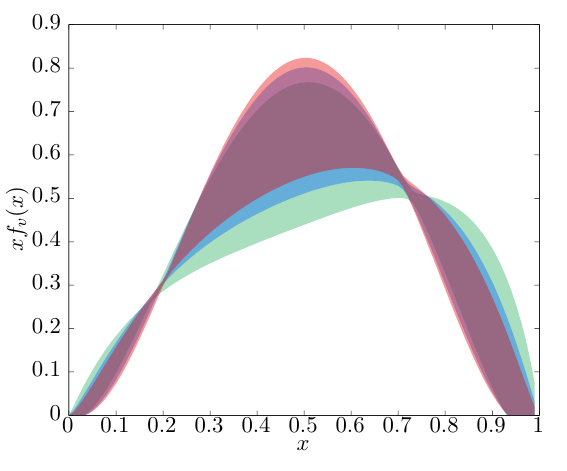}

\includegraphics[scale=0.6]{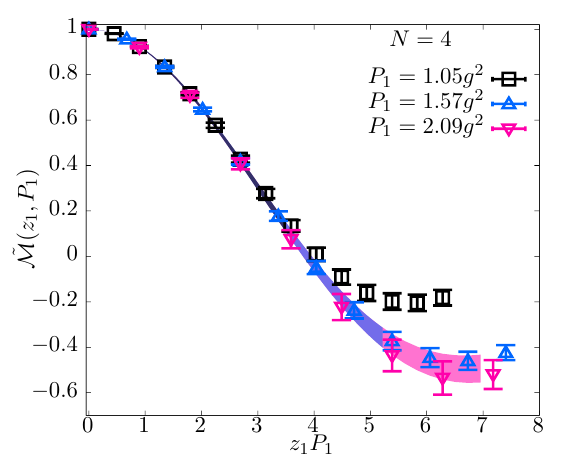}
\includegraphics[scale=0.6]{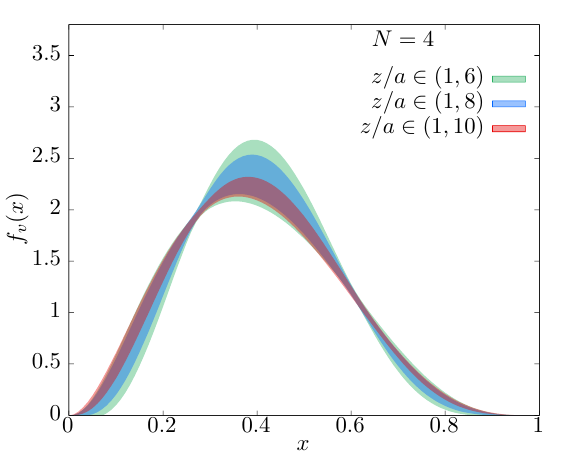}
\includegraphics[scale=0.6]{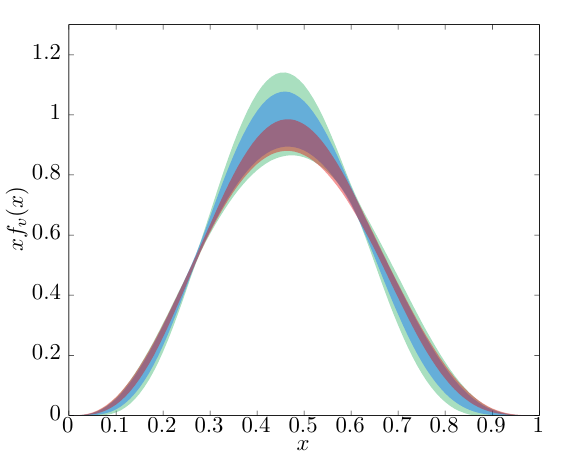}

\includegraphics[scale=0.6]{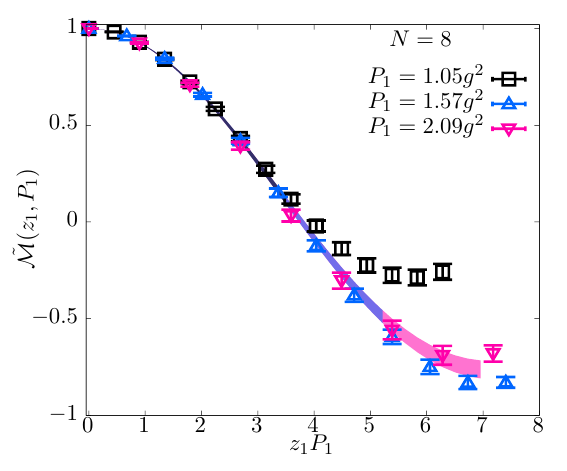}
\includegraphics[scale=0.6]{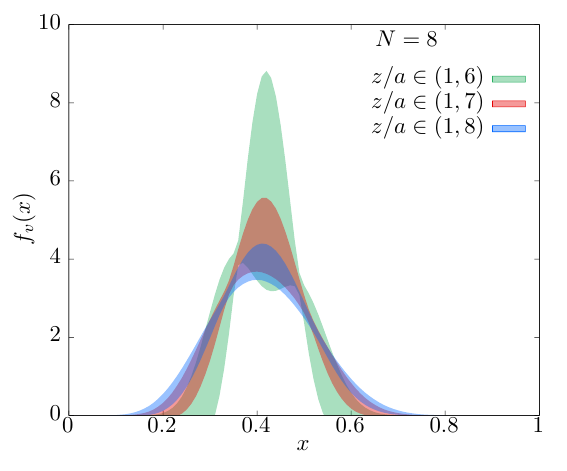}
\includegraphics[scale=0.6]{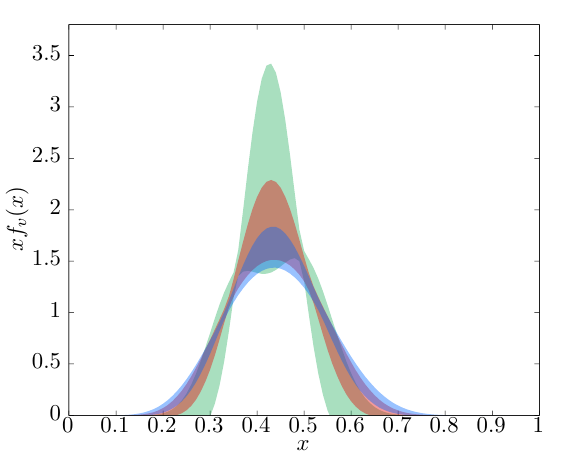}
\caption{
Reconstruction of valence PDF, $f_v(x)$, by fits to two-parameter
ansatz. The left panels show the fits (bands) to the bilocal matrix
element $\tilde{\cal M}(z_1,P_1)$ (points) via leading-twist
expression in \eqn{opework}. The middle panels show the inferred
valence PDF, $f_v(x)$. The different colored bands correspond to
different fit ranges $[0,z_1]$. The right panels show $x f_v(x)$.
Top to bottom are $N=0,2,4,8$ theories respectively.
}
\eefs{pdfs}

\befs
\centering
\includegraphics[scale=0.9]{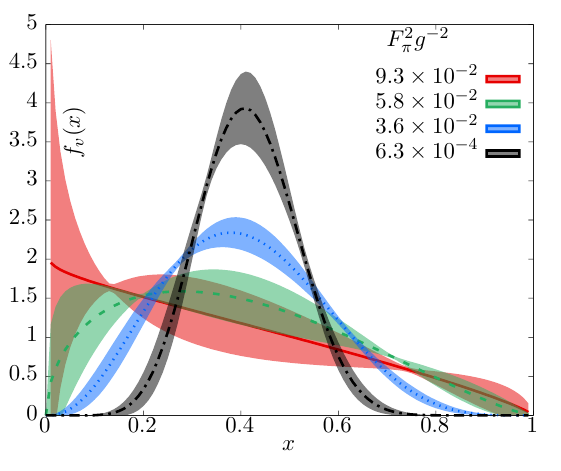}
\includegraphics[scale=0.9]{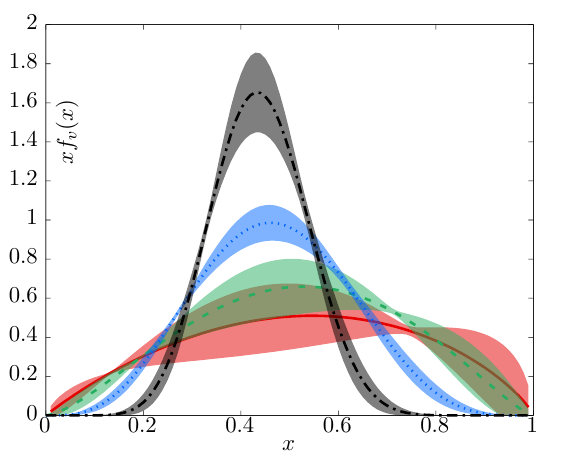}
\caption{
    (left) The reconstructed valence pion PDFs, $f_v(x)$, and (right) their
    corresponding momentum distribution, $x f_v(x)$  are
    shown as a function of pion decay constant that characterizes
    the vacuum of different $N=0,2,4,8$ flavor theories.}
\eefs{pdfpx0comp}

It is the observables that dictate the shape of the full $x$-dependent
PDF that are quite sensitive the infrared rather than the moments
themselves. One such observable is the
$\log$-derivatives~\cite{Gao:2020ito} of moments $\beta_{\rm
eff}(k)=-1-\partial\log(\langle x^k\rangle_v)/\partial\log(k)$ which
approaches the large-$x$ exponent $\beta$ for $k\to\infty$. As
explained in~\cite{Gao:2020ito}, we define the discretized version of the 
effective $\beta$
for any $k$ as
\beq
\beta_{\rm eff}(k)=\frac{\langle x^{k-2}\rangle_v - \langle x^{k+2}\rangle_v}{\langle x^{k}\rangle_v} \frac{k}{4}-1.
\eeq{effbeta}
Using $k=4$, we find $\beta_{\rm eff}$ for $N=0,2,4,8$ to be $0.8(5),
1.0(4), 2.1(2), 3.3(2)$ respectively. As expected, this quantity
shows a sharp increase as the theory moves from being strongly
confined to being infrared conformal. The $u-d$ PDF, $f_{u-d}(x)$
being a positive quantity and having a probabilistic interpretation,
also admits the canonical shape observables, cumulants $\kappa_n$,
\beqa
\kappa_n &\equiv& \frac{\partial^n}{\partial s^n}\log\left(\int_{-1}^1 f_{u-d}(x) e^{s x}dx\right)\bigg{|}_{s=0}\qquad{\rm with},\cr
\kappa_4 &=& \langle x^4\rangle_v - 3 \langle x^2\rangle_v^2,\cr
\kappa_6 &=& \langle x^6\rangle_v - 15 \langle x^2\rangle_v \langle x^4\rangle_v +30 \langle x^2\rangle_v^3.
\eeqa{cumul}
Using the model independent estimates of the even moments up to
$\langle x^6\rangle_v$, we find the fourth and sixth cumulants,
$[\kappa_4,\kappa_6]$, for $N=0,2,4,8$ flavors to be\quad [0.01(3),
-0.02(2)],\quad [0.05(2), -0.04(1)],\quad [0.071(8), -0.054(4)],\quad
[0.02(8), -0.059(4)] respectively. This variation is shown in the
bottom panel of \fgn{fpicorrel}, and it can seen to be very sensitive
to the IR changes. One could do a similar analysis by including the
non-vanishing odd valence moments, but we specifically chose the
cumulants of $u-d$ PDF so as to keep the analysis fully model-independent.
The aim of this exercise was to point to some good observables of the
pion PDF that seem to be sensitive about the IR, and consequently,
we were able to deduce simply from the model independent analysis
that the shape of the PDF will show sharp changes as the theory
morphs. Next, we will see these inferences concretely arise in the 
reconstructed $x$-dependent valence PDFs.

\subsubsection{Model dependent analysis: PDF reconstruction}

Now we reconstruct the $x$-dependent valence PDFs that best describe
the real space data for $\tilde{\cal M}(z_1,P_1)$. For this we use
the two-parameter functional form of the PDF in \eqn{pdfmodel} that
was completely sufficient to describe $\tilde{\cal M}(z_1,P_1)$ at
all $N$ and in the range of $z_1$ and $P_1$ described before; in
fact, when we tried to make the ansatz more complex by adding
subleading small-$x$ terms $x^{\alpha+0.5}$ and $x^{\alpha+1}$, the
fits became quite unstable and hence we resort to the simpler
two-parameter ansatz.  Essentially, the parametrized PDF enters
through its corresponding moments $\langle x^{2k}\rangle_v(\alpha,\beta)$,
that is then input into the leading twist OPE in \eqn{opework} to
get the best values of $\alpha$ and $\beta$. In the left panels of
\fgn{pdfs}, we have shown the resulting curves for $\tilde{\cal
M}(z_1,P_1)$ from such two-parameter fits superimposed on the data.
The quality of fits are as good as the one from model-independent
fits shown in \fgn{itdfit}.

The valence PDFs, $f_v(x)$, corresponding to the best fits are shown
in the middle panels of \fgn{pdfs}, and the rightmost panels are
simply the same data replotted as the momentum distribution, $x
f_v(x)$. We checked that the reconstructed PDFs were robust against
variations in the fit ranges by changing the maximum of the fit
range from $z_1=6a$ to $z_1=10a$. These variations are shown as the
bands of different colors in the middle and right panels of \fgn{pdfs}.
Since the data points fall on universal curves well to begin with,
the reconstructed PDFs also show almost no variations; so we simply
take the estimate with $z_1=8a$ for further discussions. It should
be noted that the scales in the different panels in \fgn{pdfs} are
different, but it is already clear that the PDFs get narrower as
$N$ increases. In terms of the exponents $[\alpha,\beta]$ of the
PDFs, they change as $[0.0(7), 0.8(8)]$,\quad $[0.5(6),1.4(8)]$,\quad
$[2.5(6),4.1(8)]$,\quad $[9(2), 13(3)]$ for $N=0,2,4,8$
respectively. The values of the large-$x$ exponent $\beta$ for the
strongly broken phase for $N=0$ and 2, are the typical value around
1 and 2 as in 3+1 dimensions.  The exponent subsequently gets larger
as the theory is pushed into the near-conformal and conformal
regimes. Perhaps it is of interest to note that numerically, 
$\beta\approx \alpha+1$ for these PDFs, 
which makes $x f_v(x)$ appear almost symmetrical around 
their peak positions. As a cross-check, we also plot the 
values of moments from this analysis using PDF ansatz in 
the top panel of \fgn{fpicorrel} as the open symbols, which 
nicely agrees with the moments obtained from the model-independent 
analysis.

We summarize the PDF determination in \fgn{pdfpx0comp} by putting
together the PDFs from all $N$, and showing it as a function of the
induced dependence on the infrared scale $F_\pi^2$. The left and
right panels show $f_v(x)$ and $x f_v(x)$ respectively. The depletion
of the IR scales can be seen to have visible effect on the pion
PDF. The effect of strong scale-breaking is to broaden the pion PDF
over the entire range of $x$; implying indirectly, the increased
importance of gluons and the sea quarks.  This is the case for
$F_\pi^2=9.3\times 10^{-2} g^2, 5.8\times 10^{-2} g^2$. As the
symmetry-breaking is made about three-times weaker with $F_\pi^2=3.6\times
10^{-2} g^2$, we start seeing the PDF get sharper around the middle
values $x\approx 0.4$ to 0.5, pointing to less important role of
the gluons, as well as of instances of valence quarks that carry
all of the pion momentum. As the theory enters a phase with
$F_\pi^2\approx 0$ which is most likely to be conformal in the
infrared, which is made gapped simply by finite quark mass and
finite box size, the PDF gets sharply peaked around $\langle
x\rangle_v \approx 0.42$, pointing to a near dominance of the valence
quarks.  This extreme case can be seen as a control in this
calculation; that is, the quark structure of an artificial pion-like state emerging simply
because of finite mass and volume,
being not consistent with the quark structure of an actual pion state in the scale-broken theories points
to the important causal role of the infrared vacuum structure in shaping
the valence quark structure of the Nambu-Goldstone boson.

\section{Conclusions and discussion}\label{sec:concl}

We presented a lattice calculation of the valence quark structure
of the Nambu-Goldstone boson (which we refer to as the pion) of the
flavor symmetry breaking in 2+1 dimensional SU(2) gauge theory
coupled to many massless flavors of fermions. The motivation for
this work was to first of all see if the quark structure of the
pion is sensitive to the long-distance vacuum structure, as one
would expect; and secondly to understand precisely how much this
dependence is and in what observables this shows up. For this work,
we used $N=0,2,4$ and 8 flavors of nearly massless dynamical
Wilson-Dirac fermions in the sea, and the valence fermion mass tuned
such that the pion mass stayed the same at $0.53 g^2$ for all
flavors. We studied the theories at a fixed lattice spacing and
fixed finite box size. We used the pion decay constant $F_\pi$ as
a measure of the strength of scale-breaking in the infrared, and
correlated its decrease as a function of $N$ with other infrared
quantities and to the short-distance quark structure of the pion to
$F_\pi$.

We showed that as the strength of the infrared scale breaking
decreases, the pion Ioffe-time distribution (ITD) or bilocal quark
bilinear matrix element on the light-cone becomes sensitive to this
effect for Ioffe-time (or light-front distance) $\nu>3$ with an
almost near-universal behavior for $\nu<3$; the effect is seen by
a slower fall-off of the ITD at $\nu>3$ as the theory gets more
broken. We found that the individual moments of the valence pion
PDF themselves show only a weak dependence to the changes in the
infrared. However, the effect gets amplified when one constructs
observables appropriately from the moments, such that they underlie
the shape of the $x$-dependent valence PDF and equivalently of the
$u-d$ PDF; we demonstrated this in terms of the first few cumulants
of the $u-d$ PDF and in terms of the log-derivative of the moments
with respect to the order of the moment that determine the large-$x$
behavior. We reconstructed the valence PDF of the pion based on a
two-parameter ansatz. The above behavior of the ITD resulted in a
broadening of the valence PDF over small and large $x$ regions when
the value of the $F_\pi$ increased.  When the $F_\pi$ was near zero
in the near-conformal region, one could see a sharp localization
of the PDF around $\langle x\rangle_v\approx 0.42$. The key results
in this paper are shown in \fgn{allitd} and \fgn{pdfpx0comp}.

As an outcome of this work, we established the 2+1 dimensional QCD
as a good model system to perform {\sl computational experiments} on 
the nonperturbative aspects of
the internal structure of hadrons using the recent developments in 
leading-twist matching frameworks. The short-coming of the
present work is that we do not compare and contrast the behavior of the
pion PDF with that of another {\sl ordinary} non-Goldstone boson,
such the axial-vector or the diquark states. We intend to perform
these comparative studies in future computations, especially by
using lattices which have larger extents in the direction of boost
so as to reduce the effect of Lorentz contraction (rather an
expansion) of the lattice extent longitudinal to the boost. Another
improvement one could do is to extend this calculation to SU(3)
theory in 2+1 dimensions; this will extend the range of flavor $N$
where the theory is scale-broken, thereby making the changes to the
PDFs more gradual and easier to study than done here. Owing to the
lower-dimension used, performing an exact massless overlap fermion
computation to improve on this work will be feasible. Understanding
the observations made in this paper in terms of simplistic model
calculations will also shed more light on how the UV is correlated
to the IR. With the availability of many-flavor theory computations 
in 3+1 dimensions (e.g.,~\cite{Appelquist:2020xua}), 
performed due to its relevance to composite Higgs
models, it would be interesting to use them to understand the evolution
of quark structures with scale depletion as  $N_f$ is changed from 2 to the near-conformal 
point near 8 or 10; especially, ask how does large-$x$ exponent 
$\beta$ change for 3+1 dimensional pion?

It would also be amusing to study the properties of the bilocal
bilinear matrix element (Ioffe time distribution) in the long
distance limit of the quark-antiquark separation when the theory
is in the conformal phase for $N>6$, such that the {\sl higher-twist}
effects now are actually going to be due to operators with  non-trivial
infrared scaling dimensions, and thereby shed new light into the
higher-spin operators of fixed twist
in the infrared CFT and its conformal blocks, possibly corresponding
to scalar-vector-vector-scalar four point function.  A recent
study~\cite{Braun:2020zjm} of conformal QCD in 4-$\epsilon$ dimensions
might be helpful in this endeavor, by carefully extrapolating the
results to $\epsilon=1$.

\section*{Acknowledgments}
N.K. would like to thank Chris Monahan, Swagato Mukherjee, Rajamani Narayanan,
Kostas Orginos, Peter Petreczky, Jianwei Qiu and Raza Sufian for
helpful comments and discussions. N.K. thanks all the members of 
the BNL-SBU collaboration and the HadStruc collaboration for fruitful discussions
which greatly helped the present work. N.K. is supported by Jefferson
Science Associates, LLC under U.S. DOE Contract \#DE-AC05-06OR23177
and in part by U.S. DOE grant \#DE-FG02-04ER41302.  N.K. acknowledges
the William \& Mary Research Computing for providing computational
resources and technical support that have contributed to the results
reported within this paper. (\url{https://www.wm.edu/it/rc}).

\appendix

\section{The behavior of $P_1=0$ matrix element ${\cal M}^B$}\label{sec:px0apx}

In \scn{pi3ptext}, we described the extrapolation of three-point
function to obtain the ``bare" matrix element, ${\cal M}^B(z_1,P_1)$.
The nomenclature ``bare" here simply means the matrix element
obtained before taking the ratio in \eqn{ritd}, as there are no
truly divergent behaviors in 2+1 dimensions due to its
super-renormalizability, and even for the Wilson-line, we expect
it to contribute only a benign $\exp{-c' g^2 z}$ nonperturbative
higher twist effect. In this appendix, we look at ${\cal M}^B(z_1,P_1=0)$
itself.

\bef[b]
\centering
\includegraphics[scale=0.7]{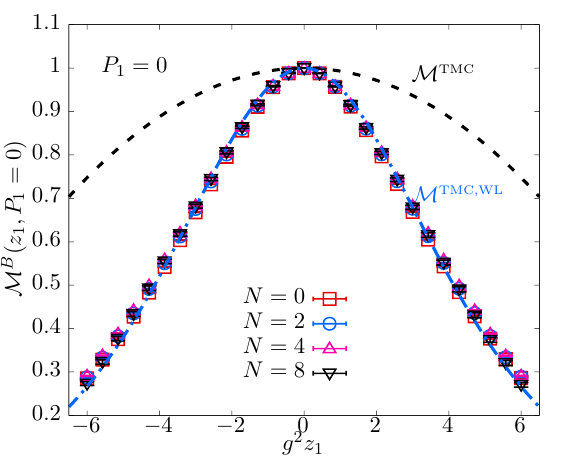}
\caption{The matrix element ${\cal M}^B(z_1,P_1=0)$ (before forming
ratios) is shown as a function of quark-antiquark separation $g^2
z_1$ in units of coupling.  The different colored symbols are the
data taken from $N=0,2,4,8$ flavor theories. The black dashed line
is the expectation based on target mass corrections from leading-twist
trace terms. The blue dashed line is the behavior modeled by
\eqn{modeltmc} to take into account the screening behavior of the Wilson-line.}
\eef{px0me}

From \fgn{hb} in the main text, we can notice that the $P_1=0$
matrix element with pion at rest shows a $z_1$ dependence.  To see
why it is interesting, for argument-sake, if we assume that the
leading-twist term was the only piece at $P_1=0$ OPE, then one
simply does not expect any $z_1$ dependence.  In \fgn{px0me}, we
have put together the $P_1=0$ data (shown as the points) for ${\cal
M}^B$ at all flavors $N$ as a function of lattice separation,
$z_1/a$. For the data shown, the Wilson-line entering the bilocal
operator was smeared using 2-steps of Stout. It is quite surprising
that the $P_1=0$ matrix element shows absolutely no dependence on
the flavor or changing $F_\pi$ equivalently.  Due to the finite
valence pion mass, there can be $z_1$ dependence from the target
mass corrections (TMC) that arises due to trace terms at leading
twist~\cite{Chen:2016utp,Radyushkin:2017ffo}.  We expect this to
be described by
\beq
{\cal M}^{\rm TMC}(z_1) = 1-\frac{(M_\pi^{\rm val}z_1)^2}{8}\langle x^2\rangle_v+{\cal O}\left((M^{\rm val}_\pi z_1)^4\right).
\eeq{tmc}
To see if this arises because of the TMC, we have plotted \eqn{tmc}
as the dashed black curve, using the value of $\langle x^2\rangle\approx
0.2$ that we observed in \fgn{fpicorrel}. This behavior is definitely
not sufficient to describe the data. The other $z_1$ dependence
should, of course, be from the Wilson-line due to its $\exp(-c'g^2
z)$ behavior for larger $|z_1|$ with some $c'$. Since it should be
even with respect to $z_1\to -z_1$, we model this behavior as
\beq
{\cal M}^{\rm TMC,WL}(z_1)\equiv \frac{{\cal M}^{\rm TMC}(z_1)}{\cosh(c' g^2 z_1)}.
\eeq{modeltmc}
In \fgn{px0me}, we plot ${\cal M}^{\rm TMC,WL}$ as the blue dashed
curve using $c'=0.281$. The value of $c'$ will be dependent on the
construction of Wilson-line itself, such as the steps of smearing
(in our case, the value of $c'$ decreases to 0.206 when 6-step stout
was used). We see that ${\cal M}^{\rm TMC,WL}$ nicely describes the
data at all $z_1$ and for all $N$. Thus, through this exercise, we
first understand that the non-perturbative screening behavior of
Wilson-line is important in ${\cal M}^B$, and therefore, it is very
important to form ratios, like in 3+1 dimensions, to get rid of
this trivial $z_1$ dependence.  In \fgn{stout}, we showed how this
cancellation works well by using Wilson-lines with two-different
smearing parameters, thereby justifying the application of leading
twist framework to the ratio $\tilde{\cal M}$.  Secondly, due to
universal behavior of ${\cal M}^B(P_1=0)$ at all $N$, the screening
behavior does not care about the infrared physics at all, pointing
to the fact that it arises due to the Wilson-line self-interaction
at ultraviolet scales.

\bibliography{pap.bib}

\begin{thebibliography}{103}%
\makeatletter
\providecommand \@ifxundefined [1]{%
 \@ifx{#1\undefined}
}%
\providecommand \@ifnum [1]{%
 \ifnum #1\expandafter \@firstoftwo
 \else \expandafter \@secondoftwo
 \fi
}%
\providecommand \@ifx [1]{%
 \ifx #1\expandafter \@firstoftwo
 \else \expandafter \@secondoftwo
 \fi
}%
\providecommand \natexlab [1]{#1}%
\providecommand \enquote  [1]{``#1''}%
\providecommand \bibnamefont  [1]{#1}%
\providecommand \bibfnamefont [1]{#1}%
\providecommand \citenamefont [1]{#1}%
\providecommand \href@noop [0]{\@secondoftwo}%
\providecommand \href [0]{\begingroup \@sanitize@url \@href}%
\providecommand \@href[1]{\@@startlink{#1}\@@href}%
\providecommand \@@href[1]{\endgroup#1\@@endlink}%
\providecommand \@sanitize@url [0]{\catcode `\\12\catcode `\$12\catcode
  `\&12\catcode `\#12\catcode `\^12\catcode `\_12\catcode `\%12\relax}%
\providecommand \@@startlink[1]{}%
\providecommand \@@endlink[0]{}%
\providecommand \url  [0]{\begingroup\@sanitize@url \@url }%
\providecommand \@url [1]{\endgroup\@href {#1}{\urlprefix }}%
\providecommand \urlprefix  [0]{URL }%
\providecommand \Eprint [0]{\href }%
\providecommand \doibase [0]{http://dx.doi.org/}%
\providecommand \selectlanguage [0]{\@gobble}%
\providecommand \bibinfo  [0]{\@secondoftwo}%
\providecommand \bibfield  [0]{\@secondoftwo}%
\providecommand \translation [1]{[#1]}%
\providecommand \BibitemOpen [0]{}%
\providecommand \bibitemStop [0]{}%
\providecommand \bibitemNoStop [0]{.\EOS\space}%
\providecommand \EOS [0]{\spacefactor3000\relax}%
\providecommand \BibitemShut  [1]{\csname bibitem#1\endcsname}%
\let\auto@bib@innerbib\@empty
\bibitem [{\citenamefont {Dissertori}\ \emph {et~al.}(2003)\citenamefont
  {Dissertori}, \citenamefont {Knowles},\ and\ \citenamefont
  {Schmelling}}]{dissertori2003quantum}%
  \BibitemOpen
  \bibfield  {author} {\bibinfo {author} {\bibfnamefont {G.}~\bibnamefont
  {Dissertori}}, \bibinfo {author} {\bibfnamefont {I.~G.}\ \bibnamefont
  {Knowles}}, \ and\ \bibinfo {author} {\bibfnamefont {M.}~\bibnamefont
  {Schmelling}},\ }\href@noop {} {\emph {\bibinfo {title} {Quantum
  chromodynamics: high energy experiments and theory}}},\ Vol.\ \bibinfo
  {volume} {115}\ (\bibinfo  {publisher} {Oxford University Press},\ \bibinfo
  {year} {2003})\BibitemShut {NoStop}%
\bibitem [{\citenamefont {Liu}(2017)}]{Liu:2016kbb}%
  \BibitemOpen
  \bibfield  {author} {\bibinfo {author} {\bibfnamefont {C.}~\bibnamefont
  {Liu}},\ }\href {\doibase 10.22323/1.256.0006} {\bibfield  {journal}
  {\bibinfo  {journal} {PoS}\ }\textbf {\bibinfo {volume} {LATTICE2016}},\
  \bibinfo {pages} {006} (\bibinfo {year} {2017})},\ \Eprint
  {http://arxiv.org/abs/1612.00103} {arXiv:1612.00103 [hep-lat]} \BibitemShut
  {NoStop}%
\bibitem [{\citenamefont {Padmanath}(2018)}]{Padmanath:2019wid}%
  \BibitemOpen
  \bibfield  {author} {\bibinfo {author} {\bibfnamefont {M.}~\bibnamefont
  {Padmanath}},\ }\href {\doibase 10.22323/1.334.0013} {\bibfield  {journal}
  {\bibinfo  {journal} {PoS}\ }\textbf {\bibinfo {volume} {LATTICE2018}},\
  \bibinfo {pages} {013} (\bibinfo {year} {2018})},\ \Eprint
  {http://arxiv.org/abs/1905.09651} {arXiv:1905.09651 [hep-lat]} \BibitemShut
  {NoStop}%
\bibitem [{\citenamefont {Briceno}\ \emph {et~al.}(2018)\citenamefont
  {Briceno}, \citenamefont {Dudek},\ and\ \citenamefont
  {Young}}]{Briceno:2017max}%
  \BibitemOpen
  \bibfield  {author} {\bibinfo {author} {\bibfnamefont {R.~A.}\ \bibnamefont
  {Briceno}}, \bibinfo {author} {\bibfnamefont {J.~J.}\ \bibnamefont {Dudek}},
  \ and\ \bibinfo {author} {\bibfnamefont {R.~D.}\ \bibnamefont {Young}},\
  }\href {\doibase 10.1103/RevModPhys.90.025001} {\bibfield  {journal}
  {\bibinfo  {journal} {Rev. Mod. Phys.}\ }\textbf {\bibinfo {volume} {90}},\
  \bibinfo {pages} {025001} (\bibinfo {year} {2018})},\ \Eprint
  {http://arxiv.org/abs/1706.06223} {arXiv:1706.06223 [hep-lat]} \BibitemShut
  {NoStop}%
\bibitem [{\citenamefont {Accardi}\ \emph {et~al.}(2016)\citenamefont {Accardi}
  \emph {et~al.}}]{Accardi:2012qut}%
  \BibitemOpen
  \bibfield  {author} {\bibinfo {author} {\bibfnamefont {A.}~\bibnamefont
  {Accardi}} \emph {et~al.},\ }\href {\doibase 10.1140/epja/i2016-16268-9}
  {\bibfield  {journal} {\bibinfo  {journal} {Eur. Phys. J. A}\ }\textbf
  {\bibinfo {volume} {52}},\ \bibinfo {pages} {268} (\bibinfo {year} {2016})},\
  \Eprint {http://arxiv.org/abs/1212.1701} {arXiv:1212.1701 [nucl-ex]}
  \BibitemShut {NoStop}%
\bibitem [{\citenamefont {Ji}(1995)}]{Ji:1994av}%
  \BibitemOpen
  \bibfield  {author} {\bibinfo {author} {\bibfnamefont {X.-D.}\ \bibnamefont
  {Ji}},\ }\href {\doibase 10.1103/PhysRevLett.74.1071} {\bibfield  {journal}
  {\bibinfo  {journal} {Phys. Rev. Lett.}\ }\textbf {\bibinfo {volume} {74}},\
  \bibinfo {pages} {1071} (\bibinfo {year} {1995})},\ \Eprint
  {http://arxiv.org/abs/hep-ph/9410274} {arXiv:hep-ph/9410274} \BibitemShut
  {NoStop}%
\bibitem [{\citenamefont {Hatta}\ and\ \citenamefont
  {Zhao}(2020)}]{Hatta:2020iin}%
  \BibitemOpen
  \bibfield  {author} {\bibinfo {author} {\bibfnamefont {Y.}~\bibnamefont
  {Hatta}}\ and\ \bibinfo {author} {\bibfnamefont {Y.}~\bibnamefont {Zhao}},\
  }\href {\doibase 10.1103/PhysRevD.102.034004} {\bibfield  {journal} {\bibinfo
   {journal} {Phys. Rev. D}\ }\textbf {\bibinfo {volume} {102}},\ \bibinfo
  {pages} {034004} (\bibinfo {year} {2020})},\ \Eprint
  {http://arxiv.org/abs/2006.02798} {arXiv:2006.02798 [hep-ph]} \BibitemShut
  {NoStop}%
\bibitem [{\citenamefont {Alexandrou}\ \emph {et~al.}(2017)\citenamefont
  {Alexandrou}, \citenamefont {Constantinou}, \citenamefont {Hadjiyiannakou},
  \citenamefont {Jansen}, \citenamefont {Kallidonis}, \citenamefont {Koutsou},
  \citenamefont {Vaquero Avil\'es-Casco},\ and\ \citenamefont
  {Wiese}}]{Alexandrou:2017oeh}%
  \BibitemOpen
  \bibfield  {author} {\bibinfo {author} {\bibfnamefont {C.}~\bibnamefont
  {Alexandrou}}, \bibinfo {author} {\bibfnamefont {M.}~\bibnamefont
  {Constantinou}}, \bibinfo {author} {\bibfnamefont {K.}~\bibnamefont
  {Hadjiyiannakou}}, \bibinfo {author} {\bibfnamefont {K.}~\bibnamefont
  {Jansen}}, \bibinfo {author} {\bibfnamefont {C.}~\bibnamefont {Kallidonis}},
  \bibinfo {author} {\bibfnamefont {G.}~\bibnamefont {Koutsou}}, \bibinfo
  {author} {\bibfnamefont {A.}~\bibnamefont {Vaquero Avil\'es-Casco}}, \ and\
  \bibinfo {author} {\bibfnamefont {C.}~\bibnamefont {Wiese}},\ }\href
  {\doibase 10.1103/PhysRevLett.119.142002} {\bibfield  {journal} {\bibinfo
  {journal} {Phys. Rev. Lett.}\ }\textbf {\bibinfo {volume} {119}},\ \bibinfo
  {pages} {142002} (\bibinfo {year} {2017})},\ \Eprint
  {http://arxiv.org/abs/1706.02973} {arXiv:1706.02973 [hep-lat]} \BibitemShut
  {NoStop}%
\bibitem [{\citenamefont {Alexandrou}\ \emph {et~al.}(2020)\citenamefont
  {Alexandrou}, \citenamefont {Bacchio}, \citenamefont {Constantinou},
  \citenamefont {Finkenrath}, \citenamefont {Hadjiyiannakou}, \citenamefont
  {Jansen}, \citenamefont {Koutsou}, \citenamefont {Panagopoulos},\ and\
  \citenamefont {Spanoudes}}]{Alexandrou:2020sml}%
  \BibitemOpen
  \bibfield  {author} {\bibinfo {author} {\bibfnamefont {C.}~\bibnamefont
  {Alexandrou}}, \bibinfo {author} {\bibfnamefont {S.}~\bibnamefont {Bacchio}},
  \bibinfo {author} {\bibfnamefont {M.}~\bibnamefont {Constantinou}}, \bibinfo
  {author} {\bibfnamefont {J.}~\bibnamefont {Finkenrath}}, \bibinfo {author}
  {\bibfnamefont {K.}~\bibnamefont {Hadjiyiannakou}}, \bibinfo {author}
  {\bibfnamefont {K.}~\bibnamefont {Jansen}}, \bibinfo {author} {\bibfnamefont
  {G.}~\bibnamefont {Koutsou}}, \bibinfo {author} {\bibfnamefont
  {H.}~\bibnamefont {Panagopoulos}}, \ and\ \bibinfo {author} {\bibfnamefont
  {G.}~\bibnamefont {Spanoudes}},\ }\href {\doibase
  10.1103/PhysRevD.101.094513} {\bibfield  {journal} {\bibinfo  {journal}
  {Phys. Rev. D}\ }\textbf {\bibinfo {volume} {101}},\ \bibinfo {pages}
  {094513} (\bibinfo {year} {2020})},\ \Eprint
  {http://arxiv.org/abs/2003.08486} {arXiv:2003.08486 [hep-lat]} \BibitemShut
  {NoStop}%
\bibitem [{\citenamefont {Badier}\ \emph {et~al.}(1983)\citenamefont {Badier}
  \emph {et~al.}}]{Badier:1983mj}%
  \BibitemOpen
  \bibfield  {author} {\bibinfo {author} {\bibfnamefont {J.}~\bibnamefont
  {Badier}} \emph {et~al.} (\bibinfo {collaboration} {NA3}),\ }\href {\doibase
  10.1007/BF01573728} {\bibfield  {journal} {\bibinfo  {journal} {Z. Phys. C}\
  }\textbf {\bibinfo {volume} {18}},\ \bibinfo {pages} {281} (\bibinfo {year}
  {1983})}\BibitemShut {NoStop}%
\bibitem [{\citenamefont {Betev}\ \emph {et~al.}(1985)\citenamefont {Betev}
  \emph {et~al.}}]{Betev:1985pf}%
  \BibitemOpen
  \bibfield  {author} {\bibinfo {author} {\bibfnamefont {B.}~\bibnamefont
  {Betev}} \emph {et~al.} (\bibinfo {collaboration} {NA10}),\ }\href {\doibase
  10.1007/BF01550243} {\bibfield  {journal} {\bibinfo  {journal} {Z. Phys. C}\
  }\textbf {\bibinfo {volume} {28}},\ \bibinfo {pages} {9} (\bibinfo {year}
  {1985})}\BibitemShut {NoStop}%
\bibitem [{\citenamefont {Conway}\ \emph {et~al.}(1989)\citenamefont {Conway}
  \emph {et~al.}}]{Conway:1989fs}%
  \BibitemOpen
  \bibfield  {author} {\bibinfo {author} {\bibfnamefont {J.}~\bibnamefont
  {Conway}} \emph {et~al.},\ }\href {\doibase 10.1103/PhysRevD.39.92}
  {\bibfield  {journal} {\bibinfo  {journal} {Phys. Rev. D}\ }\textbf {\bibinfo
  {volume} {39}},\ \bibinfo {pages} {92} (\bibinfo {year} {1989})}\BibitemShut
  {NoStop}%
\bibitem [{\citenamefont {Owens}(1984)}]{Owens:1984zj}%
  \BibitemOpen
  \bibfield  {author} {\bibinfo {author} {\bibfnamefont {J.}~\bibnamefont
  {Owens}},\ }\href {\doibase 10.1103/PhysRevD.30.943} {\bibfield  {journal}
  {\bibinfo  {journal} {Phys. Rev. D}\ }\textbf {\bibinfo {volume} {30}},\
  \bibinfo {pages} {943} (\bibinfo {year} {1984})}\BibitemShut {NoStop}%
\bibitem [{\citenamefont {Sutton}\ \emph {et~al.}(1992)\citenamefont {Sutton},
  \citenamefont {Martin}, \citenamefont {Roberts},\ and\ \citenamefont
  {Stirling}}]{Sutton:1991ay}%
  \BibitemOpen
  \bibfield  {author} {\bibinfo {author} {\bibfnamefont {P.}~\bibnamefont
  {Sutton}}, \bibinfo {author} {\bibfnamefont {A.~D.}\ \bibnamefont {Martin}},
  \bibinfo {author} {\bibfnamefont {R.}~\bibnamefont {Roberts}}, \ and\
  \bibinfo {author} {\bibfnamefont {W.}~\bibnamefont {Stirling}},\ }\href
  {\doibase 10.1103/PhysRevD.45.2349} {\bibfield  {journal} {\bibinfo
  {journal} {Phys. Rev. D}\ }\textbf {\bibinfo {volume} {45}},\ \bibinfo
  {pages} {2349} (\bibinfo {year} {1992})}\BibitemShut {NoStop}%
\bibitem [{\citenamefont {Gluck}\ \emph {et~al.}(1992)\citenamefont {Gluck},
  \citenamefont {Reya},\ and\ \citenamefont {Vogt}}]{Gluck:1991ey}%
  \BibitemOpen
  \bibfield  {author} {\bibinfo {author} {\bibfnamefont {M.}~\bibnamefont
  {Gluck}}, \bibinfo {author} {\bibfnamefont {E.}~\bibnamefont {Reya}}, \ and\
  \bibinfo {author} {\bibfnamefont {A.}~\bibnamefont {Vogt}},\ }\href {\doibase
  10.1007/BF01559743} {\bibfield  {journal} {\bibinfo  {journal} {Z. Phys. C}\
  }\textbf {\bibinfo {volume} {53}},\ \bibinfo {pages} {651} (\bibinfo {year}
  {1992})}\BibitemShut {NoStop}%
\bibitem [{\citenamefont {Gluck}\ \emph {et~al.}(1999)\citenamefont {Gluck},
  \citenamefont {Reya},\ and\ \citenamefont {Schienbein}}]{Gluck:1999xe}%
  \BibitemOpen
  \bibfield  {author} {\bibinfo {author} {\bibfnamefont {M.}~\bibnamefont
  {Gluck}}, \bibinfo {author} {\bibfnamefont {E.}~\bibnamefont {Reya}}, \ and\
  \bibinfo {author} {\bibfnamefont {I.}~\bibnamefont {Schienbein}},\ }\href
  {\doibase 10.1007/s100529900124} {\bibfield  {journal} {\bibinfo  {journal}
  {Eur. Phys. J. C}\ }\textbf {\bibinfo {volume} {10}},\ \bibinfo {pages} {313}
  (\bibinfo {year} {1999})},\ \Eprint {http://arxiv.org/abs/hep-ph/9903288}
  {arXiv:hep-ph/9903288} \BibitemShut {NoStop}%
\bibitem [{\citenamefont {Wijesooriya}\ \emph {et~al.}(2005)\citenamefont
  {Wijesooriya}, \citenamefont {Reimer},\ and\ \citenamefont
  {Holt}}]{Wijesooriya:2005ir}%
  \BibitemOpen
  \bibfield  {author} {\bibinfo {author} {\bibfnamefont {K.}~\bibnamefont
  {Wijesooriya}}, \bibinfo {author} {\bibfnamefont {P.}~\bibnamefont {Reimer}},
  \ and\ \bibinfo {author} {\bibfnamefont {R.}~\bibnamefont {Holt}},\ }\href
  {\doibase 10.1103/PhysRevC.72.065203} {\bibfield  {journal} {\bibinfo
  {journal} {Phys. Rev. C}\ }\textbf {\bibinfo {volume} {72}},\ \bibinfo
  {pages} {065203} (\bibinfo {year} {2005})},\ \Eprint
  {http://arxiv.org/abs/nucl-ex/0509012} {arXiv:nucl-ex/0509012} \BibitemShut
  {NoStop}%
\bibitem [{\citenamefont {Aicher}\ \emph {et~al.}(2010)\citenamefont {Aicher},
  \citenamefont {Schafer},\ and\ \citenamefont {Vogelsang}}]{Aicher:2010cb}%
  \BibitemOpen
  \bibfield  {author} {\bibinfo {author} {\bibfnamefont {M.}~\bibnamefont
  {Aicher}}, \bibinfo {author} {\bibfnamefont {A.}~\bibnamefont {Schafer}}, \
  and\ \bibinfo {author} {\bibfnamefont {W.}~\bibnamefont {Vogelsang}},\ }\href
  {\doibase 10.1103/PhysRevLett.105.252003} {\bibfield  {journal} {\bibinfo
  {journal} {Phys. Rev. Lett.}\ }\textbf {\bibinfo {volume} {105}},\ \bibinfo
  {pages} {252003} (\bibinfo {year} {2010})},\ \Eprint
  {http://arxiv.org/abs/1009.2481} {arXiv:1009.2481 [hep-ph]} \BibitemShut
  {NoStop}%
\bibitem [{\citenamefont {Brodsky}\ and\ \citenamefont
  {Farrar}(1973)}]{Brodsky:1973kr}%
  \BibitemOpen
  \bibfield  {author} {\bibinfo {author} {\bibfnamefont {S.~J.}\ \bibnamefont
  {Brodsky}}\ and\ \bibinfo {author} {\bibfnamefont {G.~R.}\ \bibnamefont
  {Farrar}},\ }\href {\doibase 10.1103/PhysRevLett.31.1153} {\bibfield
  {journal} {\bibinfo  {journal} {Phys. Rev. Lett.}\ }\textbf {\bibinfo
  {volume} {31}},\ \bibinfo {pages} {1153} (\bibinfo {year}
  {1973})}\BibitemShut {NoStop}%
\bibitem [{\citenamefont {Nguyen}\ \emph {et~al.}(2011)\citenamefont {Nguyen},
  \citenamefont {Bashir}, \citenamefont {Roberts},\ and\ \citenamefont
  {Tandy}}]{Nguyen:2011jy}%
  \BibitemOpen
  \bibfield  {author} {\bibinfo {author} {\bibfnamefont {T.}~\bibnamefont
  {Nguyen}}, \bibinfo {author} {\bibfnamefont {A.}~\bibnamefont {Bashir}},
  \bibinfo {author} {\bibfnamefont {C.~D.}\ \bibnamefont {Roberts}}, \ and\
  \bibinfo {author} {\bibfnamefont {P.~C.}\ \bibnamefont {Tandy}},\ }\href
  {\doibase 10.1103/PhysRevC.83.062201} {\bibfield  {journal} {\bibinfo
  {journal} {Phys. Rev. C}\ }\textbf {\bibinfo {volume} {83}},\ \bibinfo
  {pages} {062201} (\bibinfo {year} {2011})},\ \Eprint
  {http://arxiv.org/abs/1102.2448} {arXiv:1102.2448 [nucl-th]} \BibitemShut
  {NoStop}%
\bibitem [{\citenamefont {Chen}\ \emph
  {et~al.}(2016{\natexlab{a}})\citenamefont {Chen}, \citenamefont {Chang},
  \citenamefont {Roberts}, \citenamefont {Wan},\ and\ \citenamefont
  {Zong}}]{Chen:2016sno}%
  \BibitemOpen
  \bibfield  {author} {\bibinfo {author} {\bibfnamefont {C.}~\bibnamefont
  {Chen}}, \bibinfo {author} {\bibfnamefont {L.}~\bibnamefont {Chang}},
  \bibinfo {author} {\bibfnamefont {C.~D.}\ \bibnamefont {Roberts}}, \bibinfo
  {author} {\bibfnamefont {S.}~\bibnamefont {Wan}}, \ and\ \bibinfo {author}
  {\bibfnamefont {H.-S.}\ \bibnamefont {Zong}},\ }\href {\doibase
  10.1103/PhysRevD.93.074021} {\bibfield  {journal} {\bibinfo  {journal} {Phys.
  Rev. D}\ }\textbf {\bibinfo {volume} {93}},\ \bibinfo {pages} {074021}
  (\bibinfo {year} {2016}{\natexlab{a}})},\ \Eprint
  {http://arxiv.org/abs/1602.01502} {arXiv:1602.01502 [nucl-th]} \BibitemShut
  {NoStop}%
\bibitem [{\citenamefont {Cui}\ \emph {et~al.}(2020)\citenamefont {Cui},
  \citenamefont {Ding}, \citenamefont {Gao}, \citenamefont {Raya},
  \citenamefont {Binosi}, \citenamefont {Chang}, \citenamefont {Roberts},
  \citenamefont {Rodr\'\i{}guez-Quintero},\ and\ \citenamefont
  {Schmidt}}]{Cui:2020tdf}%
  \BibitemOpen
  \bibfield  {author} {\bibinfo {author} {\bibfnamefont {Z.-F.}\ \bibnamefont
  {Cui}}, \bibinfo {author} {\bibfnamefont {M.}~\bibnamefont {Ding}}, \bibinfo
  {author} {\bibfnamefont {F.}~\bibnamefont {Gao}}, \bibinfo {author}
  {\bibfnamefont {K.}~\bibnamefont {Raya}}, \bibinfo {author} {\bibfnamefont
  {D.}~\bibnamefont {Binosi}}, \bibinfo {author} {\bibfnamefont
  {L.}~\bibnamefont {Chang}}, \bibinfo {author} {\bibfnamefont {C.~D.}\
  \bibnamefont {Roberts}}, \bibinfo {author} {\bibfnamefont {J.}~\bibnamefont
  {Rodr\'\i{}guez-Quintero}}, \ and\ \bibinfo {author} {\bibfnamefont {S.~M.}\
  \bibnamefont {Schmidt}},\ }\href {\doibase 10.1140/epjc/s10052-020-08578-4}
  {\bibfield  {journal} {\bibinfo  {journal} {Eur. Phys. J. C}\ }\textbf
  {\bibinfo {volume} {80}},\ \bibinfo {pages} {1064} (\bibinfo {year}
  {2020})}\BibitemShut {NoStop}%
\bibitem [{\citenamefont {Roberts}\ and\ \citenamefont
  {Schmidt}(2020)}]{Roberts:2020udq}%
  \BibitemOpen
  \bibfield  {author} {\bibinfo {author} {\bibfnamefont {C.~D.}\ \bibnamefont
  {Roberts}}\ and\ \bibinfo {author} {\bibfnamefont {S.~M.}\ \bibnamefont
  {Schmidt}}\ }(\bibinfo {year} {2020})\ \Eprint
  {http://arxiv.org/abs/2006.08782} {arXiv:2006.08782 [hep-ph]} \BibitemShut
  {NoStop}%
\bibitem [{\citenamefont {de~Teramond}\ \emph {et~al.}(2018)\citenamefont
  {de~Teramond}, \citenamefont {Liu}, \citenamefont {Sufian}, \citenamefont
  {Dosch}, \citenamefont {Brodsky},\ and\ \citenamefont
  {Deur}}]{deTeramond:2018ecg}%
  \BibitemOpen
  \bibfield  {author} {\bibinfo {author} {\bibfnamefont {G.~F.}\ \bibnamefont
  {de~Teramond}}, \bibinfo {author} {\bibfnamefont {T.}~\bibnamefont {Liu}},
  \bibinfo {author} {\bibfnamefont {R.~S.}\ \bibnamefont {Sufian}}, \bibinfo
  {author} {\bibfnamefont {H.~G.}\ \bibnamefont {Dosch}}, \bibinfo {author}
  {\bibfnamefont {S.~J.}\ \bibnamefont {Brodsky}}, \ and\ \bibinfo {author}
  {\bibfnamefont {A.}~\bibnamefont {Deur}} (\bibinfo {collaboration} {HLFHS}),\
  }\href {\doibase 10.1103/PhysRevLett.120.182001} {\bibfield  {journal}
  {\bibinfo  {journal} {Phys. Rev. Lett.}\ }\textbf {\bibinfo {volume} {120}},\
  \bibinfo {pages} {182001} (\bibinfo {year} {2018})},\ \Eprint
  {http://arxiv.org/abs/1801.09154} {arXiv:1801.09154 [hep-ph]} \BibitemShut
  {NoStop}%
\bibitem [{\citenamefont {Ruiz~Arriola}(2002)}]{RuizArriola:2002wr}%
  \BibitemOpen
  \bibfield  {author} {\bibinfo {author} {\bibfnamefont {E.}~\bibnamefont
  {Ruiz~Arriola}},\ }\href@noop {} {\bibfield  {journal} {\bibinfo  {journal}
  {Acta Phys. Polon. B}\ }\textbf {\bibinfo {volume} {33}},\ \bibinfo {pages}
  {4443} (\bibinfo {year} {2002})},\ \Eprint
  {http://arxiv.org/abs/hep-ph/0210007} {arXiv:hep-ph/0210007} \BibitemShut
  {NoStop}%
\bibitem [{\citenamefont {Broniowski}\ and\ \citenamefont
  {Ruiz~Arriola}(2017)}]{Broniowski:2017wbr}%
  \BibitemOpen
  \bibfield  {author} {\bibinfo {author} {\bibfnamefont {W.}~\bibnamefont
  {Broniowski}}\ and\ \bibinfo {author} {\bibfnamefont {E.}~\bibnamefont
  {Ruiz~Arriola}},\ }\href {\doibase 10.1016/j.physletb.2017.08.055} {\bibfield
   {journal} {\bibinfo  {journal} {Phys. Lett. B}\ }\textbf {\bibinfo {volume}
  {773}},\ \bibinfo {pages} {385} (\bibinfo {year} {2017})},\ \Eprint
  {http://arxiv.org/abs/1707.09588} {arXiv:1707.09588 [hep-ph]} \BibitemShut
  {NoStop}%
\bibitem [{\citenamefont {Lan}\ \emph {et~al.}(2020)\citenamefont {Lan},
  \citenamefont {Mondal}, \citenamefont {Jia}, \citenamefont {Zhao},\ and\
  \citenamefont {Vary}}]{Lan:2019rba}%
  \BibitemOpen
  \bibfield  {author} {\bibinfo {author} {\bibfnamefont {J.}~\bibnamefont
  {Lan}}, \bibinfo {author} {\bibfnamefont {C.}~\bibnamefont {Mondal}},
  \bibinfo {author} {\bibfnamefont {S.}~\bibnamefont {Jia}}, \bibinfo {author}
  {\bibfnamefont {X.}~\bibnamefont {Zhao}}, \ and\ \bibinfo {author}
  {\bibfnamefont {J.~P.}\ \bibnamefont {Vary}},\ }\href {\doibase
  10.1103/PhysRevD.101.034024} {\bibfield  {journal} {\bibinfo  {journal}
  {Phys. Rev. D}\ }\textbf {\bibinfo {volume} {101}},\ \bibinfo {pages}
  {034024} (\bibinfo {year} {2020})},\ \Eprint
  {http://arxiv.org/abs/1907.01509} {arXiv:1907.01509 [nucl-th]} \BibitemShut
  {NoStop}%
\bibitem [{\citenamefont {Barry}\ \emph {et~al.}(2018)\citenamefont {Barry},
  \citenamefont {Sato}, \citenamefont {Melnitchouk},\ and\ \citenamefont
  {Ji}}]{Barry:2018ort}%
  \BibitemOpen
  \bibfield  {author} {\bibinfo {author} {\bibfnamefont {P.}~\bibnamefont
  {Barry}}, \bibinfo {author} {\bibfnamefont {N.}~\bibnamefont {Sato}},
  \bibinfo {author} {\bibfnamefont {W.}~\bibnamefont {Melnitchouk}}, \ and\
  \bibinfo {author} {\bibfnamefont {C.-R.}\ \bibnamefont {Ji}},\ }\href
  {\doibase 10.1103/PhysRevLett.121.152001} {\bibfield  {journal} {\bibinfo
  {journal} {Phys. Rev. Lett.}\ }\textbf {\bibinfo {volume} {121}},\ \bibinfo
  {pages} {152001} (\bibinfo {year} {2018})},\ \Eprint
  {http://arxiv.org/abs/1804.01965} {arXiv:1804.01965 [hep-ph]} \BibitemShut
  {NoStop}%
\bibitem [{\citenamefont {Novikov}\ \emph {et~al.}(2020)\citenamefont {Novikov}
  \emph {et~al.}}]{Novikov:2020snp}%
  \BibitemOpen
  \bibfield  {author} {\bibinfo {author} {\bibfnamefont {I.}~\bibnamefont
  {Novikov}} \emph {et~al.},\ }\href {\doibase 10.1103/PhysRevD.102.014040}
  {\bibfield  {journal} {\bibinfo  {journal} {Phys. Rev. D}\ }\textbf {\bibinfo
  {volume} {102}},\ \bibinfo {pages} {014040} (\bibinfo {year} {2020})},\
  \Eprint {http://arxiv.org/abs/2002.02902} {arXiv:2002.02902 [hep-ph]}
  \BibitemShut {NoStop}%
\bibitem [{\citenamefont {Bednar}\ \emph {et~al.}(2020)\citenamefont {Bednar},
  \citenamefont {Clo\"et},\ and\ \citenamefont {Tandy}}]{Bednar:2018mtf}%
  \BibitemOpen
  \bibfield  {author} {\bibinfo {author} {\bibfnamefont {K.~D.}\ \bibnamefont
  {Bednar}}, \bibinfo {author} {\bibfnamefont {I.~C.}\ \bibnamefont {Clo\"et}},
  \ and\ \bibinfo {author} {\bibfnamefont {P.~C.}\ \bibnamefont {Tandy}},\
  }\href {\doibase 10.1103/PhysRevLett.124.042002} {\bibfield  {journal}
  {\bibinfo  {journal} {Phys. Rev. Lett.}\ }\textbf {\bibinfo {volume} {124}},\
  \bibinfo {pages} {042002} (\bibinfo {year} {2020})},\ \Eprint
  {http://arxiv.org/abs/1811.12310} {arXiv:1811.12310 [nucl-th]} \BibitemShut
  {NoStop}%
\bibitem [{\citenamefont {Ji}(2013)}]{Ji:2013dva}%
  \BibitemOpen
  \bibfield  {author} {\bibinfo {author} {\bibfnamefont {X.}~\bibnamefont
  {Ji}},\ }\href {\doibase 10.1103/PhysRevLett.110.262002} {\bibfield
  {journal} {\bibinfo  {journal} {Phys. Rev. Lett.}\ }\textbf {\bibinfo
  {volume} {110}},\ \bibinfo {pages} {262002} (\bibinfo {year} {2013})},\
  \Eprint {http://arxiv.org/abs/1305.1539} {arXiv:1305.1539 [hep-ph]}
  \BibitemShut {NoStop}%
\bibitem [{\citenamefont {Ji}(2014)}]{Ji:2014gla}%
  \BibitemOpen
  \bibfield  {author} {\bibinfo {author} {\bibfnamefont {X.}~\bibnamefont
  {Ji}},\ }\href {\doibase 10.1007/s11433-014-5492-3} {\bibfield  {journal}
  {\bibinfo  {journal} {Sci. China Phys. Mech. Astron.}\ }\textbf {\bibinfo
  {volume} {57}},\ \bibinfo {pages} {1407} (\bibinfo {year} {2014})},\ \Eprint
  {http://arxiv.org/abs/1404.6680} {arXiv:1404.6680 [hep-ph]} \BibitemShut
  {NoStop}%
\bibitem [{\citenamefont
  {Radyushkin}(2017{\natexlab{a}})}]{Radyushkin:2017cyf}%
  \BibitemOpen
  \bibfield  {author} {\bibinfo {author} {\bibfnamefont {A.}~\bibnamefont
  {Radyushkin}},\ }\href {\doibase 10.1103/PhysRevD.96.034025} {\bibfield
  {journal} {\bibinfo  {journal} {Phys. Rev. D}\ }\textbf {\bibinfo {volume}
  {96}},\ \bibinfo {pages} {034025} (\bibinfo {year} {2017}{\natexlab{a}})},\
  \Eprint {http://arxiv.org/abs/1705.01488} {arXiv:1705.01488 [hep-ph]}
  \BibitemShut {NoStop}%
\bibitem [{\citenamefont {Orginos}\ \emph {et~al.}(2017)\citenamefont
  {Orginos}, \citenamefont {Radyushkin}, \citenamefont {Karpie},\ and\
  \citenamefont {Zafeiropoulos}}]{Orginos:2017kos}%
  \BibitemOpen
  \bibfield  {author} {\bibinfo {author} {\bibfnamefont {K.}~\bibnamefont
  {Orginos}}, \bibinfo {author} {\bibfnamefont {A.}~\bibnamefont {Radyushkin}},
  \bibinfo {author} {\bibfnamefont {J.}~\bibnamefont {Karpie}}, \ and\ \bibinfo
  {author} {\bibfnamefont {S.}~\bibnamefont {Zafeiropoulos}},\ }\href {\doibase
  10.1103/PhysRevD.96.094503} {\bibfield  {journal} {\bibinfo  {journal} {Phys.
  Rev. D}\ }\textbf {\bibinfo {volume} {96}},\ \bibinfo {pages} {094503}
  (\bibinfo {year} {2017})},\ \Eprint {http://arxiv.org/abs/1706.05373}
  {arXiv:1706.05373 [hep-ph]} \BibitemShut {NoStop}%
\bibitem [{\citenamefont {Braun}\ and\ \citenamefont
  {M\"uller}(2008)}]{Braun:2007wv}%
  \BibitemOpen
  \bibfield  {author} {\bibinfo {author} {\bibfnamefont {V.}~\bibnamefont
  {Braun}}\ and\ \bibinfo {author} {\bibfnamefont {D.}~\bibnamefont
  {M\"uller}},\ }\href {\doibase 10.1140/epjc/s10052-008-0608-4} {\bibfield
  {journal} {\bibinfo  {journal} {Eur. Phys. J. C}\ }\textbf {\bibinfo {volume}
  {55}},\ \bibinfo {pages} {349} (\bibinfo {year} {2008})},\ \Eprint
  {http://arxiv.org/abs/0709.1348} {arXiv:0709.1348 [hep-ph]} \BibitemShut
  {NoStop}%
\bibitem [{\citenamefont {Ma}\ and\ \citenamefont
  {Qiu}(2018{\natexlab{a}})}]{Ma:2014jla}%
  \BibitemOpen
  \bibfield  {author} {\bibinfo {author} {\bibfnamefont {Y.-Q.}\ \bibnamefont
  {Ma}}\ and\ \bibinfo {author} {\bibfnamefont {J.-W.}\ \bibnamefont {Qiu}},\
  }\href {\doibase 10.1103/PhysRevD.98.074021} {\bibfield  {journal} {\bibinfo
  {journal} {Phys. Rev. D}\ }\textbf {\bibinfo {volume} {98}},\ \bibinfo
  {pages} {074021} (\bibinfo {year} {2018}{\natexlab{a}})},\ \Eprint
  {http://arxiv.org/abs/1404.6860} {arXiv:1404.6860 [hep-ph]} \BibitemShut
  {NoStop}%
\bibitem [{\citenamefont {Ma}\ and\ \citenamefont
  {Qiu}(2018{\natexlab{b}})}]{Ma:2017pxb}%
  \BibitemOpen
  \bibfield  {author} {\bibinfo {author} {\bibfnamefont {Y.-Q.}\ \bibnamefont
  {Ma}}\ and\ \bibinfo {author} {\bibfnamefont {J.-W.}\ \bibnamefont {Qiu}},\
  }\href {\doibase 10.1103/PhysRevLett.120.022003} {\bibfield  {journal}
  {\bibinfo  {journal} {Phys. Rev. Lett.}\ }\textbf {\bibinfo {volume} {120}},\
  \bibinfo {pages} {022003} (\bibinfo {year} {2018}{\natexlab{b}})},\ \Eprint
  {http://arxiv.org/abs/1709.03018} {arXiv:1709.03018 [hep-ph]} \BibitemShut
  {NoStop}%
\bibitem [{\citenamefont {Constantinou}(2020)}]{Constantinou:2020pek}%
  \BibitemOpen
  \bibfield  {author} {\bibinfo {author} {\bibfnamefont {M.}~\bibnamefont
  {Constantinou}},\ }in\ \href@noop {} {\emph {\bibinfo {booktitle} {{38th
  International Symposium on Lattice Field Theory}}}}\ (\bibinfo {year}
  {2020})\ \Eprint {http://arxiv.org/abs/2010.02445} {arXiv:2010.02445
  [hep-lat]} \BibitemShut {NoStop}%
\bibitem [{\citenamefont {Zhao}(2019)}]{Zhao:2018fyu}%
  \BibitemOpen
  \bibfield  {author} {\bibinfo {author} {\bibfnamefont {Y.}~\bibnamefont
  {Zhao}},\ }\href {\doibase 10.1142/S0217751X18300338} {\bibfield  {journal}
  {\bibinfo  {journal} {Int. J. Mod. Phys. A}\ }\textbf {\bibinfo {volume}
  {33}},\ \bibinfo {pages} {1830033} (\bibinfo {year} {2019})},\ \Eprint
  {http://arxiv.org/abs/1812.07192} {arXiv:1812.07192 [hep-ph]} \BibitemShut
  {NoStop}%
\bibitem [{\citenamefont {Cichy}\ and\ \citenamefont
  {Constantinou}(2019)}]{Cichy:2018mum}%
  \BibitemOpen
  \bibfield  {author} {\bibinfo {author} {\bibfnamefont {K.}~\bibnamefont
  {Cichy}}\ and\ \bibinfo {author} {\bibfnamefont {M.}~\bibnamefont
  {Constantinou}},\ }\href {\doibase 10.1155/2019/3036904} {\bibfield
  {journal} {\bibinfo  {journal} {Adv. High Energy Phys.}\ }\textbf {\bibinfo
  {volume} {2019}},\ \bibinfo {pages} {3036904} (\bibinfo {year} {2019})},\
  \Eprint {http://arxiv.org/abs/1811.07248} {arXiv:1811.07248 [hep-lat]}
  \BibitemShut {NoStop}%
\bibitem [{\citenamefont {Monahan}(2018)}]{Monahan:2018euv}%
  \BibitemOpen
  \bibfield  {author} {\bibinfo {author} {\bibfnamefont {C.}~\bibnamefont
  {Monahan}},\ }\href {\doibase 10.22323/1.334.0018} {\bibfield  {journal}
  {\bibinfo  {journal} {PoS}\ }\textbf {\bibinfo {volume} {LATTICE2018}},\
  \bibinfo {pages} {018} (\bibinfo {year} {2018})},\ \Eprint
  {http://arxiv.org/abs/1811.00678} {arXiv:1811.00678 [hep-lat]} \BibitemShut
  {NoStop}%
\bibitem [{\citenamefont {Ji}\ \emph {et~al.}(2020)\citenamefont {Ji},
  \citenamefont {Liu}, \citenamefont {Liu}, \citenamefont {Zhang},\ and\
  \citenamefont {Zhao}}]{Ji:2020ect}%
  \BibitemOpen
  \bibfield  {author} {\bibinfo {author} {\bibfnamefont {X.}~\bibnamefont
  {Ji}}, \bibinfo {author} {\bibfnamefont {Y.-S.}\ \bibnamefont {Liu}},
  \bibinfo {author} {\bibfnamefont {Y.}~\bibnamefont {Liu}}, \bibinfo {author}
  {\bibfnamefont {J.-H.}\ \bibnamefont {Zhang}}, \ and\ \bibinfo {author}
  {\bibfnamefont {Y.}~\bibnamefont {Zhao}},\ }\href@noop {} {\  (\bibinfo
  {year} {2020})},\ \Eprint {http://arxiv.org/abs/2004.03543} {arXiv:2004.03543
  [hep-ph]} \BibitemShut {NoStop}%
\bibitem [{\citenamefont {Gao}\ \emph {et~al.}(2020)\citenamefont {Gao},
  \citenamefont {Jin}, \citenamefont {Kallidonis}, \citenamefont {Karthik},
  \citenamefont {Mukherjee}, \citenamefont {Petreczky}, \citenamefont
  {Shugert}, \citenamefont {Syritsyn},\ and\ \citenamefont
  {Zhao}}]{Gao:2020ito}%
  \BibitemOpen
  \bibfield  {author} {\bibinfo {author} {\bibfnamefont {X.}~\bibnamefont
  {Gao}}, \bibinfo {author} {\bibfnamefont {L.}~\bibnamefont {Jin}}, \bibinfo
  {author} {\bibfnamefont {C.}~\bibnamefont {Kallidonis}}, \bibinfo {author}
  {\bibfnamefont {N.}~\bibnamefont {Karthik}}, \bibinfo {author} {\bibfnamefont
  {S.}~\bibnamefont {Mukherjee}}, \bibinfo {author} {\bibfnamefont
  {P.}~\bibnamefont {Petreczky}}, \bibinfo {author} {\bibfnamefont
  {C.}~\bibnamefont {Shugert}}, \bibinfo {author} {\bibfnamefont
  {S.}~\bibnamefont {Syritsyn}}, \ and\ \bibinfo {author} {\bibfnamefont
  {Y.}~\bibnamefont {Zhao}},\ }\href {\doibase 10.1103/PhysRevD.102.094513}
  {\bibfield  {journal} {\bibinfo  {journal} {Phys. Rev. D}\ }\textbf {\bibinfo
  {volume} {102}},\ \bibinfo {pages} {094513} (\bibinfo {year} {2020})},\
  \Eprint {http://arxiv.org/abs/2007.06590} {arXiv:2007.06590 [hep-lat]}
  \BibitemShut {NoStop}%
\bibitem [{\citenamefont {Zhang}\ \emph {et~al.}(2019)\citenamefont {Zhang},
  \citenamefont {Chen}, \citenamefont {Jin}, \citenamefont {Lin}, \citenamefont
  {Sch\"afer},\ and\ \citenamefont {Zhao}}]{Chen:2018fwa}%
  \BibitemOpen
  \bibfield  {author} {\bibinfo {author} {\bibfnamefont {J.-H.}\ \bibnamefont
  {Zhang}}, \bibinfo {author} {\bibfnamefont {J.-W.}\ \bibnamefont {Chen}},
  \bibinfo {author} {\bibfnamefont {L.}~\bibnamefont {Jin}}, \bibinfo {author}
  {\bibfnamefont {H.-W.}\ \bibnamefont {Lin}}, \bibinfo {author} {\bibfnamefont
  {A.}~\bibnamefont {Sch\"afer}}, \ and\ \bibinfo {author} {\bibfnamefont
  {Y.}~\bibnamefont {Zhao}},\ }\href {\doibase 10.1103/PhysRevD.100.034505}
  {\bibfield  {journal} {\bibinfo  {journal} {Phys. Rev. D}\ }\textbf {\bibinfo
  {volume} {100}},\ \bibinfo {pages} {034505} (\bibinfo {year} {2019})},\
  \Eprint {http://arxiv.org/abs/1804.01483} {arXiv:1804.01483 [hep-lat]}
  \BibitemShut {NoStop}%
\bibitem [{\citenamefont {Izubuchi}\ \emph {et~al.}(2019)\citenamefont
  {Izubuchi}, \citenamefont {Jin}, \citenamefont {Kallidonis}, \citenamefont
  {Karthik}, \citenamefont {Mukherjee}, \citenamefont {Petreczky},
  \citenamefont {Shugert},\ and\ \citenamefont {Syritsyn}}]{Izubuchi:2019lyk}%
  \BibitemOpen
  \bibfield  {author} {\bibinfo {author} {\bibfnamefont {T.}~\bibnamefont
  {Izubuchi}}, \bibinfo {author} {\bibfnamefont {L.}~\bibnamefont {Jin}},
  \bibinfo {author} {\bibfnamefont {C.}~\bibnamefont {Kallidonis}}, \bibinfo
  {author} {\bibfnamefont {N.}~\bibnamefont {Karthik}}, \bibinfo {author}
  {\bibfnamefont {S.}~\bibnamefont {Mukherjee}}, \bibinfo {author}
  {\bibfnamefont {P.}~\bibnamefont {Petreczky}}, \bibinfo {author}
  {\bibfnamefont {C.}~\bibnamefont {Shugert}}, \ and\ \bibinfo {author}
  {\bibfnamefont {S.}~\bibnamefont {Syritsyn}},\ }\href {\doibase
  10.1103/PhysRevD.100.034516} {\bibfield  {journal} {\bibinfo  {journal}
  {Phys. Rev. D}\ }\textbf {\bibinfo {volume} {100}},\ \bibinfo {pages}
  {034516} (\bibinfo {year} {2019})},\ \Eprint
  {http://arxiv.org/abs/1905.06349} {arXiv:1905.06349 [hep-lat]} \BibitemShut
  {NoStop}%
\bibitem [{\citenamefont {Jo\'o}\ \emph {et~al.}(2019)\citenamefont {Jo\'o},
  \citenamefont {Karpie}, \citenamefont {Orginos}, \citenamefont {Radyushkin},
  \citenamefont {Richards}, \citenamefont {Sufian},\ and\ \citenamefont
  {Zafeiropoulos}}]{Joo:2019bzr}%
  \BibitemOpen
  \bibfield  {author} {\bibinfo {author} {\bibfnamefont {B.}~\bibnamefont
  {Jo\'o}}, \bibinfo {author} {\bibfnamefont {J.}~\bibnamefont {Karpie}},
  \bibinfo {author} {\bibfnamefont {K.}~\bibnamefont {Orginos}}, \bibinfo
  {author} {\bibfnamefont {A.~V.}\ \bibnamefont {Radyushkin}}, \bibinfo
  {author} {\bibfnamefont {D.~G.}\ \bibnamefont {Richards}}, \bibinfo {author}
  {\bibfnamefont {R.~S.}\ \bibnamefont {Sufian}}, \ and\ \bibinfo {author}
  {\bibfnamefont {S.}~\bibnamefont {Zafeiropoulos}},\ }\href {\doibase
  10.1103/PhysRevD.100.114512} {\bibfield  {journal} {\bibinfo  {journal}
  {Phys. Rev. D}\ }\textbf {\bibinfo {volume} {100}},\ \bibinfo {pages}
  {114512} (\bibinfo {year} {2019})},\ \Eprint
  {http://arxiv.org/abs/1909.08517} {arXiv:1909.08517 [hep-lat]} \BibitemShut
  {NoStop}%
\bibitem [{\citenamefont {Lin}\ \emph {et~al.}(2020)\citenamefont {Lin},
  \citenamefont {Chen}, \citenamefont {Fan}, \citenamefont {Zhang},\ and\
  \citenamefont {Zhang}}]{Lin:2020ssv}%
  \BibitemOpen
  \bibfield  {author} {\bibinfo {author} {\bibfnamefont {H.-W.}\ \bibnamefont
  {Lin}}, \bibinfo {author} {\bibfnamefont {J.-W.}\ \bibnamefont {Chen}},
  \bibinfo {author} {\bibfnamefont {Z.}~\bibnamefont {Fan}}, \bibinfo {author}
  {\bibfnamefont {J.-H.}\ \bibnamefont {Zhang}}, \ and\ \bibinfo {author}
  {\bibfnamefont {R.}~\bibnamefont {Zhang}},\ }\href@noop {} {\  (\bibinfo
  {year} {2020})},\ \Eprint {http://arxiv.org/abs/2003.14128} {arXiv:2003.14128
  [hep-lat]} \BibitemShut {NoStop}%
\bibitem [{\citenamefont {Sufian}\ \emph {et~al.}(2019)\citenamefont {Sufian},
  \citenamefont {Karpie}, \citenamefont {Egerer}, \citenamefont {Orginos},
  \citenamefont {Qiu},\ and\ \citenamefont {Richards}}]{Sufian:2019bol}%
  \BibitemOpen
  \bibfield  {author} {\bibinfo {author} {\bibfnamefont {R.~S.}\ \bibnamefont
  {Sufian}}, \bibinfo {author} {\bibfnamefont {J.}~\bibnamefont {Karpie}},
  \bibinfo {author} {\bibfnamefont {C.}~\bibnamefont {Egerer}}, \bibinfo
  {author} {\bibfnamefont {K.}~\bibnamefont {Orginos}}, \bibinfo {author}
  {\bibfnamefont {J.-W.}\ \bibnamefont {Qiu}}, \ and\ \bibinfo {author}
  {\bibfnamefont {D.~G.}\ \bibnamefont {Richards}},\ }\href {\doibase
  10.1103/PhysRevD.99.074507} {\bibfield  {journal} {\bibinfo  {journal} {Phys.
  Rev. D}\ }\textbf {\bibinfo {volume} {99}},\ \bibinfo {pages} {074507}
  (\bibinfo {year} {2019})},\ \Eprint {http://arxiv.org/abs/1901.03921}
  {arXiv:1901.03921 [hep-lat]} \BibitemShut {NoStop}%
\bibitem [{\citenamefont {Sufian}\ \emph {et~al.}(2020)\citenamefont {Sufian},
  \citenamefont {Egerer}, \citenamefont {Karpie}, \citenamefont {Edwards},
  \citenamefont {Jo\'o}, \citenamefont {Ma}, \citenamefont {Orginos},
  \citenamefont {Qiu},\ and\ \citenamefont {Richards}}]{Sufian:2020vzb}%
  \BibitemOpen
  \bibfield  {author} {\bibinfo {author} {\bibfnamefont {R.~S.}\ \bibnamefont
  {Sufian}}, \bibinfo {author} {\bibfnamefont {C.}~\bibnamefont {Egerer}},
  \bibinfo {author} {\bibfnamefont {J.}~\bibnamefont {Karpie}}, \bibinfo
  {author} {\bibfnamefont {R.~G.}\ \bibnamefont {Edwards}}, \bibinfo {author}
  {\bibfnamefont {B.}~\bibnamefont {Jo\'o}}, \bibinfo {author} {\bibfnamefont
  {Y.-Q.}\ \bibnamefont {Ma}}, \bibinfo {author} {\bibfnamefont
  {K.}~\bibnamefont {Orginos}}, \bibinfo {author} {\bibfnamefont {J.-W.}\
  \bibnamefont {Qiu}}, \ and\ \bibinfo {author} {\bibfnamefont {D.~G.}\
  \bibnamefont {Richards}},\ }\href {\doibase 10.1103/PhysRevD.102.054508}
  {\bibfield  {journal} {\bibinfo  {journal} {Phys. Rev. D}\ }\textbf {\bibinfo
  {volume} {102}},\ \bibinfo {pages} {054508} (\bibinfo {year} {2020})},\
  \Eprint {http://arxiv.org/abs/2001.04960} {arXiv:2001.04960 [hep-lat]}
  \BibitemShut {NoStop}%
\bibitem [{\citenamefont {Aguilar}\ \emph {et~al.}(2019)\citenamefont {Aguilar}
  \emph {et~al.}}]{Aguilar:2019teb}%
  \BibitemOpen
  \bibfield  {author} {\bibinfo {author} {\bibfnamefont {A.~C.}\ \bibnamefont
  {Aguilar}} \emph {et~al.},\ }\href {\doibase 10.1140/epja/i2019-12885-0}
  {\bibfield  {journal} {\bibinfo  {journal} {Eur. Phys. J. A}\ }\textbf
  {\bibinfo {volume} {55}},\ \bibinfo {pages} {190} (\bibinfo {year} {2019})},\
  \Eprint {http://arxiv.org/abs/1907.08218} {arXiv:1907.08218 [nucl-ex]}
  \BibitemShut {NoStop}%
\bibitem [{\citenamefont {Adams}\ \emph {et~al.}(2018)\citenamefont {Adams}
  \emph {et~al.}}]{Denisov:2018unj}%
  \BibitemOpen
  \bibfield  {author} {\bibinfo {author} {\bibfnamefont {B.}~\bibnamefont
  {Adams}} \emph {et~al.},\ }\href@noop {} {\  (\bibinfo {year} {2018})},\
  \Eprint {http://arxiv.org/abs/1808.00848} {arXiv:1808.00848 [hep-ex]}
  \BibitemShut {NoStop}%
\bibitem [{\citenamefont {Karthik}\ and\ \citenamefont
  {Narayanan}(2018)}]{Karthik:2018nzf}%
  \BibitemOpen
  \bibfield  {author} {\bibinfo {author} {\bibfnamefont {N.}~\bibnamefont
  {Karthik}}\ and\ \bibinfo {author} {\bibfnamefont {R.}~\bibnamefont
  {Narayanan}},\ }\href {\doibase 10.1103/PhysRevD.97.054510} {\bibfield
  {journal} {\bibinfo  {journal} {Phys. Rev. D}\ }\textbf {\bibinfo {volume}
  {97}},\ \bibinfo {pages} {054510} (\bibinfo {year} {2018})},\ \Eprint
  {http://arxiv.org/abs/1801.02637} {arXiv:1801.02637 [hep-th]} \BibitemShut
  {NoStop}%
\bibitem [{\citenamefont {Seiberg}\ \emph {et~al.}(2016)\citenamefont
  {Seiberg}, \citenamefont {Senthil}, \citenamefont {Wang},\ and\ \citenamefont
  {Witten}}]{Seiberg:2016gmd}%
  \BibitemOpen
  \bibfield  {author} {\bibinfo {author} {\bibfnamefont {N.}~\bibnamefont
  {Seiberg}}, \bibinfo {author} {\bibfnamefont {T.}~\bibnamefont {Senthil}},
  \bibinfo {author} {\bibfnamefont {C.}~\bibnamefont {Wang}}, \ and\ \bibinfo
  {author} {\bibfnamefont {E.}~\bibnamefont {Witten}},\ }\href {\doibase
  10.1016/j.aop.2016.08.007} {\bibfield  {journal} {\bibinfo  {journal} {Annals
  Phys.}\ }\textbf {\bibinfo {volume} {374}},\ \bibinfo {pages} {395} (\bibinfo
  {year} {2016})},\ \Eprint {http://arxiv.org/abs/1606.01989} {arXiv:1606.01989
  [hep-th]} \BibitemShut {NoStop}%
\bibitem [{\citenamefont {Karch}\ and\ \citenamefont
  {Tong}(2016)}]{Karch:2016sxi}%
  \BibitemOpen
  \bibfield  {author} {\bibinfo {author} {\bibfnamefont {A.}~\bibnamefont
  {Karch}}\ and\ \bibinfo {author} {\bibfnamefont {D.}~\bibnamefont {Tong}},\
  }\href {\doibase 10.1103/PhysRevX.6.031043} {\bibfield  {journal} {\bibinfo
  {journal} {Phys. Rev. X}\ }\textbf {\bibinfo {volume} {6}},\ \bibinfo {pages}
  {031043} (\bibinfo {year} {2016})},\ \Eprint
  {http://arxiv.org/abs/1606.01893} {arXiv:1606.01893 [hep-th]} \BibitemShut
  {NoStop}%
\bibitem [{\citenamefont {Choi}\ \emph {et~al.}(2020)\citenamefont {Choi},
  \citenamefont {Delmastro}, \citenamefont {Gomis},\ and\ \citenamefont
  {Komargodski}}]{Choi:2018tuh}%
  \BibitemOpen
  \bibfield  {author} {\bibinfo {author} {\bibfnamefont {C.}~\bibnamefont
  {Choi}}, \bibinfo {author} {\bibfnamefont {D.}~\bibnamefont {Delmastro}},
  \bibinfo {author} {\bibfnamefont {J.}~\bibnamefont {Gomis}}, \ and\ \bibinfo
  {author} {\bibfnamefont {Z.}~\bibnamefont {Komargodski}},\ }\href {\doibase
  10.1007/JHEP03(2020)078} {\bibfield  {journal} {\bibinfo  {journal} {JHEP}\
  }\textbf {\bibinfo {volume} {03}},\ \bibinfo {pages} {078} (\bibinfo {year}
  {2020})},\ \Eprint {http://arxiv.org/abs/1810.07720} {arXiv:1810.07720
  [hep-th]} \BibitemShut {NoStop}%
\bibitem [{\citenamefont {Sharon}(2018)}]{Sharon:2018apk}%
  \BibitemOpen
  \bibfield  {author} {\bibinfo {author} {\bibfnamefont {A.}~\bibnamefont
  {Sharon}},\ }\href {\doibase 10.1007/JHEP08(2018)078} {\bibfield  {journal}
  {\bibinfo  {journal} {JHEP}\ }\textbf {\bibinfo {volume} {08}},\ \bibinfo
  {pages} {078} (\bibinfo {year} {2018})},\ \Eprint
  {http://arxiv.org/abs/1803.06983} {arXiv:1803.06983 [hep-th]} \BibitemShut
  {NoStop}%
\bibitem [{\citenamefont {Gazit}\ \emph {et~al.}(2018)\citenamefont {Gazit},
  \citenamefont {Assaad}, \citenamefont {Sachdev}, \citenamefont {Vishwanath},\
  and\ \citenamefont {Wang}}]{Gazit:2018vsa}%
  \BibitemOpen
  \bibfield  {author} {\bibinfo {author} {\bibfnamefont {S.}~\bibnamefont
  {Gazit}}, \bibinfo {author} {\bibfnamefont {F.~F.}\ \bibnamefont {Assaad}},
  \bibinfo {author} {\bibfnamefont {S.}~\bibnamefont {Sachdev}}, \bibinfo
  {author} {\bibfnamefont {A.}~\bibnamefont {Vishwanath}}, \ and\ \bibinfo
  {author} {\bibfnamefont {C.}~\bibnamefont {Wang}},\ }\href {\doibase
  10.1073/pnas.1806338115} {\bibfield  {journal} {\bibinfo  {journal} {Proc.
  Nat. Acad. Sci.}\ }\textbf {\bibinfo {volume} {115}},\ \bibinfo {pages}
  {E6987} (\bibinfo {year} {2018})},\ \Eprint {http://arxiv.org/abs/1804.01095}
  {arXiv:1804.01095 [cond-mat.str-el]} \BibitemShut {NoStop}%
\bibitem [{\citenamefont {Song}\ \emph {et~al.}(2020)\citenamefont {Song},
  \citenamefont {He}, \citenamefont {Vishwanath},\ and\ \citenamefont
  {Wang}}]{Song:2018ial}%
  \BibitemOpen
  \bibfield  {author} {\bibinfo {author} {\bibfnamefont {X.-Y.}\ \bibnamefont
  {Song}}, \bibinfo {author} {\bibfnamefont {Y.-C.}\ \bibnamefont {He}},
  \bibinfo {author} {\bibfnamefont {A.}~\bibnamefont {Vishwanath}}, \ and\
  \bibinfo {author} {\bibfnamefont {C.}~\bibnamefont {Wang}},\ }\href {\doibase
  10.1103/PhysRevX.10.011033} {\bibfield  {journal} {\bibinfo  {journal} {Phys.
  Rev. X}\ }\textbf {\bibinfo {volume} {10}},\ \bibinfo {pages} {011033}
  (\bibinfo {year} {2020})},\ \Eprint {http://arxiv.org/abs/1811.11182}
  {arXiv:1811.11182 [cond-mat.str-el]} \BibitemShut {NoStop}%
\bibitem [{\citenamefont {Ma}\ and\ \citenamefont {He}(2020)}]{Ma:2020pjs}%
  \BibitemOpen
  \bibfield  {author} {\bibinfo {author} {\bibfnamefont {R.}~\bibnamefont
  {Ma}}\ and\ \bibinfo {author} {\bibfnamefont {Y.-C.}\ \bibnamefont {He}},\
  }\href {\doibase 10.1103/PhysRevResearch.2.033348} {\bibfield  {journal}
  {\bibinfo  {journal} {Phys. Rev. Res.}\ }\textbf {\bibinfo {volume} {2}},\
  \bibinfo {pages} {033348} (\bibinfo {year} {2020})},\ \Eprint
  {http://arxiv.org/abs/2003.05954} {arXiv:2003.05954 [cond-mat.str-el]}
  \BibitemShut {NoStop}%
\bibitem [{\citenamefont {Chester}\ and\ \citenamefont
  {Pufu}(2016{\natexlab{a}})}]{Chester:2016ref}%
  \BibitemOpen
  \bibfield  {author} {\bibinfo {author} {\bibfnamefont {S.~M.}\ \bibnamefont
  {Chester}}\ and\ \bibinfo {author} {\bibfnamefont {S.~S.}\ \bibnamefont
  {Pufu}},\ }\href {\doibase 10.1007/JHEP08(2016)069} {\bibfield  {journal}
  {\bibinfo  {journal} {JHEP}\ }\textbf {\bibinfo {volume} {08}},\ \bibinfo
  {pages} {069} (\bibinfo {year} {2016}{\natexlab{a}})},\ \Eprint
  {http://arxiv.org/abs/1603.05582} {arXiv:1603.05582 [hep-th]} \BibitemShut
  {NoStop}%
\bibitem [{\citenamefont {Chester}\ and\ \citenamefont
  {Pufu}(2016{\natexlab{b}})}]{Chester:2016wrc}%
  \BibitemOpen
  \bibfield  {author} {\bibinfo {author} {\bibfnamefont {S.~M.}\ \bibnamefont
  {Chester}}\ and\ \bibinfo {author} {\bibfnamefont {S.~S.}\ \bibnamefont
  {Pufu}},\ }\href {\doibase 10.1007/JHEP08(2016)019} {\bibfield  {journal}
  {\bibinfo  {journal} {JHEP}\ }\textbf {\bibinfo {volume} {08}},\ \bibinfo
  {pages} {019} (\bibinfo {year} {2016}{\natexlab{b}})},\ \Eprint
  {http://arxiv.org/abs/1601.03476} {arXiv:1601.03476 [hep-th]} \BibitemShut
  {NoStop}%
\bibitem [{\citenamefont {Pufu}(2014)}]{Pufu:2013vpa}%
  \BibitemOpen
  \bibfield  {author} {\bibinfo {author} {\bibfnamefont {S.~S.}\ \bibnamefont
  {Pufu}},\ }\href {\doibase 10.1103/PhysRevD.89.065016} {\bibfield  {journal}
  {\bibinfo  {journal} {Phys. Rev. D}\ }\textbf {\bibinfo {volume} {89}},\
  \bibinfo {pages} {065016} (\bibinfo {year} {2014})},\ \Eprint
  {http://arxiv.org/abs/1303.6125} {arXiv:1303.6125 [hep-th]} \BibitemShut
  {NoStop}%
\bibitem [{\citenamefont {Ji}\ \emph {et~al.}(2019)\citenamefont {Ji},
  \citenamefont {Liu},\ and\ \citenamefont {Zahed}}]{Ji:2018waw}%
  \BibitemOpen
  \bibfield  {author} {\bibinfo {author} {\bibfnamefont {X.}~\bibnamefont
  {Ji}}, \bibinfo {author} {\bibfnamefont {Y.}~\bibnamefont {Liu}}, \ and\
  \bibinfo {author} {\bibfnamefont {I.}~\bibnamefont {Zahed}},\ }\href
  {\doibase 10.1103/PhysRevD.99.054008} {\bibfield  {journal} {\bibinfo
  {journal} {Phys. Rev. D}\ }\textbf {\bibinfo {volume} {99}},\ \bibinfo
  {pages} {054008} (\bibinfo {year} {2019})},\ \Eprint
  {http://arxiv.org/abs/1807.07528} {arXiv:1807.07528 [hep-ph]} \BibitemShut
  {NoStop}%
\bibitem [{\citenamefont {Jia}\ \emph {et~al.}(2018)\citenamefont {Jia},
  \citenamefont {Liang}, \citenamefont {Xiong},\ and\ \citenamefont
  {Yu}}]{Jia:2018qee}%
  \BibitemOpen
  \bibfield  {author} {\bibinfo {author} {\bibfnamefont {Y.}~\bibnamefont
  {Jia}}, \bibinfo {author} {\bibfnamefont {S.}~\bibnamefont {Liang}}, \bibinfo
  {author} {\bibfnamefont {X.}~\bibnamefont {Xiong}}, \ and\ \bibinfo {author}
  {\bibfnamefont {R.}~\bibnamefont {Yu}},\ }\href {\doibase
  10.1103/PhysRevD.98.054011} {\bibfield  {journal} {\bibinfo  {journal} {Phys.
  Rev. D}\ }\textbf {\bibinfo {volume} {98}},\ \bibinfo {pages} {054011}
  (\bibinfo {year} {2018})},\ \Eprint {http://arxiv.org/abs/1804.04644}
  {arXiv:1804.04644 [hep-th]} \BibitemShut {NoStop}%
\bibitem [{\citenamefont {Del~Debbio}\ \emph {et~al.}(2020)\citenamefont
  {Del~Debbio}, \citenamefont {Giani},\ and\ \citenamefont
  {Monahan}}]{DelDebbio:2020cbz}%
  \BibitemOpen
  \bibfield  {author} {\bibinfo {author} {\bibfnamefont {L.}~\bibnamefont
  {Del~Debbio}}, \bibinfo {author} {\bibfnamefont {T.}~\bibnamefont {Giani}}, \
  and\ \bibinfo {author} {\bibfnamefont {C.~J.}\ \bibnamefont {Monahan}},\
  }\href {\doibase 10.1007/JHEP09(2020)021} {\bibfield  {journal} {\bibinfo
  {journal} {JHEP}\ }\textbf {\bibinfo {volume} {09}},\ \bibinfo {pages} {021}
  (\bibinfo {year} {2020})},\ \Eprint {http://arxiv.org/abs/2007.02131}
  {arXiv:2007.02131 [hep-lat]} \BibitemShut {NoStop}%
\bibitem [{\citenamefont {Kock}\ \emph {et~al.}(2020)\citenamefont {Kock},
  \citenamefont {Liu},\ and\ \citenamefont {Zahed}}]{Kock:2020frx}%
  \BibitemOpen
  \bibfield  {author} {\bibinfo {author} {\bibfnamefont {A.}~\bibnamefont
  {Kock}}, \bibinfo {author} {\bibfnamefont {Y.}~\bibnamefont {Liu}}, \ and\
  \bibinfo {author} {\bibfnamefont {I.}~\bibnamefont {Zahed}},\ }\href
  {\doibase 10.1103/PhysRevD.102.014039} {\bibfield  {journal} {\bibinfo
  {journal} {Phys. Rev. D}\ }\textbf {\bibinfo {volume} {102}},\ \bibinfo
  {pages} {014039} (\bibinfo {year} {2020})},\ \Eprint
  {http://arxiv.org/abs/2004.01595} {arXiv:2004.01595 [hep-ph]} \BibitemShut
  {NoStop}%
\bibitem [{\citenamefont {Magnea}(2000)}]{Magnea:1999iv}%
  \BibitemOpen
  \bibfield  {author} {\bibinfo {author} {\bibfnamefont {U.}~\bibnamefont
  {Magnea}},\ }\href {\doibase 10.1103/PhysRevD.61.056005} {\bibfield
  {journal} {\bibinfo  {journal} {Phys. Rev. D}\ }\textbf {\bibinfo {volume}
  {61}},\ \bibinfo {pages} {056005} (\bibinfo {year} {2000})},\ \Eprint
  {http://arxiv.org/abs/hep-th/9907096} {arXiv:hep-th/9907096} \BibitemShut
  {NoStop}%
\bibitem [{\citenamefont {Pisarski}(1984)}]{Pisarski:1984dj}%
  \BibitemOpen
  \bibfield  {author} {\bibinfo {author} {\bibfnamefont {R.~D.}\ \bibnamefont
  {Pisarski}},\ }\href {\doibase 10.1103/PhysRevD.29.2423} {\bibfield
  {journal} {\bibinfo  {journal} {Phys. Rev. D}\ }\textbf {\bibinfo {volume}
  {29}},\ \bibinfo {pages} {2423} (\bibinfo {year} {1984})}\BibitemShut
  {NoStop}%
\bibitem [{\citenamefont {Vafa}\ and\ \citenamefont
  {Witten}(1984)}]{Vafa:1984xh}%
  \BibitemOpen
  \bibfield  {author} {\bibinfo {author} {\bibfnamefont {C.}~\bibnamefont
  {Vafa}}\ and\ \bibinfo {author} {\bibfnamefont {E.}~\bibnamefont {Witten}},\
  }\href {\doibase 10.1007/BF01212397} {\bibfield  {journal} {\bibinfo
  {journal} {Commun. Math. Phys.}\ }\textbf {\bibinfo {volume} {95}},\ \bibinfo
  {pages} {257} (\bibinfo {year} {1984})}\BibitemShut {NoStop}%
\bibitem [{\citenamefont {Appelquist}\ and\ \citenamefont
  {Nash}(1990)}]{Appelquist:1989tc}%
  \BibitemOpen
  \bibfield  {author} {\bibinfo {author} {\bibfnamefont {T.}~\bibnamefont
  {Appelquist}}\ and\ \bibinfo {author} {\bibfnamefont {D.}~\bibnamefont
  {Nash}},\ }\href {\doibase 10.1103/PhysRevLett.64.721} {\bibfield  {journal}
  {\bibinfo  {journal} {Phys. Rev. Lett.}\ }\textbf {\bibinfo {volume} {64}},\
  \bibinfo {pages} {721} (\bibinfo {year} {1990})}\BibitemShut {NoStop}%
\bibitem [{\citenamefont {Goldman}\ and\ \citenamefont
  {Mulligan}(2016)}]{Goldman:2016wwk}%
  \BibitemOpen
  \bibfield  {author} {\bibinfo {author} {\bibfnamefont {H.}~\bibnamefont
  {Goldman}}\ and\ \bibinfo {author} {\bibfnamefont {M.}~\bibnamefont
  {Mulligan}},\ }\href {\doibase 10.1103/PhysRevD.94.065031} {\bibfield
  {journal} {\bibinfo  {journal} {Phys. Rev. D}\ }\textbf {\bibinfo {volume}
  {94}},\ \bibinfo {pages} {065031} (\bibinfo {year} {2016})},\ \Eprint
  {http://arxiv.org/abs/1606.07067} {arXiv:1606.07067 [hep-th]} \BibitemShut
  {NoStop}%
\bibitem [{\citenamefont {Niemi}\ and\ \citenamefont
  {Semenoff}(1983)}]{Niemi:1983rq}%
  \BibitemOpen
  \bibfield  {author} {\bibinfo {author} {\bibfnamefont {A.}~\bibnamefont
  {Niemi}}\ and\ \bibinfo {author} {\bibfnamefont {G.}~\bibnamefont
  {Semenoff}},\ }\href {\doibase 10.1103/PhysRevLett.51.2077} {\bibfield
  {journal} {\bibinfo  {journal} {Phys. Rev. Lett.}\ }\textbf {\bibinfo
  {volume} {51}},\ \bibinfo {pages} {2077} (\bibinfo {year}
  {1983})}\BibitemShut {NoStop}%
\bibitem [{\citenamefont {Redlich}(1984)}]{Redlich:1983dv}%
  \BibitemOpen
  \bibfield  {author} {\bibinfo {author} {\bibfnamefont {A.}~\bibnamefont
  {Redlich}},\ }\href {\doibase 10.1103/PhysRevD.29.2366} {\bibfield  {journal}
  {\bibinfo  {journal} {Phys. Rev. D}\ }\textbf {\bibinfo {volume} {29}},\
  \bibinfo {pages} {2366} (\bibinfo {year} {1984})}\BibitemShut {NoStop}%
\bibitem [{\citenamefont {Thomson}\ and\ \citenamefont
  {Sachdev}(2017)}]{Thomson:2016ttt}%
  \BibitemOpen
  \bibfield  {author} {\bibinfo {author} {\bibfnamefont {A.}~\bibnamefont
  {Thomson}}\ and\ \bibinfo {author} {\bibfnamefont {S.}~\bibnamefont
  {Sachdev}},\ }\href {\doibase 10.1103/PhysRevB.95.205128} {\bibfield
  {journal} {\bibinfo  {journal} {Phys. Rev. B}\ }\textbf {\bibinfo {volume}
  {95}},\ \bibinfo {pages} {205128} (\bibinfo {year} {2017})},\ \Eprint
  {http://arxiv.org/abs/1607.05279} {arXiv:1607.05279 [cond-mat.str-el]}
  \BibitemShut {NoStop}%
\bibitem [{\citenamefont {Del~Debbio}\ and\ \citenamefont
  {Zwicky}(2010)}]{DelDebbio:2010ze}%
  \BibitemOpen
  \bibfield  {author} {\bibinfo {author} {\bibfnamefont {L.}~\bibnamefont
  {Del~Debbio}}\ and\ \bibinfo {author} {\bibfnamefont {R.}~\bibnamefont
  {Zwicky}},\ }\href {\doibase 10.1103/PhysRevD.82.014502} {\bibfield
  {journal} {\bibinfo  {journal} {Phys. Rev. D}\ }\textbf {\bibinfo {volume}
  {82}},\ \bibinfo {pages} {014502} (\bibinfo {year} {2010})},\ \Eprint
  {http://arxiv.org/abs/1005.2371} {arXiv:1005.2371 [hep-ph]} \BibitemShut
  {NoStop}%
\bibitem [{\citenamefont {Rummukainen}\ and\ \citenamefont
  {Gottlieb}(1995)}]{Rummukainen:1995vs}%
  \BibitemOpen
  \bibfield  {author} {\bibinfo {author} {\bibfnamefont {K.}~\bibnamefont
  {Rummukainen}}\ and\ \bibinfo {author} {\bibfnamefont {S.~A.}\ \bibnamefont
  {Gottlieb}},\ }\href {\doibase 10.1016/0550-3213(95)00313-H} {\bibfield
  {journal} {\bibinfo  {journal} {Nucl. Phys. B}\ }\textbf {\bibinfo {volume}
  {450}},\ \bibinfo {pages} {397} (\bibinfo {year} {1995})},\ \Eprint
  {http://arxiv.org/abs/hep-lat/9503028} {arXiv:hep-lat/9503028} \BibitemShut
  {NoStop}%
\bibitem [{\citenamefont {Braun}\ \emph {et~al.}(1995)\citenamefont {Braun},
  \citenamefont {Gornicki},\ and\ \citenamefont {Mankiewicz}}]{Braun:1994jq}%
  \BibitemOpen
  \bibfield  {author} {\bibinfo {author} {\bibfnamefont {V.}~\bibnamefont
  {Braun}}, \bibinfo {author} {\bibfnamefont {P.}~\bibnamefont {Gornicki}}, \
  and\ \bibinfo {author} {\bibfnamefont {L.}~\bibnamefont {Mankiewicz}},\
  }\href {\doibase 10.1103/PhysRevD.51.6036} {\bibfield  {journal} {\bibinfo
  {journal} {Phys. Rev. D}\ }\textbf {\bibinfo {volume} {51}},\ \bibinfo
  {pages} {6036} (\bibinfo {year} {1995})},\ \Eprint
  {http://arxiv.org/abs/hep-ph/9410318} {arXiv:hep-ph/9410318} \BibitemShut
  {NoStop}%
\bibitem [{\citenamefont {Carlson}\ and\ \citenamefont
  {Freid}(2017)}]{Carlson:2017gpk}%
  \BibitemOpen
  \bibfield  {author} {\bibinfo {author} {\bibfnamefont {C.~E.}\ \bibnamefont
  {Carlson}}\ and\ \bibinfo {author} {\bibfnamefont {M.}~\bibnamefont
  {Freid}},\ }\href {\doibase 10.1103/PhysRevD.95.094504} {\bibfield  {journal}
  {\bibinfo  {journal} {Phys. Rev. D}\ }\textbf {\bibinfo {volume} {95}},\
  \bibinfo {pages} {094504} (\bibinfo {year} {2017})},\ \Eprint
  {http://arxiv.org/abs/1702.05775} {arXiv:1702.05775 [hep-ph]} \BibitemShut
  {NoStop}%
\bibitem [{\citenamefont {Brice\~no}\ \emph {et~al.}(2017)\citenamefont
  {Brice\~no}, \citenamefont {Hansen},\ and\ \citenamefont
  {Monahan}}]{Briceno:2017cpo}%
  \BibitemOpen
  \bibfield  {author} {\bibinfo {author} {\bibfnamefont {R.~A.}\ \bibnamefont
  {Brice\~no}}, \bibinfo {author} {\bibfnamefont {M.~T.}\ \bibnamefont
  {Hansen}}, \ and\ \bibinfo {author} {\bibfnamefont {C.~J.}\ \bibnamefont
  {Monahan}},\ }\href {\doibase 10.1103/PhysRevD.96.014502} {\bibfield
  {journal} {\bibinfo  {journal} {Phys. Rev. D}\ }\textbf {\bibinfo {volume}
  {96}},\ \bibinfo {pages} {014502} (\bibinfo {year} {2017})},\ \Eprint
  {http://arxiv.org/abs/1703.06072} {arXiv:1703.06072 [hep-lat]} \BibitemShut
  {NoStop}%
\bibitem [{\citenamefont {Martinelli}\ and\ \citenamefont
  {Sachrajda}(1987)}]{Martinelli:1987zd}%
  \BibitemOpen
  \bibfield  {author} {\bibinfo {author} {\bibfnamefont {G.}~\bibnamefont
  {Martinelli}}\ and\ \bibinfo {author} {\bibfnamefont {C.~T.}\ \bibnamefont
  {Sachrajda}},\ }\href {\doibase 10.1016/0370-2693(87)90601-0} {\bibfield
  {journal} {\bibinfo  {journal} {Phys. Lett. B}\ }\textbf {\bibinfo {volume}
  {196}},\ \bibinfo {pages} {184} (\bibinfo {year} {1987})}\BibitemShut
  {NoStop}%
\bibitem [{\citenamefont {Izubuchi}\ \emph {et~al.}(2018)\citenamefont
  {Izubuchi}, \citenamefont {Ji}, \citenamefont {Jin}, \citenamefont
  {Stewart},\ and\ \citenamefont {Zhao}}]{Izubuchi:2018srq}%
  \BibitemOpen
  \bibfield  {author} {\bibinfo {author} {\bibfnamefont {T.}~\bibnamefont
  {Izubuchi}}, \bibinfo {author} {\bibfnamefont {X.}~\bibnamefont {Ji}},
  \bibinfo {author} {\bibfnamefont {L.}~\bibnamefont {Jin}}, \bibinfo {author}
  {\bibfnamefont {I.~W.}\ \bibnamefont {Stewart}}, \ and\ \bibinfo {author}
  {\bibfnamefont {Y.}~\bibnamefont {Zhao}},\ }\href {\doibase
  10.1103/PhysRevD.98.056004} {\bibfield  {journal} {\bibinfo  {journal} {Phys.
  Rev. D}\ }\textbf {\bibinfo {volume} {98}},\ \bibinfo {pages} {056004}
  (\bibinfo {year} {2018})},\ \Eprint {http://arxiv.org/abs/1801.03917}
  {arXiv:1801.03917 [hep-ph]} \BibitemShut {NoStop}%
\bibitem [{\citenamefont {Chen}\ \emph
  {et~al.}(2016{\natexlab{b}})\citenamefont {Chen}, \citenamefont {Cohen},
  \citenamefont {Ji}, \citenamefont {Lin},\ and\ \citenamefont
  {Zhang}}]{Chen:2016utp}%
  \BibitemOpen
  \bibfield  {author} {\bibinfo {author} {\bibfnamefont {J.-W.}\ \bibnamefont
  {Chen}}, \bibinfo {author} {\bibfnamefont {S.~D.}\ \bibnamefont {Cohen}},
  \bibinfo {author} {\bibfnamefont {X.}~\bibnamefont {Ji}}, \bibinfo {author}
  {\bibfnamefont {H.-W.}\ \bibnamefont {Lin}}, \ and\ \bibinfo {author}
  {\bibfnamefont {J.-H.}\ \bibnamefont {Zhang}},\ }\href {\doibase
  10.1016/j.nuclphysb.2016.07.033} {\bibfield  {journal} {\bibinfo  {journal}
  {Nucl. Phys. B}\ }\textbf {\bibinfo {volume} {911}},\ \bibinfo {pages} {246}
  (\bibinfo {year} {2016}{\natexlab{b}})},\ \Eprint
  {http://arxiv.org/abs/1603.06664} {arXiv:1603.06664 [hep-ph]} \BibitemShut
  {NoStop}%
\bibitem [{\citenamefont
  {Radyushkin}(2017{\natexlab{b}})}]{Radyushkin:2017ffo}%
  \BibitemOpen
  \bibfield  {author} {\bibinfo {author} {\bibfnamefont {A.}~\bibnamefont
  {Radyushkin}},\ }\href {\doibase 10.1016/j.physletb.2017.05.024} {\bibfield
  {journal} {\bibinfo  {journal} {Phys. Lett. B}\ }\textbf {\bibinfo {volume}
  {770}},\ \bibinfo {pages} {514} (\bibinfo {year} {2017}{\natexlab{b}})},\
  \Eprint {http://arxiv.org/abs/1702.01726} {arXiv:1702.01726 [hep-ph]}
  \BibitemShut {NoStop}%
\bibitem [{\citenamefont {Dotsenko}\ and\ \citenamefont
  {Vergeles}(1980)}]{Dotsenko:1979wb}%
  \BibitemOpen
  \bibfield  {author} {\bibinfo {author} {\bibfnamefont {V.}~\bibnamefont
  {Dotsenko}}\ and\ \bibinfo {author} {\bibfnamefont {S.}~\bibnamefont
  {Vergeles}},\ }\href {\doibase 10.1016/0550-3213(80)90103-0} {\bibfield
  {journal} {\bibinfo  {journal} {Nucl. Phys. B}\ }\textbf {\bibinfo {volume}
  {169}},\ \bibinfo {pages} {527} (\bibinfo {year} {1980})}\BibitemShut
  {NoStop}%
\bibitem [{\citenamefont {Ishikawa}\ \emph {et~al.}(2017)\citenamefont
  {Ishikawa}, \citenamefont {Ma}, \citenamefont {Qiu},\ and\ \citenamefont
  {Yoshida}}]{Ishikawa:2017faj}%
  \BibitemOpen
  \bibfield  {author} {\bibinfo {author} {\bibfnamefont {T.}~\bibnamefont
  {Ishikawa}}, \bibinfo {author} {\bibfnamefont {Y.-Q.}\ \bibnamefont {Ma}},
  \bibinfo {author} {\bibfnamefont {J.-W.}\ \bibnamefont {Qiu}}, \ and\
  \bibinfo {author} {\bibfnamefont {S.}~\bibnamefont {Yoshida}},\ }\href
  {\doibase 10.1103/PhysRevD.96.094019} {\bibfield  {journal} {\bibinfo
  {journal} {Phys. Rev. D}\ }\textbf {\bibinfo {volume} {96}},\ \bibinfo
  {pages} {094019} (\bibinfo {year} {2017})},\ \Eprint
  {http://arxiv.org/abs/1707.03107} {arXiv:1707.03107 [hep-ph]} \BibitemShut
  {NoStop}%
\bibitem [{\citenamefont {Ji}\ \emph {et~al.}(2018)\citenamefont {Ji},
  \citenamefont {Zhang},\ and\ \citenamefont {Zhao}}]{Ji:2017oey}%
  \BibitemOpen
  \bibfield  {author} {\bibinfo {author} {\bibfnamefont {X.}~\bibnamefont
  {Ji}}, \bibinfo {author} {\bibfnamefont {J.-H.}\ \bibnamefont {Zhang}}, \
  and\ \bibinfo {author} {\bibfnamefont {Y.}~\bibnamefont {Zhao}},\ }\href
  {\doibase 10.1103/PhysRevLett.120.112001} {\bibfield  {journal} {\bibinfo
  {journal} {Phys. Rev. Lett.}\ }\textbf {\bibinfo {volume} {120}},\ \bibinfo
  {pages} {112001} (\bibinfo {year} {2018})},\ \Eprint
  {http://arxiv.org/abs/1706.08962} {arXiv:1706.08962 [hep-ph]} \BibitemShut
  {NoStop}%
\bibitem [{\citenamefont {Teper}(1999)}]{Teper:1998te}%
  \BibitemOpen
  \bibfield  {author} {\bibinfo {author} {\bibfnamefont {M.~J.}\ \bibnamefont
  {Teper}},\ }\href {\doibase 10.1103/PhysRevD.59.014512} {\bibfield  {journal}
  {\bibinfo  {journal} {Phys. Rev. D}\ }\textbf {\bibinfo {volume} {59}},\
  \bibinfo {pages} {014512} (\bibinfo {year} {1999})},\ \Eprint
  {http://arxiv.org/abs/hep-lat/9804008} {arXiv:hep-lat/9804008} \BibitemShut
  {NoStop}%
\bibitem [{\citenamefont {Karthik}\ and\ \citenamefont
  {Narayanan}(2016{\natexlab{a}})}]{Karthik:2015sgq}%
  \BibitemOpen
  \bibfield  {author} {\bibinfo {author} {\bibfnamefont {N.}~\bibnamefont
  {Karthik}}\ and\ \bibinfo {author} {\bibfnamefont {R.}~\bibnamefont
  {Narayanan}},\ }\href {\doibase 10.1103/PhysRevD.93.045020} {\bibfield
  {journal} {\bibinfo  {journal} {Phys. Rev. D}\ }\textbf {\bibinfo {volume}
  {93}},\ \bibinfo {pages} {045020} (\bibinfo {year} {2016}{\natexlab{a}})},\
  \Eprint {http://arxiv.org/abs/1512.02993} {arXiv:1512.02993 [hep-lat]}
  \BibitemShut {NoStop}%
\bibitem [{\citenamefont {Karthik}\ and\ \citenamefont
  {Narayanan}(2015)}]{Karthik:2015sza}%
  \BibitemOpen
  \bibfield  {author} {\bibinfo {author} {\bibfnamefont {N.}~\bibnamefont
  {Karthik}}\ and\ \bibinfo {author} {\bibfnamefont {R.}~\bibnamefont
  {Narayanan}},\ }\href {\doibase 10.1103/PhysRevD.92.025003} {\bibfield
  {journal} {\bibinfo  {journal} {Phys. Rev. D}\ }\textbf {\bibinfo {volume}
  {92}},\ \bibinfo {pages} {025003} (\bibinfo {year} {2015})},\ \Eprint
  {http://arxiv.org/abs/1505.01051} {arXiv:1505.01051 [hep-th]} \BibitemShut
  {NoStop}%
\bibitem [{\citenamefont {Morningstar}\ and\ \citenamefont
  {Peardon}(2004)}]{Morningstar:2003gk}%
  \BibitemOpen
  \bibfield  {author} {\bibinfo {author} {\bibfnamefont {C.}~\bibnamefont
  {Morningstar}}\ and\ \bibinfo {author} {\bibfnamefont {M.~J.}\ \bibnamefont
  {Peardon}},\ }\href {\doibase 10.1103/PhysRevD.69.054501} {\bibfield
  {journal} {\bibinfo  {journal} {Phys. Rev. D}\ }\textbf {\bibinfo {volume}
  {69}},\ \bibinfo {pages} {054501} (\bibinfo {year} {2004})},\ \Eprint
  {http://arxiv.org/abs/hep-lat/0311018} {arXiv:hep-lat/0311018} \BibitemShut
  {NoStop}%
\bibitem [{\citenamefont {Capitani}\ \emph {et~al.}(2006)\citenamefont
  {Capitani}, \citenamefont {Durr},\ and\ \citenamefont
  {Hoelbling}}]{Capitani:2006ni}%
  \BibitemOpen
  \bibfield  {author} {\bibinfo {author} {\bibfnamefont {S.}~\bibnamefont
  {Capitani}}, \bibinfo {author} {\bibfnamefont {S.}~\bibnamefont {Durr}}, \
  and\ \bibinfo {author} {\bibfnamefont {C.}~\bibnamefont {Hoelbling}},\ }\href
  {\doibase 10.1088/1126-6708/2006/11/028} {\bibfield  {journal} {\bibinfo
  {journal} {JHEP}\ }\textbf {\bibinfo {volume} {11}},\ \bibinfo {pages} {028}
  (\bibinfo {year} {2006})},\ \Eprint {http://arxiv.org/abs/hep-lat/0607006}
  {arXiv:hep-lat/0607006} \BibitemShut {NoStop}%
\bibitem [{\citenamefont {Gupta}\ and\ \citenamefont
  {Karthik}(2013)}]{Gupta:2013vha}%
  \BibitemOpen
  \bibfield  {author} {\bibinfo {author} {\bibfnamefont {S.}~\bibnamefont
  {Gupta}}\ and\ \bibinfo {author} {\bibfnamefont {N.}~\bibnamefont
  {Karthik}},\ }\href {\doibase 10.1103/PhysRevD.87.094001} {\bibfield
  {journal} {\bibinfo  {journal} {Phys. Rev. D}\ }\textbf {\bibinfo {volume}
  {87}},\ \bibinfo {pages} {094001} (\bibinfo {year} {2013})},\ \Eprint
  {http://arxiv.org/abs/1302.4917} {arXiv:1302.4917 [hep-lat]} \BibitemShut
  {NoStop}%
\bibitem [{\citenamefont {Duane}\ \emph {et~al.}(1987)\citenamefont {Duane},
  \citenamefont {Kennedy}, \citenamefont {Pendleton},\ and\ \citenamefont
  {Roweth}}]{Duane:1987de}%
  \BibitemOpen
  \bibfield  {author} {\bibinfo {author} {\bibfnamefont {S.}~\bibnamefont
  {Duane}}, \bibinfo {author} {\bibfnamefont {A.}~\bibnamefont {Kennedy}},
  \bibinfo {author} {\bibfnamefont {B.}~\bibnamefont {Pendleton}}, \ and\
  \bibinfo {author} {\bibfnamefont {D.}~\bibnamefont {Roweth}},\ }\href
  {\doibase 10.1016/0370-2693(87)91197-X} {\bibfield  {journal} {\bibinfo
  {journal} {Phys. Lett. B}\ }\textbf {\bibinfo {volume} {195}},\ \bibinfo
  {pages} {216} (\bibinfo {year} {1987})}\BibitemShut {NoStop}%
\bibitem [{\citenamefont {Omelyan}\ \emph {et~al.}(2002)\citenamefont
  {Omelyan}, \citenamefont {Mryglod},\ and\ \citenamefont
  {Folk}}]{PhysRevE.65.056706}%
  \BibitemOpen
  \bibfield  {author} {\bibinfo {author} {\bibfnamefont {I.~P.}\ \bibnamefont
  {Omelyan}}, \bibinfo {author} {\bibfnamefont {I.~M.}\ \bibnamefont
  {Mryglod}}, \ and\ \bibinfo {author} {\bibfnamefont {R.}~\bibnamefont
  {Folk}},\ }\href {\doibase 10.1103/PhysRevE.65.056706} {\bibfield  {journal}
  {\bibinfo  {journal} {Phys. Rev. E}\ }\textbf {\bibinfo {volume} {65}},\
  \bibinfo {pages} {056706} (\bibinfo {year} {2002})}\BibitemShut {NoStop}%
\bibitem [{\citenamefont {Karthik}(2014)}]{Karthik:2014nta}%
  \BibitemOpen
  \bibfield  {author} {\bibinfo {author} {\bibfnamefont {N.}~\bibnamefont
  {Karthik}},\ }\href@noop {} {\  (\bibinfo {year} {2014})},\ \Eprint
  {http://arxiv.org/abs/1401.1072} {arXiv:1401.1072 [hep-lat]} \BibitemShut
  {NoStop}%
\bibitem [{\citenamefont {Gusken}\ \emph {et~al.}(1989)\citenamefont {Gusken},
  \citenamefont {Low}, \citenamefont {Mutter}, \citenamefont {Sommer},
  \citenamefont {Patel},\ and\ \citenamefont {Schilling}}]{Gusken:1989ad}%
  \BibitemOpen
  \bibfield  {author} {\bibinfo {author} {\bibfnamefont {S.}~\bibnamefont
  {Gusken}}, \bibinfo {author} {\bibfnamefont {U.}~\bibnamefont {Low}},
  \bibinfo {author} {\bibfnamefont {K.}~\bibnamefont {Mutter}}, \bibinfo
  {author} {\bibfnamefont {R.}~\bibnamefont {Sommer}}, \bibinfo {author}
  {\bibfnamefont {A.}~\bibnamefont {Patel}}, \ and\ \bibinfo {author}
  {\bibfnamefont {K.}~\bibnamefont {Schilling}},\ }\href {\doibase
  10.1016/S0370-2693(89)80034-6} {\bibfield  {journal} {\bibinfo  {journal}
  {Phys. Lett. B}\ }\textbf {\bibinfo {volume} {227}},\ \bibinfo {pages} {266}
  (\bibinfo {year} {1989})}\BibitemShut {NoStop}%
\bibitem [{\citenamefont {Bali}\ \emph {et~al.}(2016)\citenamefont {Bali},
  \citenamefont {Lang}, \citenamefont {Musch},\ and\ \citenamefont
  {Sch\"afer}}]{Bali:2016lva}%
  \BibitemOpen
  \bibfield  {author} {\bibinfo {author} {\bibfnamefont {G.~S.}\ \bibnamefont
  {Bali}}, \bibinfo {author} {\bibfnamefont {B.}~\bibnamefont {Lang}}, \bibinfo
  {author} {\bibfnamefont {B.~U.}\ \bibnamefont {Musch}}, \ and\ \bibinfo
  {author} {\bibfnamefont {A.}~\bibnamefont {Sch\"afer}},\ }\href {\doibase
  10.1103/PhysRevD.93.094515} {\bibfield  {journal} {\bibinfo  {journal} {Phys.
  Rev. D}\ }\textbf {\bibinfo {volume} {93}},\ \bibinfo {pages} {094515}
  (\bibinfo {year} {2016})},\ \Eprint {http://arxiv.org/abs/1602.05525}
  {arXiv:1602.05525 [hep-lat]} \BibitemShut {NoStop}%
\bibitem [{\citenamefont {Karthik}\ and\ \citenamefont
  {Narayanan}(2016{\natexlab{b}})}]{Karthik:2016ppr}%
  \BibitemOpen
  \bibfield  {author} {\bibinfo {author} {\bibfnamefont {N.}~\bibnamefont
  {Karthik}}\ and\ \bibinfo {author} {\bibfnamefont {R.}~\bibnamefont
  {Narayanan}},\ }\href {\doibase 10.1103/PhysRevD.94.065026} {\bibfield
  {journal} {\bibinfo  {journal} {Phys. Rev. D}\ }\textbf {\bibinfo {volume}
  {94}},\ \bibinfo {pages} {065026} (\bibinfo {year} {2016}{\natexlab{b}})},\
  \Eprint {http://arxiv.org/abs/1606.04109} {arXiv:1606.04109 [hep-th]}
  \BibitemShut {NoStop}%
\bibitem [{\citenamefont {Mastropas}\ and\ \citenamefont
  {Richards}(2014)}]{Mastropas:2014fsa}%
  \BibitemOpen
  \bibfield  {author} {\bibinfo {author} {\bibfnamefont {E.~V.}\ \bibnamefont
  {Mastropas}}\ and\ \bibinfo {author} {\bibfnamefont {D.~G.}\ \bibnamefont
  {Richards}} (\bibinfo {collaboration} {Hadron Spectrum}),\ }\href {\doibase
  10.1103/PhysRevD.90.014511} {\bibfield  {journal} {\bibinfo  {journal} {Phys.
  Rev. D}\ }\textbf {\bibinfo {volume} {90}},\ \bibinfo {pages} {014511}
  (\bibinfo {year} {2014})},\ \Eprint {http://arxiv.org/abs/1403.5575}
  {arXiv:1403.5575 [hep-lat]} \BibitemShut {NoStop}%
\bibitem [{\citenamefont {Verbaarschot}\ and\ \citenamefont
  {Zahed}(1994)}]{Verbaarschot:1994ip}%
  \BibitemOpen
  \bibfield  {author} {\bibinfo {author} {\bibfnamefont {J.}~\bibnamefont
  {Verbaarschot}}\ and\ \bibinfo {author} {\bibfnamefont {I.}~\bibnamefont
  {Zahed}},\ }\href {\doibase 10.1103/PhysRevLett.73.2288} {\bibfield
  {journal} {\bibinfo  {journal} {Phys. Rev. Lett.}\ }\textbf {\bibinfo
  {volume} {73}},\ \bibinfo {pages} {2288} (\bibinfo {year} {1994})},\ \Eprint
  {http://arxiv.org/abs/hep-th/9405005} {arXiv:hep-th/9405005} \BibitemShut
  {NoStop}%
\bibitem [{\citenamefont {Karpie}\ \emph {et~al.}(2018)\citenamefont {Karpie},
  \citenamefont {Orginos},\ and\ \citenamefont
  {Zafeiropoulos}}]{Karpie:2018zaz}%
  \BibitemOpen
  \bibfield  {author} {\bibinfo {author} {\bibfnamefont {J.}~\bibnamefont
  {Karpie}}, \bibinfo {author} {\bibfnamefont {K.}~\bibnamefont {Orginos}}, \
  and\ \bibinfo {author} {\bibfnamefont {S.}~\bibnamefont {Zafeiropoulos}},\
  }\href {\doibase 10.1007/JHEP11(2018)178} {\bibfield  {journal} {\bibinfo
  {journal} {JHEP}\ }\textbf {\bibinfo {volume} {11}},\ \bibinfo {pages} {178}
  (\bibinfo {year} {2018})},\ \Eprint {http://arxiv.org/abs/1807.10933}
  {arXiv:1807.10933 [hep-lat]} \BibitemShut {NoStop}%
\bibitem [{\citenamefont {Appelquist}\ \emph {et~al.}(2020)\citenamefont
  {Appelquist} \emph {et~al.}}]{Appelquist:2020xua}%
  \BibitemOpen
  \bibfield  {author} {\bibinfo {author} {\bibfnamefont {T.}~\bibnamefont
  {Appelquist}} \emph {et~al.} (\bibinfo {collaboration} {Lattice Strong
  Dynamics}),\ }\href@noop {} {\  (\bibinfo {year} {2020})},\ \Eprint
  {http://arxiv.org/abs/2007.01810} {arXiv:2007.01810 [hep-ph]} \BibitemShut
  {NoStop}%
\bibitem [{\citenamefont {Braun}\ \emph {et~al.}(2020)\citenamefont {Braun},
  \citenamefont {Ji},\ and\ \citenamefont {Manashov}}]{Braun:2020zjm}%
  \BibitemOpen
  \bibfield  {author} {\bibinfo {author} {\bibfnamefont {V.}~\bibnamefont
  {Braun}}, \bibinfo {author} {\bibfnamefont {Y.}~\bibnamefont {Ji}}, \ and\
  \bibinfo {author} {\bibfnamefont {A.}~\bibnamefont {Manashov}},\ }\href@noop
  {} {\  (\bibinfo {year} {2020})},\ \Eprint {http://arxiv.org/abs/2011.04533}
  {arXiv:2011.04533 [hep-ph]} \BibitemShut {NoStop}%
\end{thebibliography}%

\end{document}